\documentclass[12pt,english]{article}

\newcommand{\dS}{{\rm dS }}
\newcommand{\AdS}{{\rm AdS }}
\newcommand{\de}{\text{d}}
\newcommand{\so}{{\mathfrak {so}}}

\newcommand{\SO}{{\rm {SO}}}

\newcommand{\SU}{{\rm {SU}}}
\newcommand{\su}{{\mathfrak {su}}}
\newcommand{\spp}{{\mathfrak {sp}}}

\newcommand{\ket}[1]{\bigl| #1 \bigr\rangle}
\newcommand{\bra}[1]{\bigl\langle #1 \bigr|}

\usepackage[latin9]{inputenc}
\usepackage{geometry}
\usepackage{babel}

\usepackage[cbgreek]{textgreek}
\usepackage{verbatim}
\usepackage{prettyref}
\usepackage{amsmath}
\usepackage{amssymb}
\usepackage{mathtools}
\usepackage{mathabx}
\usepackage{mathrsfs}
\usepackage{graphicx}
\usepackage{esint}
\usepackage{physics}
 \usepackage{slashed}
 \usepackage{subcaption}
 \usepackage{tikz}
 \usepackage{pgfplots}
 \pgfplotsset{compat=1.18} 

\usepackage[compress]{cite}
\usepackage{dsfont}
\usepackage{xcolor}
\usepackage{color}
\definecolor{darkgreen}{rgb}{0,0.5,0}
\definecolor{darkblue}{rgb}{0,0,0.6}
\definecolor{purple2}{rgb}{0.4,.2,0.7}
\usepackage[colorlinks=true,citecolor=purple,linkcolor=purple2,urlcolor=black]{hyperref}

\numberwithin{equation}{section}
\numberwithin{figure}{section}
\numberwithin{table}{section}

\def\CG{{\cal G}} 
\def\CD{{\cal D}} 
\def\CH{{\cal H}} 
\def\CQ{{\cal Q}}

\def\CP{{\cal P}}
\def\CK{{\cal K}}
\def\CS{{\cal S}}
\def\CG{{\cal G}} 
 
\def\CH{{\cal H}}

\def\CI{{\cal I}} 
 
\def\IC{{\mathbb C}}
\def\IR{{\mathbb R}}

\def\IN{{\mathbb N}}

\def\one{{\mathds 1}}

\def\tr{\,{\rm tr}\,}

\def\IR{{\mathbb R}}

\newcommand{\rme}{\mathrm e}
\newcommand{\rmd}{\mathrm d}
\newcommand{\rmi}{\mathrm i}

\usepackage{fancyvrb}

\title{\textbf{Coherent spin states and emergent \\de Sitter quasinormal modes}}
\author{\,\href{mailto:k.parmentier@columbia.edu}{Klaas Parmentier}\vspace{0.2cm} \\
\it \small \;Department of Physics, Columbia University,\\
\it \small 538 West 120th Street, New York, NY 10027, USA}
\date{}
\begin{document}
\fontseries{mx}\selectfont	
\maketitle
\begin{abstract}
As a toy model for the microscopic description of matter in de Sitter space, we consider a Hamiltonian acting on the spin-$j$ representation of $\SU(2)$. This is a model with a finite-dimensional Hilbert space, from which quasinormal modes emerge in the large-spin limit. The path integral over coherent spin states can be evaluated at the semiclassical level and from it we find the single-particle de Sitter density of states, including $1/j$ corrections. Along the way, we discuss the use of quasinormal modes in quantum mechanics, starting from the paradigmatic upside-down harmonic oscillator.   
\end{abstract}
\thispagestyle{empty}
\newpage
\thispagestyle{empty}
\addtocounter{page}{-2}
\setcounter{tocdepth}{2}
{\hypersetup{linkcolor=black}\tableofcontents}

\def\nn{\nonumber}

\newpage

\section{Quasinormal microscopy}\label{sec:intro}
The time is near, in which the cosmological constant, minute but strictly positive, will come to dominate the evolution of our cosmos. The resulting accelerated expansion will drive the observable universe towards a semiclassical equilibrium state: our cosmic habitat will become asymptotically indistinguishable from a de Sitter ($\dS$) static patch \cite{gibbons77, loeb, Krauss:2007nt, Anninos:2020hfj}. Interpreting the Gibbons-Hawking entropy \cite{Gibbons:1976ue} to imply a finite Hilbert space, the universe as we know it may well be a small coherent fluctuation in an enormous but finite quantum system \cite{Banks:2000fe}. 


A rather pertinent question is: which finite quantum system? In the absence of a precise formulation of quantum gravity in $\dS$, this question remains open, although various proposals have been outlined in the literature \cite{Li:2001ky, Volovich:2001rt, Parikh:2004wh, Banks:2006rx, Susskind:2021dfc}. Often, but not always, spinors serve as pixel operators on the (stretched) horizon, which then appears as a discrete quantum system to an outside observer \cite{Susskind:1993if, Shaghoulian:2021cef, Shaghoulian:2022fop}. Other works have focused more directly on this  observer, rather than the horizon, arguing that their static patch experience may be reproduced by a (gauged) worldline quantum mechanics at large $N$  \cite{Anninos:2011af, Anninos:2021ydw, Chandrasekaran:2022cip}. Both perspectives are expected to be connected by RG flow, with the observer worldline and cosmological horizon respectively resembling the boundary and black hole horizon in an $\AdS$ setup \cite{Anninos:2011af, Coleman:2021nor, Svesko:2022txo}.

An equally crucial question regards the type of output one can even expect from such a microscopic $\dS$ description \cite{Witten:2001kn}. Precise observables typically require measurements at the boundary of spacetime, but in $\dS$ one cannot take the detector beyond the cosmological horizon of the experimental system  \cite{Banks:2005bm}. It then seems like a standard S-matrix is excluded, even in QFT. At least for light fields, wave packets simply do not separate as they would in flat space, but instead freeze out on super-horizon scales. That the precision of measurements is fundamentally limited also follows from finiteness of the entropy, which requires to give up the exact $\SO(1,D)$ symmetry of $\dS_{D}$ and limits its lifetime to the Poincar\'e recurrence time \cite{Dyson:2002nt, Goheer:2002vf}. One can then imagine a universality class of $\dS$ theories which may make different predictions for unmeasurable quantities (beyond Poincar\'e), but which all agree, within the fundamental bounds on precision, on the early-time local behavior \cite{Banks:2005bm}.

Within this broader context, we want to ask the following simple question: how would a finite quantum system reproduce the $\dS$ quasinormal modes (QNMs)? In this paper we will therefore explore non-relativistic quantum systems with a finite number of degrees of freedom and the following striking feature: the emergence of $\dS$-like QNMs in the limit where the Hilbert space dimension $N$ grows large. At finite $N$, these exponentially decaying modes cannot really exist as eigenfunctions of a Hermitian Hamiltonian. Nonetheless, they do capture the macroscopic effective dynamics with increasing accuracy as $N \to \infty$.

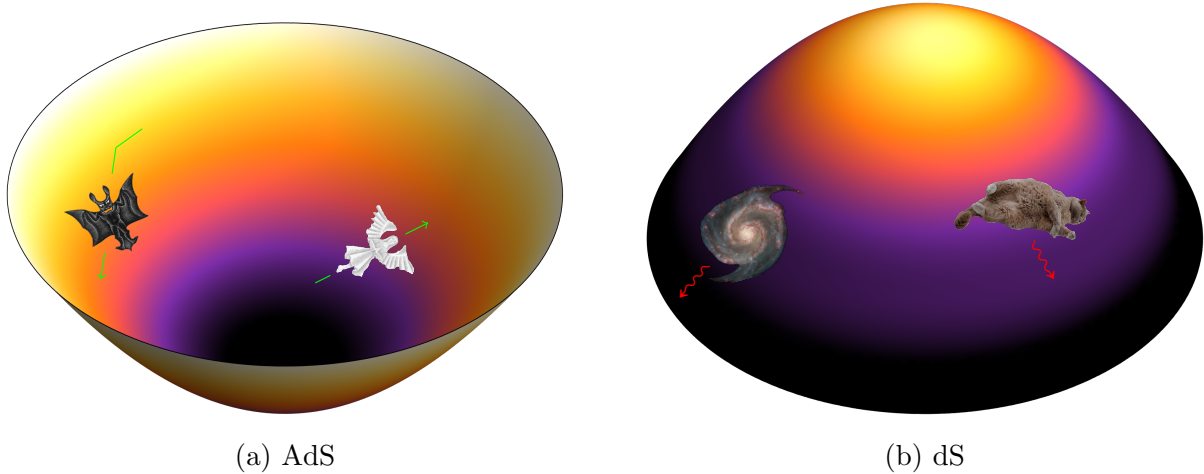
\begin{figure}
    \centering
  \begin{subfigure}{0.45\textwidth}
            \centering
            \begin{tikzpicture}
              \node[inner sep = 0pt] at (7,3) {\includegraphics[width=\textwidth]{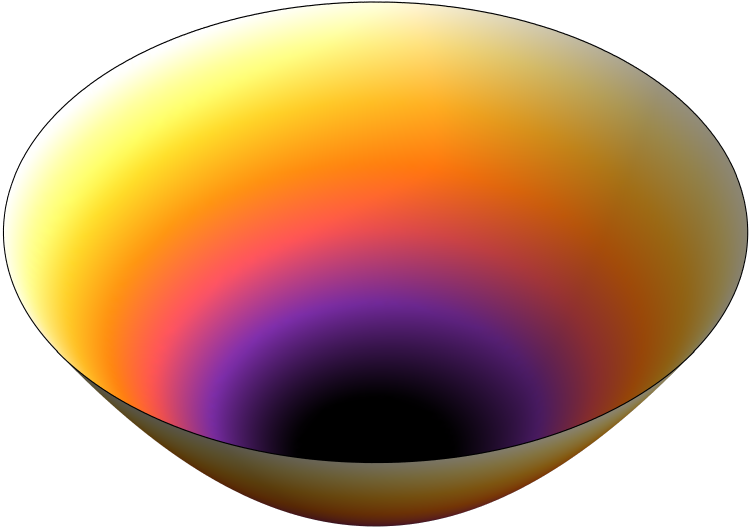}};
               \node at (4.6,3) {\includegraphics[width=0.22\textwidth]{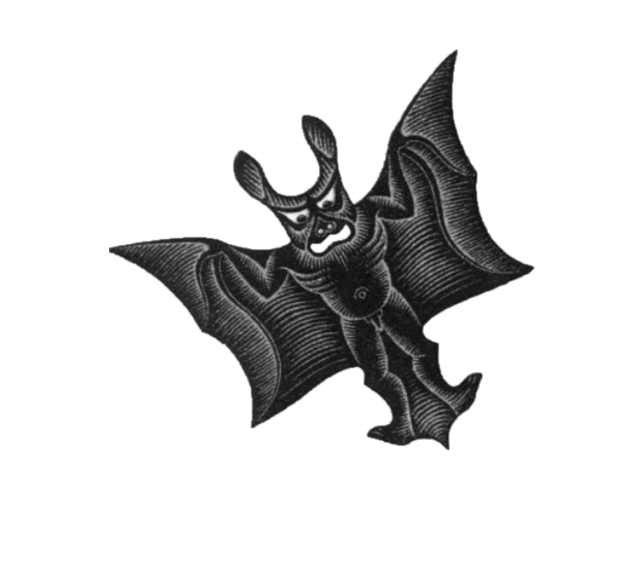}};\node at (8.2,2.7) {\includegraphics[width=0.15\textwidth]{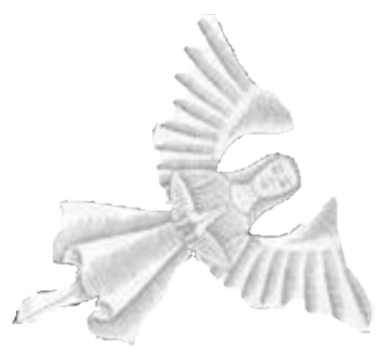}};
               \draw[green]  (5.1,4.2)   -- (4.75,3.95);
               \draw[green]  (4.75,3.95)   -- (4.7,3.6);
               \draw[green,->]  (4.6,2.55)   -- (4.55,2.2);
               \draw[green,->]  (8.6,2.8)   -- (8.9,2.95);
               \draw[green]  (7.4,2.15)   -- (7.6,2.25);
            \end{tikzpicture}
            \caption[]%
            {{\small  $\AdS$}} 
        \end{subfigure}
        \hspace{0.8cm}
        \begin{subfigure}{0.45\textwidth}  
            \centering 
            \begin{tikzpicture}
              \node[inner sep = 0pt] at (7,3) {\includegraphics[width=\textwidth]{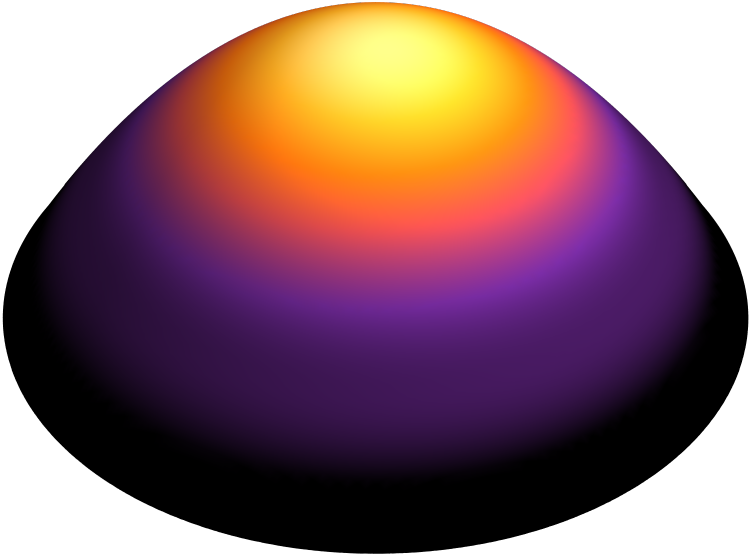}};
               \node at (8.3,3) {\includegraphics[width=0.26\textwidth]{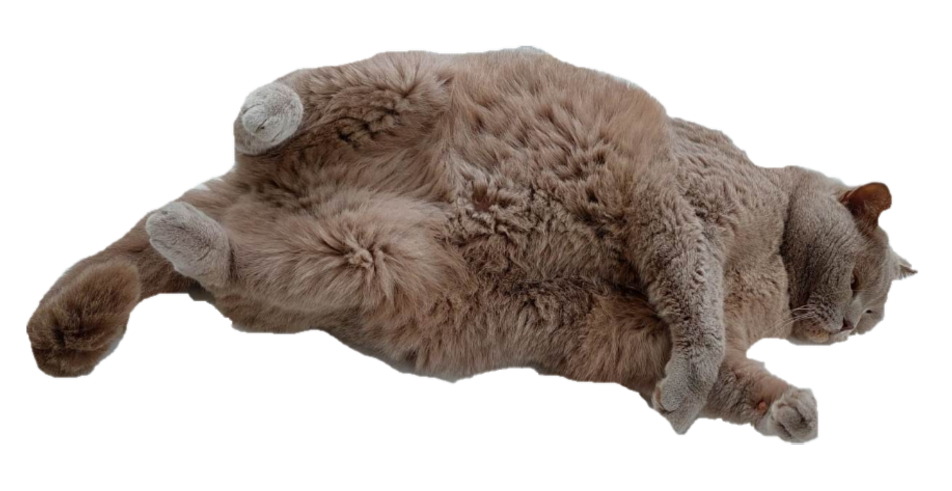}};\node[rotate=15] at (4.75,2.7) {\includegraphics[width=0.2\textwidth]{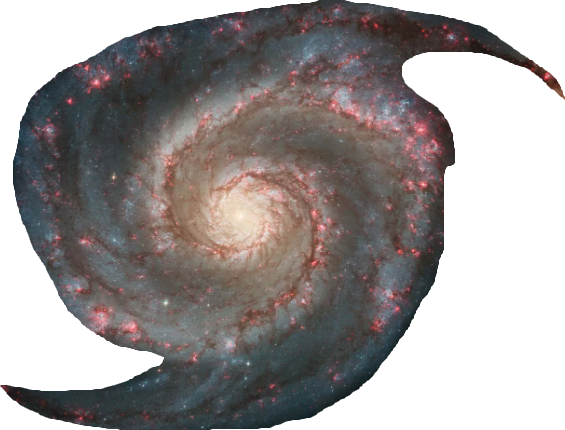}};
               \node[red, rotate=300] at (8.55,2.3) {\text{$\rightsquigarrow$}};
               \node[red, rotate=230] at (3.9,2.05) {\text{$\rightsquigarrow$}};
            \end{tikzpicture}
            \caption[]%
            {{\small  $\dS$}}
        \end{subfigure}
    \caption{Small excitations in $\AdS$ are trapped in a harmonic oscillator potential and described by oscillating normal modes. Fluctuations in $\dS$ instead experience an inverted potential, that of an upside-down harmonic oscillator. Here, quasinormal modes play a more natural role, describing dissipation towards the cosmological horizon.}
    \label{fig:idea}
\end{figure}
\newpage
Our focus on QNMs originates with static patch excitations being at most resonances, due to the presence of the cosmological horizon \cite{Banks:2005bm}. As fig.\,\ref{fig:idea} illustrates, fluctuations in $\AdS$ and $\dS$ lead rather different lives. The same is expected to hold for the bulk manifestation of simple microscopic operators. While in $\AdS$ holography they describe boundary perturbations, in $\dS$ they are likely to describe horizon disturbances: the low-lying QNMs \cite{Susskind:2022dfz}. Moreover, the QNM basis is manifestly $\dS$-invariant, and it is hard to imagine a microscopic model reproducing their spectrum without also to some extent the dynamics, and vice versa. 

The issues with defining a $\dS$ S-matrix are reminiscent of the situation in CFT, since both involve a continuous density of states. What sets the $\dS$ spectrum apart are its particular QNM poles \cite{Anninos:2020hfj, Loparco:2023rug}. Although these do not fit into a standard Hilbert space framework, it is not inconceivable for them to take over the role usually played by asympotic single-particle states. Quantizing QNMs requires a non-standard norm -- the KG-norm being infinite \cite{Ng:2012xp} -- but does lead to a natural definition of the Euclidean vacuum \cite{Jafferis:2013qia}. This was used in \cite{Cotler:2023xku} to argue for $\dS$ emerging from two entangled boundary CFTs, as already advocated for in \cite{Balasubramanian:2002zh}. 

QNM-oriented approaches have indeed been fruitful before. For instance, as Fourier transform of the density of states, the character counts QNM degeneracies, and efficiently encodes the 1-loop partition function \cite{Anninos:2020hfj}. There is of course a vast literature on QNMs \cite{Berti:2009kk, Konoplya:2011qq}, but in the current context let us just emphasize that their role in organizing the (thermo)dynamics is not limited to $\dS$ \cite{Denef:2009kn, Law:2022zdq} or partition functions alone \cite{Son:2007vk, Hui:2019aox, Dodelson:2023vrw}.  

\newpage

{\textbf{Plan of the paper:}} In sec.\,\ref{sec:qm}, the aim is to demystify QNMs a bit more, explaining their role even in ordinary quantum mechanics. We will see that in the simplest example, the upside-down (or inverted) harmonic oscillator, inserting the QNM completeness relation is equivalent to doing a Taylor series expansion. It also allows for a most convenient evaluation of the character $\tr \rme^{-\rmi t H}$. Another application consists of perturbatively determining resonance spectra. Next, in sec.\,\ref{sec:desitter}, we describe the phase space of a massive particle moving in $\dS$, deferring details to app.\,\ref{app:phase}. From it, we are naturally led to a conformal boundary quantum mechanics that reproduces two copies of the inverted oscillator QNMs; one each for the northern and southern observers. The combination of these yields the principal series character. In sec.\,\ref{sec:spin}, we present a toy model for the microscopic description of matter in $\dS$.\newline The model consists of a Hermitian Hamiltonian acting on the spin-$j$ representation of $\SU(2)$. It shares key features with the boundary Hamiltonian, and at large spin we expect it to yield the $\dS$ spectrum. This convergence is confirmed by numerical diagonalization. In sec.\,\ref{sec:semicl}, we proceed analytically, using coherent spin states and holomorphic wave functions. These let us characterize the exact spin-model spectrum in terms of Heun polynomials, and shed light on the emergent QNMs. Moreover, they allow us to understand the large-$j$ limit as a semiclassical one, in which we can evaluate the path integral for the coarse-grained character. From it, we retrieve both the leading $\dS$ result and first $1/j$ correction. Finally, in sec.\,\ref{sec:disc}, we mention several directions for future research, including possible generalizations towards interacting multi-particle microscopic models for matter in $\dS$. \\

\section{Quasinormal quantum mechanics}\label{sec:qm}
The upshot of this section is that resonances are as fundamental to the quantum mechanics of the upside-down harmonic oscillator as energy eigenstates are to the ordinary oscillator.
Despite early studies of its propagator and density of states \cite{Barton:1984ey}, as well as several general results on QNM completeness \cite{Parravicini:1979xd, Ching:1995rt, Ching:1998mxl, Nollert:1998ys, Beyer:1998nu}, it seems like the upside-down oscillator has remained somewhat underappreciated. This point was also made a few years ago by \cite{Hegde:2018xub, Subramanyan:2020fmx}, with emphasis on its broad relevance as a prototype to study physics near unstable equilibrium, and with applications ranging from black holes to quantum Hall physics. Time-dependent variations, which arise for instance in the description of superhorizon modes of light fields in Mukhanov-Sasaki variables, were studied in \cite{Rajeev:2017uwk}. The upside-down oscillator also appeared very recently as toy model in the context of double-cone wormholes \cite{Chen:2023hra}. 

Let us begin our discussion by defining the {\it character}, a quantity which will be of interest throughout the rest of the paper. For 
any given Hamiltonian $H$, it is defined as
\begin{align}
  \chi(t) \equiv {\rm tr} \, \rme^{-\rmi t H} \, .
\end{align}
For {\it elliptic}/stable systems like an ordinary harmonic oscillator it can be calculated simply by summing over the discrete energy spectrum (with a suitable $\rmi\epsilon$ prescription if needed). For a single oscillator with frequency $\omega=1$, taking $\omega\to \omega - \rmi \epsilon$ for $t>0$, one finds
\begin{align} \label{harmoscchi}
\chi(t)=  \frac{\rme^{-(\epsilon + \rmi) t/2}}{1-\rme^{-(\epsilon + \rmi) t}} \; \qquad (t>0) \, .
\end{align}

For ordinary oscillators, this Fourier transformed density of states is not particularly  convenient. On the other hand, {\it hyperbolic}/unstable systems, like the upside-down oscillator, have a continuous energy spectrum, and calculating the trace by integrating over the spectral density seems trickier. However, the final expression will simply take the form of \eqref{harmoscchi} with {\it imaginary} frequency $\omega \to -\rmi \omega$. Indeed, the main result of this section is an orthonormal pairing of resonances $\ket{\psi_n}$ with anti-resonances $|\tilde{\psi}_n\rangle$, leading to completeness relation \eqref{invhodecompid}. This trivializes the calculation of the character, as follows, for $t > 0$:
\begin{align} \label{tr1}
 \chi(t) &= \sum_n {\tr} \bigl( \rme^{-\rmi t  H} |\psi_n\rangle \langle \tilde\psi_ n| \bigr)
 = \sum_n\, \langle \tilde\psi_n| \rme^{-\rmi t  H} |\psi_n\rangle  = \sum_n \,\rme^{- (\frac{1}{2}+n)t} = \frac{\rme^{-t/2}}{|1-\rme^{-t}|} \, , 
\end{align}
and similarly for $t<0$:
\begin{align} \label{tr2}
 \chi(t) = \sum_n {\tr} \bigl( \rme^{-\rmi t  H} |\tilde\psi_n\rangle \langle \psi_ n| \bigr)
 = \sum_n\, \langle \psi_n| \rme^{-\rmi t  H} |\tilde \psi_n\rangle  = \sum_n\, \rme^{ (\frac{1}{2}+n)t} = \frac{\rme^{-t/2}}{|1-\rme^{-t}|}\,.
\end{align}
The different insertions for $t>0$ and $t<0$ are needed because forward (positive-$t$) time evolution converges on the resonance expansion, while backward (negative-$t$) time evolution converges on the anti-resonance expansion. 

\subsection{Upside-down harmonic oscillator}\label{sec:iho}
The character of the ordinary harmonic oscillator \eqref{harmoscchi}, and its generalization to stable systems with a spectrum of {\it normal} modes $\omega_n>0$, has the obvious interpretation of counting energy eigenstates. For unstable systems like the upside-down oscillator, we expect to count {\it resonances} instead, as manifested by \eqref{tr1}. To arrive at this point, we will begin by taking a closer look at what happens to the construction of the  harmonic oscillator energy eigenstates upon analytic continuation $\omega \to \pm \rmi\omega$.

\subsubsection{Analytic continuation of the harmonic oscillator}
The upside-down harmonic oscillator has a Hamiltonian 
\begin{equation}\label{udho}
 H=\frac{1}{2} (\hat{p}^2 - \omega^2 \hat{x}^2)\, ,
\end{equation}
which can be obtained from the usual oscillator by $\omega \to \pm \rmi \omega$. Similarly, we can continue the creation and annihilation operators. If we pick $\omega \to  -\rmi\omega$, we obtain for instance
\begin{align} \label{bpmdef}
 b_\pm =  \frac{\hat p \pm \omega \hat x}{\sqrt{2\omega/\rmi}}\, , \qquad [b_-,\,b_+] = 1 \, , \qquad 
 [ H,\,b_{\pm}] = \mp \, \rmi \omega \, b_\pm \, .
\end{align}
We then define the primary resonance $|\psi_0\rangle$ to be the state annihilated by $b_-$:
\begin{equation}\label{eq:ihopr}
  b_- |\psi_0\rangle = 0\, , \qquad \psi_0(x) = \frac{ \rme^{\rmi\omega x^2/2}}{(\rmi \pi)^{1/4}}  \,,
\end{equation}
where we fixed the prefactor by continuation from the harmonic oscillator. On top of $|\psi_0\rangle$, we can build a tower of excitations by acting repeatedly with $b_+$:
\begin{align}\label{eq:res}
 |\psi_n\rangle = \frac{b^n_+}{\sqrt{n!}} \, |\psi_0\rangle \,, \qquad   H |\psi_n\rangle = \omega_n \, |\psi_n \rangle \, , \qquad \omega_n =  -\rmi\omega \, (\tfrac{1}{2} + n) \, .
\end{align}

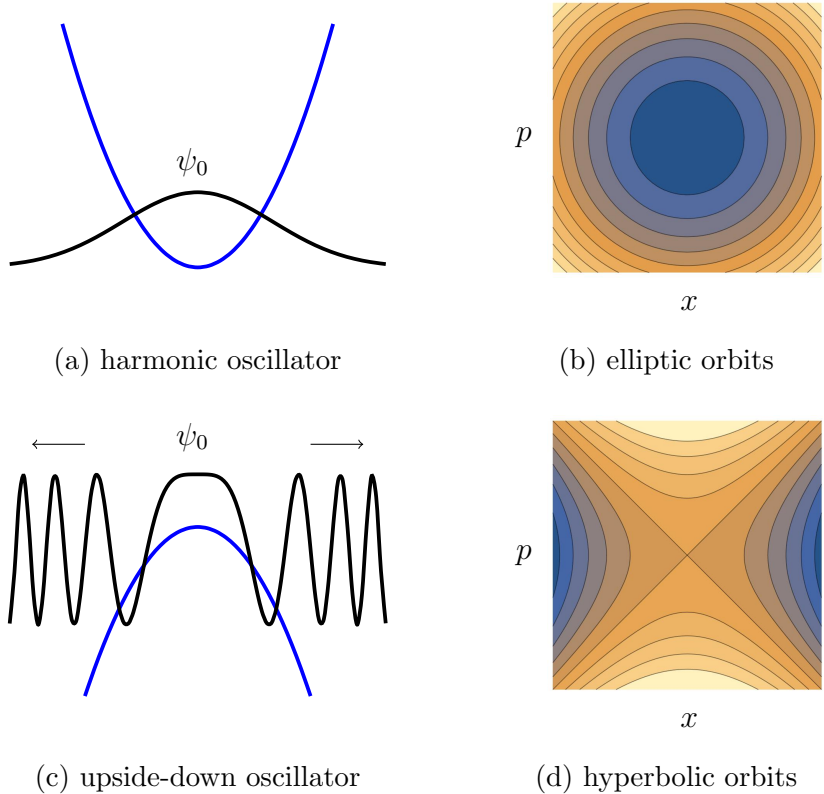
\begin{figure}[ht]
    \centering
      \begin{subfigure}{0.3\textwidth}
            \centering
            \begin{tikzpicture}[remember picture, overlay]
        \draw[blue, line width = 0.50mm]   plot[smooth,domain=-1.8:1.8] (\x, {0.7+(\x)^2});
        \draw[black, line width = 0.50mm]   plot[smooth,domain=-2.5:2.5] (\x, {0.7+exp(-0.5*(\x)^2)});
        \node at (0,2.1) {\text{$\psi_0$
        }};
        \end{tikzpicture}
            \caption[]%
            {{\small  harmonic oscillator }} 
        \end{subfigure}
        \hspace{1cm}
        \begin{subfigure}{0.3\textwidth}  
            \centering 
            \begin{tikzpicture}
              \node[inner sep = 0pt] at (7,3) {\includegraphics[width=0.75\textwidth]{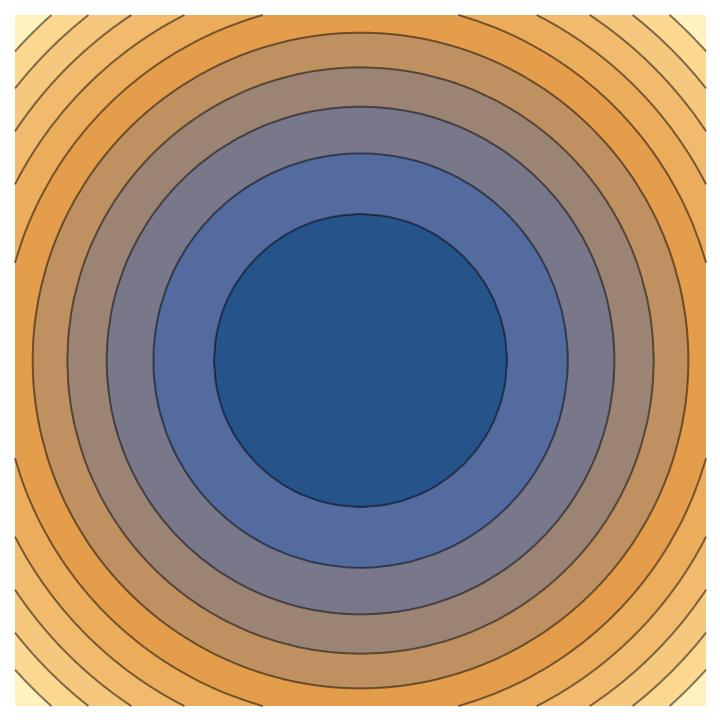}};
               \node at (4.9,3) {\text{$p$
        }};\node at (7.1,0.8) {\text{$x$
        }};
            \end{tikzpicture}
            \caption[]%
            {{\small elliptic orbits}}
        \end{subfigure}\\\vspace{0.4cm}
          \begin{subfigure}{0.3\textwidth}
            \centering
            \begin{tikzpicture}[remember picture, overlay]
        \draw[blue, line width = 0.50mm]   plot[smooth,domain=-1.5:1.5] (\x, {2.8-(\x)^2});
        \draw[black, line width = 0.50mm,samples=80]   plot[smooth,domain=-2.5:2.5] (\x, {2.5+cos(200*(\x)^2)});
        \node at (0,4) {\text{$\psi_0$
        }};
        \draw[->]  (1.5,3.9)   -- (2.2,3.9);
        \draw[<-]  (-2.2,3.9) -- (-1.5,3.9);
        \end{tikzpicture}
            \caption[]%
            {{\small  upside-down oscillator}} 
        \end{subfigure}
        \hspace{1cm}
        \begin{subfigure}{0.3\textwidth}  
            \centering 
            \begin{tikzpicture}
              \node[inner sep = 0pt] at (7,3) {\includegraphics[width=0.75\textwidth]{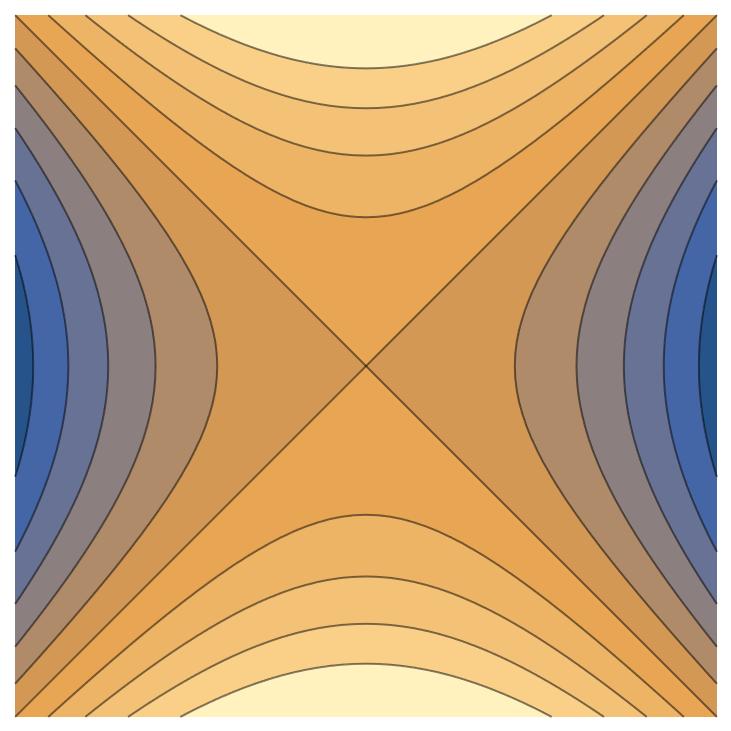}};
               \node at (4.9,3) {\text{$p$
        }};\node at (7.1,0.8) {\text{$x$
        }};
            \end{tikzpicture}
            \caption[]%
            {{\small  hyperbolic orbits}}\label{fig:hyporb}
        \end{subfigure}
    \caption{The standard harmonic oscillator (a) has a normalizable ground state, and contours of constant energy in phase space (b) correspond to elliptic orbits. The inverted oscillator (c) on the other hand, has a non-normalizable primary resonance \eqref{eq:ihopr}, the real part of which is plotted above. It is purely outgoing on either side of the potential hill. The contours of constant energy in phase space (d) correspond to hyperbolic orbits.}
    \label{fig:oscillators}
\end{figure}

Evidently, the $\psi_n(x)$, being eigenfunctions of the Hermitian operator $ H$ with {\it imaginary} eigenvalues, or more directly, being polynomials multiplied by a pure phase factor, are not in the Hilbert space; they are non-normalizable. Instead of bound state wave functions with increasingly fast oscillatory time evolution, they represent a tower of {\it resonance} wave functions with increasingly exponentially damped time evolution: 
\begin{align}\label{eq:restime}
 \rme^{-\rmi t  H } |\psi_n\rangle = \rme^{-(\frac12 + n)\omega t} \, |\psi_n\rangle \,.
\end{align}
Note that since $\hat p = \omega \hat x$ when acting on $\psi_0(x) \;\propto \;\, \rme^{\rmi \omega x^2/2}$, these resonances have purely {\it outgoing} asymptotic momentum, pointing away from the origin, on both sides of the potential hill. This feature is also the defining property of quasinormal modes. The primary resonance together with the classical phase space orbits are shown in fig.\,\ref{fig:oscillators}.

Alternatively, we could have considered the continuation $\omega \to + \rmi \omega$. Repeating the steps above then leads to tower of excitations $|\tilde \psi_n\rangle$ which represent {\it anti-resonances}, exponentially damped when evolving {\it backwards} in time:
\begin{align}\label{eq:anti}
  \langle \tilde\psi_0|x\rangle =\langle x|\psi_0\rangle\, ,\quad \langle \tilde\psi_n| =  \langle \tilde\psi_0| \, \frac{b_-^n}{\sqrt{n!}}\, ,\quad \rme^{-\rmi t  H } |\tilde\psi_n\rangle = \rme^{(\frac{1}{2} + n)\omega t} \, |\tilde\psi_n\rangle  \, .
\end{align}
Since $\hat p = - \omega \hat x$ when acting on $\tilde \psi_ 0(x) \;\propto \;\, \rme^{-\rmi \omega x^2/2}$, these anti-resonances have purely {\it ingoing} asymptotic momentum, pointing towards the origin on both sides of the hill. 

More explicitly, putting $\omega=1$, the position space wave functions are given by
\begin{equation}\label{eq:explicit}
\begin{split}
 \psi_n(x) &= \frac{1}{\sqrt{n!}} \, \Bigl(\frac{\rmi}{2} \Bigr)^{n/2} \, \bigl(-\rmi \partial_x+x\bigr)^n \, \frac{\rme^{\rmi x^2/2}}{(\rmi \pi)^{1/4}} =\frac{1}{\sqrt{n!}}\,(\frac{\rmi}{\sqrt{2}})^n \,\mathsf{H}_n(\frac{x}{\sqrt{\rmi}})\,\frac{\rme^{\rmi  x^2/2}}{(\rmi \pi)^{1/4}}\,,  \\
 \tilde\psi_n(x) &= \frac{1}{\sqrt{n!}} \, \Bigl(\frac{1}{2\rmi} \Bigr)^{n/2} \, \bigl(-\rmi \partial_x-x\bigr)^n \, \frac{ \rme^{-\rmi x^2/2}}{(\pi/\rmi)^{1/4}}
 = (-1)^n \, \psi_n(x)^* \;,
 \end{split}
\end{equation} 

where $\mathsf{H}_n$ is the $n$-th Hermite polynomial. For $n=1,2,3,4$ this gives 
\begin{equation}\label{eq:resonances}
 \begin{array}{c|c}
  n & \;\psi_n/\psi_0 \\
  \hline
 1 & \;\sqrt{2 \rmi} \, x \\
 2 & \;\frac{\rmi}{\sqrt{2}} (2 x^2-\rmi) \\
 3 &\; \frac{\rmi^{3/2}}{\sqrt{3}} (2 x^3-3 \rmi x) \\
 4 &\; \frac{-1}{2\sqrt{6}} (4 x^4 - 12 \rmi x^2 - 3) 
 \end{array}_{\;\;\textstyle{.}}
\end{equation}

\subsubsection{Decomposition of the identity}

Given \eqref{eq:res} and \eqref{eq:anti}, it follows that the resonance raising (lowering) operator is the anti-resonance lowering (raising) operator. The two types of modes are mapped into each other by $\omega \to -\omega$. For the ordinary oscillator, the analogous sign flip maps  $\rme^{-\omega x^2/2} \to \rme^{+\omega x^2/2}$, which is non-normalizable and therefore discarded. In our case, both towers are equally non-normalizable, so there is no reason to keep one and not the other. 

In fact, one might therefore think that we should keep neither. However, despite their non-normalizability, these resonances do govern the dynamics of normalizable wave packets moving in the upside-down potential, much like the energy eigenstates do for the ordinary oscillator. The time-evolution kernel for the Hamiltonian \eqref{eq:ihopr} with $\omega=1$
\begin{align}
  \bra{x}\rme^{-\rmi t H}\ket{y} = \frac{1}{\sqrt{2\pi \rmi \sinh t}} \,  \exp\biggl( \rmi \, \frac{\cosh t (x^2+y^2 ) - 2x y }{2\sinh t}  \biggr)  \, ,
\end{align} 
can be expanded out in powers of $\rme^{-t}$ when $t>0$:
\begin{align} \label{resexpkernelK}
 \bra{x}\rme^{-\rmi t H}\ket{y} = 
 \tfrac{1}{\sqrt{\rmi \pi }} \,  \rme^{-t/2}  \, \rme^{\rmi x^2/2} \,  \rme^{\rmi y^2/2} \bigl(
 1-2 \rmi \, \rme^{-t} \, x  y-\tfrac{1}{2} \rme^{-2 t} (2 x^2-\rmi) (2
   y^2-\rmi) + \cdots \bigr) \,.
\end{align}
 The different terms are recognized as the time-evolved resonances \eqref{eq:resonances}. Despite their non-normalizability, they do give rise to a fast-converging expansion of the late-time evolution of physically sensible quantities, such as the overlap between a Gaussian $\xi_b$ centered at $b$ and its time-evolved counterpart $\xi_a$:
\begingroup 
\addtolength\jot{5pt}
\begin{align} 
 \langle \xi_b|\rme^{-\rmi t H}|\xi_a \rangle &= \frac{1}{\sqrt{\pi}}\int \rmd x \,\rmd y \, \bra{x}\rme^{-\rmi t H}\ket{y} \,\rme^{-(x-b)^2/2} \,\rme^{-(y-a)^2/2}\nn\\
 &= \sum_{n \geq 0} \rme^{-(\frac{1}{2}+n) t} \langle \xi_b|\psi_n\rangle \langle \tilde \psi_n|\xi_a\rangle \, \qquad (t>0) \\
 & = \sqrt{2}\,\rme^{-t/2}  \rme^{\frac{\rmi-1}{4} (a^2+b^2)} 
 \bigl( 
   1
   +\rme^{-t} a  b \, 
   + \rme^{-2t} \tfrac{(a^2-\rmi)(b^2-\rmi)}{2} 
   +\rme^{-3t} \tfrac{(a^3-3 \rmi a)(b^3-3 \rmi b)}{6}  + \cdots \bigr)\,. \nn
\end{align}
\endgroup
This can be repeated for evolution backwards in time, $t<0$, now expanding the kernel in powers of $\rme^t$. This reverses the role of the (anti-)resonances, resulting in 
\begin{align} 
 \langle \xi_b|\rme^{-\rmi t H}|\xi_a \rangle &= \sum_{n \geq 0} \rme^{(\frac{1}{2}+n) t} \langle \xi_b|\tilde \psi_n\rangle \langle \psi_n|\xi_a\rangle    \qquad (t<0)\,.
\end{align}

Although from the explicit expressions \eqref{eq:explicit} it might not look like 
\begin{align}\label{eq:ortho}
 \langle \tilde \psi_n|\psi_m \rangle = \int \rmd x \, (-1)^n \psi_n(x) \, \psi_m(x) = \delta_{nm}\, ,
\end{align} 
it nonetheless does in a distributional sense. Inserting a convergence factor like $\rme^{-\epsilon\, x^2}$, calculating the integral, and then taking the limit $\epsilon \to 0$, leads unambiguously to the  finite result above. Clearly then, there is a sense in which the standard orthonormal decomposition in energy eigenstates of the ordinary oscillator generalizes to an analogous decomposition into (anti-)resonances of the upside-down oscillator:
\begin{align} \label{invhodecompid}
  \langle \tilde\psi_n|\psi_m \rangle = \delta_{nm}\,,\quad  \sum_n |\psi_n\rangle \langle \tilde \psi_n|= \one = \sum_n |\tilde \psi_n\rangle \langle \psi_n| \, ,
\end{align}  
with the decomposition on the left applicable to forward time evolution and the one on the right to backward time evolution in the setup described above. A somewhat more general argument is presented in app.\,\ref{app:gendec}. For a mathematically more precise account of resonances and rigged Hilbert spaces we refer the reader to \cite{Parravicini:1979xd}. 

Given the completeness relation \eqref{invhodecompid} and time-dependence \eqref{eq:restime}, it now becomes clear how the calculation of the character $\chi(t)=\tr \rme^{-\rmi t H}$ gets trivialized as claimed in \eqref{tr1} and \eqref{tr2} in the introduction of this section.

\subsubsection{Resonance expansion and Taylor series}\label{sec:taylor}
As an illuminating example of the more general decompositions in app.\,\ref{app:gendec}, consider
\begin{align} \label{cmpexample}
 c_- = \rmi\, \hat p \, , \qquad c_+ = \hat x \, .
\end{align}
Like the $b_\pm$ in \eqref{bpmdef}, the $c_\pm$ are proportional to Hermitian operators. In fact, they can be understood as resonance raising and lowering operators for the Hamiltonian
\begin{equation}\label{eq:hamxp}
      H = \frac12 (\hat x \hat p + \hat p \hat x)\, ,
\end{equation}
related to \eqref{udho} by a canonical transformation. The primary resonances are simply
\begin{align}
  \psi_0(x) = 1, \qquad \tilde\psi_0(x) = \delta(x)\,.
\end{align}
They satisfy \eqref{data}, and the descendant resonances \eqref{phichidef} become 
\begin{align}
 \psi_n(x) = \frac{x^n}{\sqrt{n!}} \, , \qquad \tilde\psi_n(x) 
  = \frac{(-1)^n \delta^{(n)}(x) }{\sqrt{n!}}  \, .
\end{align}
Orthonormality as in \eqref{eq:ortho} and \eqref{decompgen} is  easily checked. Moreover
\begin{align} \label{Taylor}
  \langle \alpha|\psi_n\rangle
= \frac{1}{\sqrt{n!}} \int dx \, \alpha(x)^* x^n  , \qquad 
\langle \tilde\psi_n|\beta\rangle = \frac{1}{\sqrt{n!}} \, \beta^{(n)}(0)\, , 
\end{align}
 for suitable $\alpha,\beta$. The completeness relation in \eqref{decompgen} then boils down to 
\begin{align} \label{eq:taylor2}
 \langle \alpha|\beta\rangle =\sum_n \,\langle \alpha|\psi_n\rangle \langle \tilde\psi_n|\beta\rangle=\sum_n \,\frac{1}{n!}  \int dx \, \alpha(x)^* \,  x^n \beta^{(n)}(0) \, ,
\end{align}
which we recognize as Taylor expanding $\beta(x)$ at $x=0$ and exchanging integral and sum. Whether this exchange is allowed and the decomposition converges depends on $\alpha$ and $\beta$. The precise conditions are more subtle than for the standard decomposition into a discrete set of normalizable states. For example, when
\begin{align} \label{examplealphabeta}
 \alpha(x) = \rme^{-\alpha \, x^2} \, , \qquad \beta(x) = \rme^{-\beta \, x^2} \, ,\qquad  \langle \alpha|\beta \rangle =  \sqrt{\frac{\pi}{\alpha+\beta}} \, ,
\end{align}
with $\alpha, \beta >0$, the decomposition suggested by \eqref{eq:taylor2} is 
\begin{align}
 \sum_n \,\langle \alpha|\psi_n \rangle \langle \tilde\psi_n|\beta\rangle = \frac{1}{\sqrt{\alpha}} \sum_{n\in 2\IN}  \, \frac{\Gamma(\frac{n+1}{2})}{\Gamma(\frac{n}{2}+1)} \left(\frac{-\beta}{\alpha}\right)^{\frac{n}2} \, ,
\end{align}
which converges to \eqref{examplealphabeta} only for $\alpha > \beta$. On the other hand, the analogous  
\begin{align}
 \sum_n \,\langle \alpha|\tilde\psi_n \rangle \langle \psi_n|\beta\rangle = \frac{1}{\sqrt{\beta}} \sum_{n\in 2\IN}  \, \frac{\Gamma(\frac{n+1}{2})}{\Gamma(\frac{n}{2}+1)} \left(\frac{-\alpha}{\beta}\right)^{\frac{n}{2}} \, ,
\end{align}
converges to \eqref{examplealphabeta} only when $\alpha < \beta$. As in \eqref{invhodecompid}, the understanding is that although these decompositions formally look equivalent, mapped into each other by Hermitian conjugation, convergence of one versus the other depends on what they are sandwiched between. For the wave packet time evolution discussed previously, the criterion was whether $t>0$ or not. Since time evolution spreads out wave packets, we can understand why in the example above this criterion translates into whether $\alpha > \beta$ or not. 

\subsubsection{Coherent state basis}\label{sec:cohstate}
As a slight detour, let us consider the coherent state basis  defined in app.\,\ref{app:cohho}, which will prove to be of use in sec.\,\ref{sec:semicl}. On holomorphic wave functions \eqref{eq:holodef}, the Hamiltonian \eqref{eq:hamxp} acts as
\begin{equation}\label{eq:cohiho}
    H \,\psi(u) = \frac{\rmi}{2}(\partial_u^2-u^2) \,\psi(u)\, ,
\end{equation}
while resonance raising and lowering operators are represented by\footnote{This corresponds to $c_- = \hat p$ and $c_+ = \rmi \hat x$. Compared to \eqref{cmpexample}, we moved the $\rmi$ for notational convenience.}
\begin{equation}\label{eq:raiselower}
    c_\pm\, \psi(u)= \frac{1}{\sqrt{2}} (u\mp \partial_u)\,\psi(u)\,.
\end{equation}
One then obtains the following towers of (anti-)resonances:
\begin{equation}\label{eq:cohresbas}
    \psi_n(u) =  \frac{\;2^{1/4}}{\sqrt{2^n n!}}\,\mathsf{H}_n(u) \,\rme^{-u^2/2}, \qquad \tilde{\psi}_n(u) =\frac{2^{1/4}(-\rmi)^n}{\sqrt{2^n n!}}\,\mathsf{H}_n(\rmi u) \,\rme^{u^2/2}\,.
\end{equation}
As before, the (anti-)resonances by themselves are non-normalizable under the inner product \eqref{eq:cohinner}. However, orthonormality in the sense  $\langle \tilde\psi_n|\psi_m \rangle = \delta_{nm}$ still holds, and the advantage of coherent states is that the required integrals 
\begin{equation}
    \langle \tilde\psi_n|\psi_m \rangle = \int \frac{\rmd^2 u}{\pi} \,\rme^{-u\bar u}\, \tilde{\psi}^*_n(\bar u) \,\psi_m (u) = \int \frac{\rmd^2 u}{\pi} \,\rme^{-u\bar u}\, (-1)^n\tilde{\psi}_n(-\bar u) \,\psi_m (u)
\end{equation}
can be calculated without need for an $\epsilon$-prescription. 

In fact, coherent states allow one to directly evaluate the character as the  trace of the time-evolution kernel for the Hamiltonian \eqref{eq:cohiho}, see also \cite{Chen:2023hra}:
\begin{equation}\label{eq:cohstatechi}
    \chi(t) = \int \frac{\rmd^2 u}{\pi} \, \rme^{-u\bar{u}} \, (u| \rme^{-\rmi t H} |\bar{u}) = \frac{\rme^{-t/2}}{|1- \rme^{-t}|}\, ,
\end{equation}
where in holomorphic polarization the kernel is given by
\begin{equation}\label{eq:prop}
    (v| \rme^{-\rmi t H} |\bar{u}) = \sqrt{\sech t}\;  \rme^{v\bar u \sech t}\,\rme^{\frac12( \bar{u}^2 - v^2)\tanh t}\,,
\end{equation}
so that \eqref{eq:cohstatechi} becomes a simple Gaussian integral.

Finally, it is instructive to obtain \eqref{eq:prop} from the path integral \eqref{eq:cohpath}: 
\begin{equation}
 (u_f|\rme^{-\rmi T H} |\bar{u}_i)   = \int \CD \bar u \CD u\,  \rme^{\rmi S}, \quad S = \frac{\rmi}{2} \int^T_0 \rmd t \,\big(\dot{\bar{u}} u- \bar u \dot{u} +u^2- \bar{u}^2  \big) -  \tfrac{\rmi}{2}u(0)\bar{u}_i -\tfrac{\rmi}{2} \bar{u}(T)u_f \, .
\end{equation}
 The classical solution satisfying $u(T) = u_f$  and $\bar{u}(0)= \bar{u}_i$ is given by
\begin{equation}
  u(t) =  -\bar{u}_i \sinh t + (u_f + \bar{u}_i \sinh T) \frac{\cosh t}{\cosh{T}} \,,  \quad \bar{u}(t) = \bar{u}_i \cosh t - (u_f + \bar{u}_i \sinh T)\frac{\sinh t}{\cosh{T}} \, . 
\end{equation}
This illustrates that paths generically get complexified: $u$ and $\bar u$ are not each other's complex conjugate. The on-shell action comes from the boundary terms alone:
\begin{equation}
     S_{\text{cl}}(u_f , \bar{u}_i| T) = -\rmi \bar{u}_i u_f \sech T - \frac{\rmi}{2}( \bar{u}^2_i - u^2_f) \tanh T \,. 
\end{equation}
Since the Hamiltonian is quadratic, the semiclassical Morette-Van-Hove formula \cite{morette, vanhove}
\begin{equation}
  (u_f|\rme^{-\rmi T H} |\bar{u}_i)_{\text{sc}} = \Big(\rmi \,\frac{\partial^2 S_{\text{cl}}}{\partial \bar{u}_i \partial u_f}\Big)^{\frac12}\,\rme^{\rmi S_{\text{cl}}}
\end{equation}
gives the exact result \eqref{eq:prop}. The semiclassical time-evolution of coherent states is rather intuitive \cite{HELLER1975321, LITTLEJOHN1986193}. The normalized overlap $|(v|\rme^{-\rmi t H}|\bar{u})|^2/(v|\bar{v})$ peaks at $v = u \cosh t  - \bar u \sinh t$, which is the classical path when both $u, \bar{u}$ are specified at the initial time.  During this Ehrenfest evolution, the wave packets are sheared along the classical hyperbolic trajectories.

\newpage
\subsection{Upside-down oscillator in a magnetic field }\label{app:poschl}
\begin{figure}[ht]
    \centering
      \begin{subfigure}{0.3\textwidth}
            \centering
            \begin{tikzpicture}[remember picture, overlay]
        \draw[blue, line width = 0.50mm]   plot[smooth,domain=0.12:3,samples=50] (\x-1.4, {0.1*(0.5/((\x)^2)-1.5*(\x)^2)+2});
        \draw[black, line width = 0.50mm] plot[smooth,domain=0.:3,samples=130] (\x-1.4, {0.3*(\x)*cos(400*(\x)^2))+2});
        \node at (0.,2.8) {\text{$\psi_{1,0}$
        }};
        \draw[->] (1.1,3.2) -- (1.5,3.2);
        \end{tikzpicture}
            \caption[]%
            {{\small radial potential}} 
        \end{subfigure}
        \hspace{1cm}
        \begin{subfigure}{0.4\textwidth}  
            \centering 
            \begin{tikzpicture}
              \node[inner sep = 0pt] at (7,3) {\includegraphics[width=0.8\textwidth]{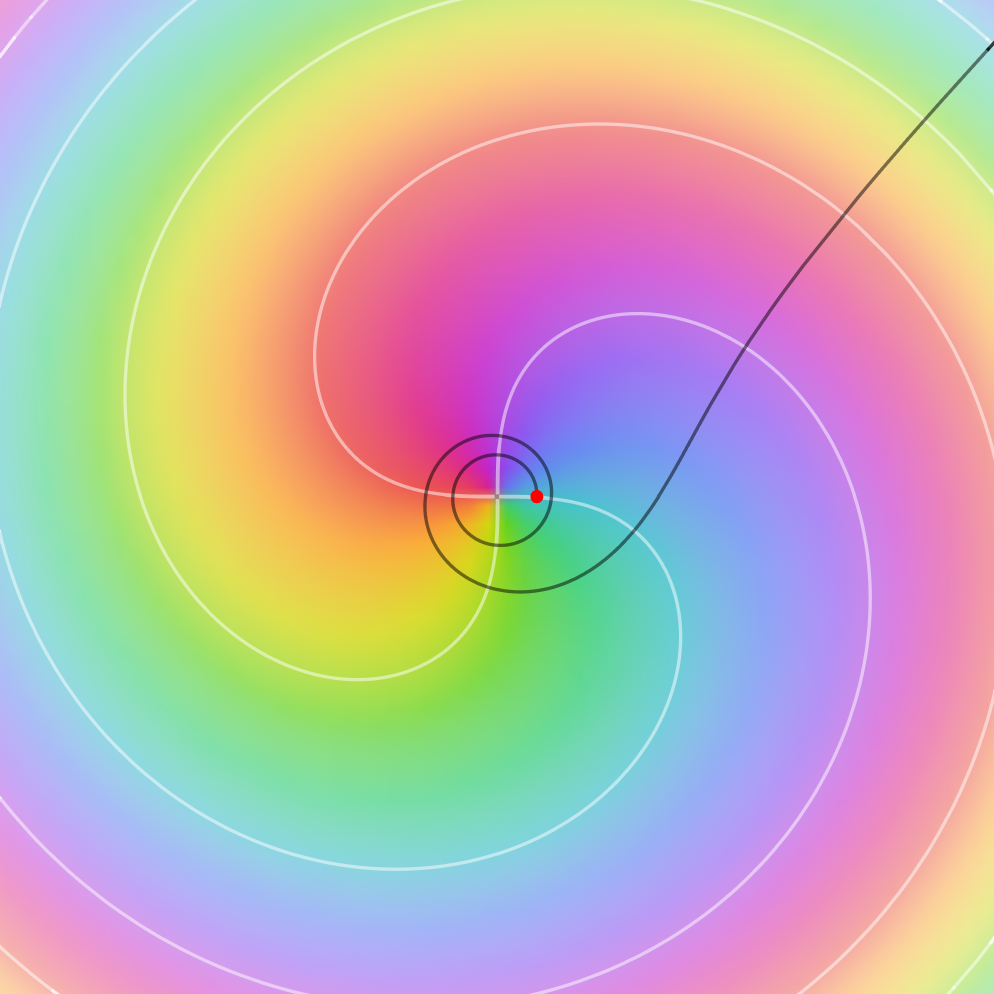}};
               \node (x2) at (4.37,5.92) {\text{$x_2$
        }};\node (x1) at (10.06,0.37) {\text{$x_1$
        }};\draw[->] (4.37,0.37) -- (x1);
        \draw[->] (4.37,0.37) -- (x2);
            \end{tikzpicture}
            \caption[]%
            {{\small  spiraling orbit}}
        \end{subfigure}
    \caption{Consider the first descendant resonance $\psi_{1,0}$ when $k^2>b^2$. In (a) we see the radial potential for angular momentum $n-n'=1$ with the real part of the resonance wave function, and in (b) its phase, together with a quantum trajectory starting at the red dot.}
    \label{fig:magres}
\end{figure}
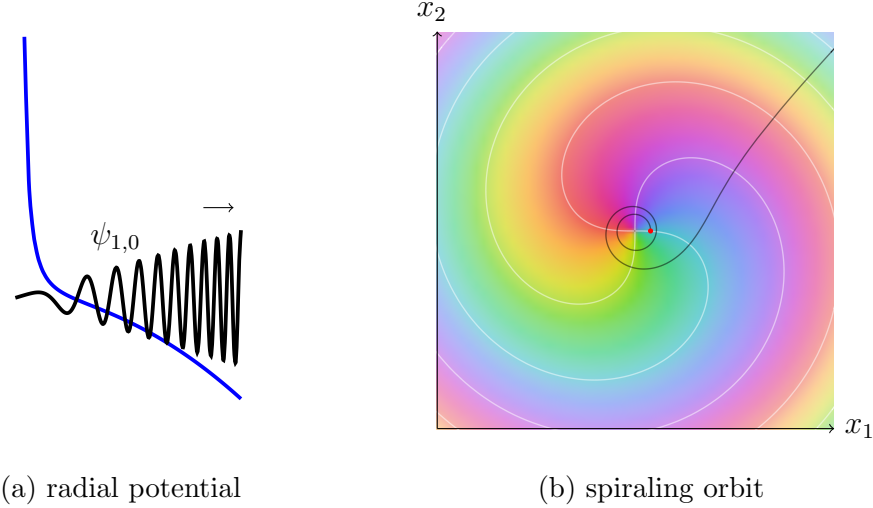

For any system with a quadratic Hamiltonian, the Heisenberg equations of motion are linear, and their solution can be written as a sum of exponential eigenmodes: 
\begin{align} \label{eigenmodesXPXP}
  \big(\hat{x}_i(t) ,\, \hat{p}_i(t)\big) = \sum_\alpha \,\rme^{-\rmi \omega_\alpha t} \, \big(\hat{v}^{(\alpha)}_i  , \, \hat{w}^{(\alpha)}_i \big)\,. 
\end{align}  
So far we considered examples with $\omega_\alpha \in \IR$ (harmonic oscillator) or $\omega_\alpha \in \rmi \IR$ (upside-down oscillator). In general, $\omega_\alpha$ will be a generic complex number.  Consider for example a particle on the plane, in a magnetic field and upside-down harmonic potential:
\begin{align}\label{eq:magham}
  H = \tfrac{1}{2}(\hat p_1 + b \hat x_2)^2 + \tfrac{1}{2}(\hat p_2 - b \hat x_1)^2 - \tfrac{1}{2}k^2 (\hat x_1^2+\hat x_2^2)  \, .
\end{align}
The system is characterized by four frequencies:
\begin{align} \label{gamgamppm}
 \omega_\pm = \pm \bigl(\sqrt{b^2-k^2} + \, b \bigr) \, , \qquad
 \omega'_{\pm} = \pm \bigl(\sqrt{b^2-k^2} - \, b \bigr) \, . 
\end{align}
The magnetic field dominates when $b^2>k^2$. In this case $\omega_\alpha \in \IR$: the classical trajectories are oscillatory, leading to normalizable energy eigenstates at the quantum level. On the other hand, when $k^2 > b^2$ we find $\omega_\alpha \in \IC$: the classical trajectories are spiraling orbits, described by a discrete set of resonances at the quantum level, see fig.\,\ref{fig:magres}. Their construction proceeds again as in app.\,\ref{app:gendec}, now with two pairs of raising and lowering operators $c_\pm$ and $c'_\pm$:
\begin{equation}
 c_\pm = \frac{\hat p_2\pm \rmi \hat p_1 - (b^2-k^2)^{1/2} \, (\hat x_1\mp \rmi \hat x_2)}{2 \, (b^2-k^2)^{1/4}}, \quad
 c_\pm' = \frac{\hat p_2 \mp \rmi \hat p_1 + (b^2-k^2)^{1/2} \, (\hat x_1 \pm \rmi \hat x_2)}{2 \, (b^2-k^2)^{1/4}}\,,
\end{equation}
in terms of which the Hamiltonian \eqref{eq:magham} can be expressed as  
\begin{align}\label{eq:hammagn}
  H = \bigl(\sqrt{b^2-k^2} + b\bigr) c_{+} c_{-} + \bigl(\sqrt{b^2-k^2} - b\bigr) c'_{+} c'_{-}  + \sqrt{b^2-k^2} \, .
\end{align} 

We can then contruct a tower of modes as before:
\begin{align}
 c_- |\psi_0 \rangle = c'_- |\psi_0 \rangle 
  = 0 
  = \langle \tilde\psi_0| c_+ = \langle \tilde \psi_0| c'_+\, ,
\end{align} 
\begin{align}
 |\psi_{n,n'} \rangle = \frac{(c_+)^n}{\sqrt{n!}} \, \frac{(c'_+)^{n'}}{\sqrt{n'!}} \, |\psi_0\rangle \, , \qquad \langle \tilde\psi_{n,n'}| = \langle \tilde\psi_0 |  \frac{(c_-)^n}{\sqrt{n!}} \, \frac{(c'_-)^{n'}}{\sqrt{n'!}} \, .
\end{align}
The $|\psi_{n,n'} \rangle$ are eigenmodes of the Hamiltonian \eqref{eq:hammagn} with eigenvalues
\begin{align}
  b^2 > k^2 :& \quad \omega_{n,n'} =  (n-n') \, b\,+ \,(n+n'+1) \, \sqrt{b^2-k^2} \,,   \\
  k^2 > b^2 :& \quad \omega_{n,n'} = (n-n') \, b\,-\,\rmi\, (n+n'+1)  \, \sqrt{k^2-b^2} \,.
\end{align}
The quantum number $n-n'$ is the angular momentum, while the combination $n+n'$ shifts either the bound state energy or resonance decay rate, depending on the relative size of $b^2$ and $k^2$. The orthonormality and completeness relations
\begin{align} \label{completenessex2}
 \langle \tilde \psi_{n,n'}|\psi_{m,m'}\rangle = \delta_{nm} \delta_{n'm'} \, , \qquad \sum_{nn'}\, |\psi_{n,n'}\rangle \langle \tilde \psi_{n,n'}| = \one = 
 \sum_{nn'} \,|\tilde\psi_{n,n'}\rangle \langle \psi_{n,n'} | 
\end{align}
can be used to calculate the character, as in \eqref{tr1}:
\begin{equation}
    \chi(t)= \tr  \rme^{-\rmi t H} = \frac{\rme^{-\rmi \omega_+ t/2}}{1-\rme^{-\rmi \omega_+ t} }\cdot  \frac{\rme^{-\rmi \omega'_+ t/2}}{1-\rme^{-\rmi \omega'_+ t} }\,,
\end{equation}
which is real and even under $t\to -t$ when $k^2>b^2$. In fact, in the most general case, finding the exponents $\omega_\alpha$ in \eqref{eigenmodesXPXP} amounts to diagonalizing a  matrix $M$ in the Lie algebra $\spp(2d,\IR)$. The spectrum of such $M$ is invariant under both $\omega \to - \omega$ and $\omega \to -\omega^*$. This means that, whenever the $\omega$ are not purely real, we can split them into two disjoint sets $\Omega_\pm$ with $\Omega_\pm = -\Omega^*_\pm = -\Omega_\mp$, each containing $d$ eigenvalues. Then we can write
\begin{align} \label{tritititi}
\chi(t)= \prod_{\omega \,\in\, \Omega_+} \frac{\rme^{-\rmi \omega t/2}}{|1-\rme^{-\rmi \omega t}| }=\prod_{\omega \,\in\, \Omega_-} \frac{\rme^{-\rmi \omega t/2}}{|1-\rme^{-\rmi \omega t} |} \, ,
\end{align} 
which is once again real and even under time reflection.

\newpage
\subsection{P\"oschl-Teller perturbatively }\label{sec:poschl}
In previous works \cite{Ching:1995rt, Ching:1998mxl, Nollert:1998ys, Beyer:1998nu}, the assumption that the potential falls off sufficiently fast allows for a more rigorous treatment of QNM completeness. In some cases though, such potentials can be approximated near their maximum by an upside-down harmonic oscillator \eqref{eq:ihopr}, and one might wonder whether their resonance spectrum can then be determined order by order in perturbation theory, similar to the standard approach for bound state spectra.  This idea will prove to be quite useful in sec.\,\ref{sec:spincorr}. To illustrate it, consider the Schr\"odinger problem for the inverted P\"oschl-Teller potential:
\begin{equation}\label{eq:ptschrod}
   \frac12(-\partial^2_x  + \sech^2 \lambda x)\, \psi(x) = E \, \psi(x)\,.
\end{equation}
Our goal now is to see how we can reproduce the exact resonance spectrum \eqref{eq:PTspectrum} in perturbation theory. Expanding the potential in \eqref{eq:ptschrod} near its maximum
\begin{equation}
    \frac12 \sech^2 \lambda x = \frac12 - \frac12 (\lambda x)^2 + \frac13 (\lambda x)^4 - \frac{17}{90}(\lambda x)^6 + \dots ,
\end{equation}
we can treat the total Hamiltonian as an unperturbed $H^{(0)}$ plus small corrections $H^{(i)}$:
\begin{equation}
    H^{(0)} = \frac12\big(-\partial^2_x + 1-(\lambda x)^2\big)\,, \quad H^{(1)} = \frac13 (\lambda x)^4\,, \quad H^{(2)}=- \frac{17}{90}(\lambda x)^6, \;\; \dots.
\end{equation}
Here, $H^{(0)}$ is the inverted oscillator analyzed in sec.\,\ref{sec:iho} with resonance spectrum 
\begin{equation}
    E^{(0)}_n = \frac12 -  \rmi \lambda (n+\frac12)\,.
\end{equation}
The unperturbed (anti-)resonances $\psi_n$ and $\tilde{\psi}_n$ are those in \eqref{eq:explicit}, up to a rescaling $x\to \sqrt{\lambda}x$. Keeping in mind this rescaling, and making use of the completeness relation \eqref{invhodecompid}, we can now determine the first perturbative correction: 
\begin{equation}
    E^{(1)}_n = \langle{\tilde\psi_n}|H^{(1)} \ket{\psi_n} =  -\frac12\lambda^2(n^2+n+\frac12)\,,
\end{equation}
which is precisely the first term in the $\lambda n \ll 1$ expansion  \eqref{eq:PTspectrum2} of the exact spectrum. The second correction is in turn given by the combination
\begin{equation}\label{eq:2ndcorrPTpert}
    E^{(2)}_n = \sum_{m\neq n} \frac{|\langle \tilde\psi_m |H^{(1)}| \psi_n\rangle|^2}{E^{(0)}_n - E^{(0)}_m}+ \langle{\tilde\psi_n}| H^{(2)} \ket{\psi_n} =\frac{\rmi}{8} \lambda^3 (n+\frac12)\, 
\end{equation}
of $H^{(1)}$, evaluated to second order in perturbation theory, and $H^{(2)}$ evaluated to first order. In calculating the matrix elements, it is useful to note that only terms with $n-m = \pm 2, \pm 4$ are non-vanishing. The result indeed matches the second correction in \eqref{eq:PTspectrum2}.

\section{Quasinormal de Sitter}\label{sec:desitter}
It is time to clarify the relevance of the upside-down harmonic oscillator to de Sitter physics. In app.\,\ref{app:phase} we obtain the phase space of a single particle in de Sitter space, described as a constrained system in embedding space. Phase space can be thought of as the space of particle trajectories. The static patch Hamiltonian acts on their asymptotic direction $p$ precisely like the inverted oscillator \eqref{eq:hamxp}. Northern and southern descriptions are related by a symplectomorphism, which flips the sign of the Hamiltonian. In this section we will see that at the quantum level, this translates into the existence of two QNM towers. Both  appear in the character, which is therefore sensitive to the global structure of phase space. 

\subsection{From phase space to boundary quantum mechanics}\label{sec:boundaryQM}

Locally, the phase space of a relativistic particle in $\dS_2$ is that of a non-relativistic particle on $\mathbb{R}$, as derived in app.\,\ref{app:phase}. The quantum description is thus in terms of wave functions $\psi(p)$. As shown in fig.\,\ref{fig:phasespace}, $p \in \IR$ is a stereographic coordinate on the future conformal boundary circle, indicating the asymptotic direction of a particle trajectory. 

It is the boost generator $\CH$  \eqref{eq:HQK}, generating time-translations in the southern static patch,  which serves as Hamiltonian for the boundary quantum mechanics. It is represented by a Hermitian operator acting on the wave functions\footnote{When acting on the states $\ket{p}$ it takes the form $H = \rmi(p\partial_p +\bar{\Delta})$ instead. }:
\begin{equation}\label{eq:ham}
     H = -\rmi\,(p \partial_p + \Delta)\, .
\end{equation}
 Note that it is essentially the same as the upside-down oscillator Hamiltonian \eqref{eq:hamxp}. Here $\Delta = \frac12 + \rmi \nu$ is the conformal weight appropriate to principal-series unitary irreps of $\SO(1,2)$. The coefficient $\nu\in \IR$ determines the mass of the particle. In a similar way, the translation and special conformal generators in \eqref{eq:HQK} become
\begin{equation}
  Q = \rmi \,\partial_{p} \,, \qquad \,  K = \rmi \,(p^2 \partial_{p}+ 2 \Delta p)\,.
\end{equation}
Together they satisfy the $\so(1,2)$ Euclidean conformal algebra
\begin{equation}
  [H, Q]= \rmi\, Q  \,, \qquad [H, K]= -\rmi\, K \,, \qquad [K,Q] = 2\rmi\, H\, .
\end{equation}
For the isometries to act unitarily, we further need to impose fall-offs 
\begin{equation}\label{eq:falloff}
    \psi(\,|p| \to \infty\,) = \big(\,\frac{1}{p^2}\,\big)^{\Delta}\,\big(\,c_\psi+ O(\frac{1}{p})\,\big)\,,
\end{equation}
with $c_\psi\in \IC$ constant. The complex vector space $\mathcal{F}_\Delta$ of infinitely differentiable functions $\psi$ satisfying the above condition then furnishes a representation of $\SO(1,2)$ \cite{Sun:2021thf}. 
Under conformal transformations such wave functions will transform as primaries
\begin{equation}\label{eq:primtrans}
    \psi'(p') = (\frac{\partial p}{\partial p'})^{\Delta} \psi(p)\,.
\end{equation}

Globally, northern and southern patches are glued together by the inversion \eqref{eq:sympcoord} which maps $p\to -1/p$. This interchanges $Q\leftrightarrow K$ and flips the sign of $H$ as in \eqref{eq:mapNS}. As argued at the end of app.\,\ref{app:phase}, we should therefore really deal with wave functions on the boundary circle. Secretly, this is already imprinted in \eqref{eq:falloff}.
Indeed, mapping between planar and global coordinates \eqref{eq:phiandJ} amounts to
\begin{equation}
    \psi(p) = (\frac{2}{1+p^2})^{\Delta} \, \psi(\varphi)\,,
\end{equation}
where we omitted the prime on the transformed wave function to avoid clutter. Then, whenever $\psi(\varphi)$ is a smooth function on the circle, with $\psi(\pi) = c_\psi$, we will have $\psi(p)$ in planar coordinates satisfying \eqref{eq:falloff}. 

In global coordinates, the conformal generators are 
\begin{equation}\label{eq:conformal}
    H = -\rmi\, (\sin\varphi \partial_\varphi + \Delta \cos\varphi)\,,\qquad 
    Q,\, K = \rmi\, \big(\, (1\pm\cos\varphi)\partial_\varphi \mp \Delta \sin \varphi\,\big) \,.
\end{equation}
Here, the inversion simply acts as $\varphi\to\varphi +\pi$. The global angular momentum $J$ is conjugate to $\varphi$, and as in \eqref{eq:phiandJ}, depending on the coordinate system, it takes the form
\begin{equation}
    J = -\rmi\, \partial_\varphi\,, \qquad J= - \rmi \,(\frac{1+p^2}{2}\partial_p + \Delta\, p)\,.
\end{equation}
Its eigenstates form a discrete basis of $\mathcal{F}_\Delta$, namely the Fourier basis:
\begin{equation}\label{eq:pbasis}
    \mathfrak f_n(\varphi) = \langle \varphi |n\rangle = \rme^{\rmi n \varphi}\,, \qquad \mathfrak f_n(p) = \big(\,\frac{2}{1+p^2}\,\big)^{\Delta} \big(\,\frac{1 +\rmi p}{1-\rmi p}\,\big)^n\,.
\end{equation}
The Hilbert space is then technically the $L^2$-completion of $\mathcal{F}_\Delta$ with the standard norm
\begin{equation}
    \bra{\xi}\ket{\psi} = \int_{S^1} \rmd \varphi\; \xi^*(\varphi)\psi(\varphi)  = \int_\IR \rmd p\; \xi^*(p)\psi(p)\,,
\end{equation}
under which the states \eqref{eq:pbasis} are of course orthogonal.

\newpage
\subsection{Two towers and the principal-series character }

Although \eqref{eq:ham} looks like a simple upside-down harmonic oscillator, we really are dealing with a different physical system, since it is $\SO(1,2)$ rather than the Heisenberg algebra which acts unitarily on it\footnote{For $J, K$ to be self-adjoint \cite{reed} fixes their domain to be $\mathcal{F}_\Delta$ and whenever $\psi$ satisfies the fall-off condition \eqref{eq:falloff} then so will $J\psi$ and $K\psi$. We cannot then restrict our attention to Gaussian wave packets as in sec.\,\ref{sec:iho}, since a finite $\rme^{\rmi a K}$ would for instance take us outside of this function space.}. An important consequence is the presence of two QNM towers in de Sitter, whereas for the upside-down oscillator we only found one. 

The two resonance towers $\psi_n$ and $\chi_n$ are found by first obtaining the primaries annihilated by $Q$, and then acting repeatedly with $K$. For the anti-resonances $\tilde\psi_n$ and $\tilde\chi_n$, we follow the same procedure with $K$ and $Q$ interchanged:
\begin{equation}\label{eq:dsaqnms}\begin{split}
   \psi_n = \frac{p^n}{\sqrt{n!}}\,,\hspace{2.3cm}&\qquad \tilde\psi_n = \frac{(-1)^n}{\sqrt{n!}}\,\partial^n_p\delta(p)\,,\\
   \chi_n =  \frac{1}{\sqrt{n!}}\,\frac{(p^2\partial_p)^n}{p^{2\Delta}}\,\delta(\frac{1}{p})\, ,& \qquad \tilde{\chi}_n =  \frac{(-1)^n}{\sqrt{n!}}\,\frac{1}{p^{2\Delta + n}}\,.
    \end{split}
\end{equation}
The (anti-)resonances of each tower have complex-conjugated frequencies
\begin{equation}\label{eq:aqnmfreq}\begin{split}
   H\psi_n &= -\rmi (\Delta+n) \psi_n\, ,\qquad  H\tilde{\psi}_n = \rmi (\bar{\Delta}+n) \tilde{\psi}_n \,, \\ 
   H \chi_n &= -\rmi (\bar\Delta+n)\chi_n\, , \qquad H \tilde{\chi}_n = \rmi (\Delta+n)\tilde{\chi}_n \,, 
   \end{split}
\end{equation}  
Since the inversion $p\to -p^{-1}$  interchanges $K\leftrightarrow Q$ and flips $H\to - H$ as in \eqref{eq:mapNS}, we have that southern resonances must be northern anti-resonances and vice versa. This inversion acts on the wave functions as \eqref{eq:primtrans}, and we can see from \eqref{eq:dsaqnms} that it indeed interchanges $\psi_n \leftrightarrow \tilde{\chi}_n$ and $\chi_n \leftrightarrow \tilde\psi_n$, which gives further intuition for the appearance of the two towers. 

Moreover, we learned in sec.\,\ref{sec:taylor} that the $\psi$-tower implements Taylor expansion at $p=0$. The role of $\chi$-tower will be to do the same at $p=\infty$. This is needed since we are allowed to consider Gaussian wave packets centered at the north pole. Having vanishing derivatives at $p=0$, these were absent from the discussion for the upside-down oscillator in sec.\,\ref{sec:taylor}. The completeness relation should then include both towers:
\begin{equation}\label{eq:dScomp}
    \one = \Theta(|p|>\mathsf{p})\,\sum_{n} |\chi_n\rangle \langle \tilde\chi_n| + \Theta(|p|< \mathsf{p})\,\sum_{n} \ket{\psi_n} \langle \tilde\psi_n| \,,
\end{equation}

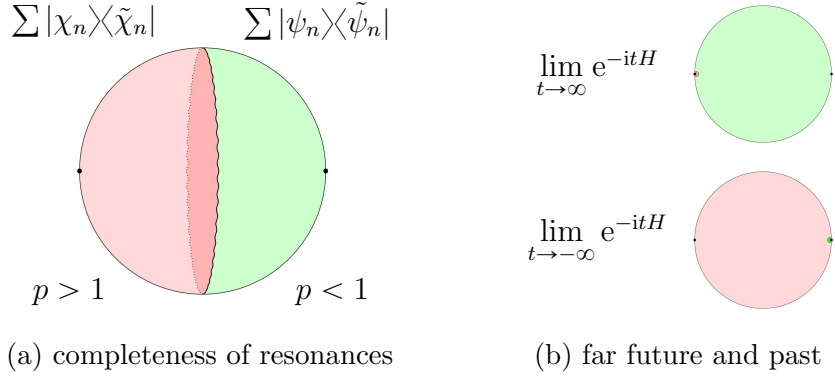
\begin{figure}
    \centering
  \begin{subfigure}{0.32\textwidth}
            \centering
            \begin{tikzpicture}
              \node[inner sep = 0pt] at (5.,5) {\includegraphics[width=0.7\textwidth]{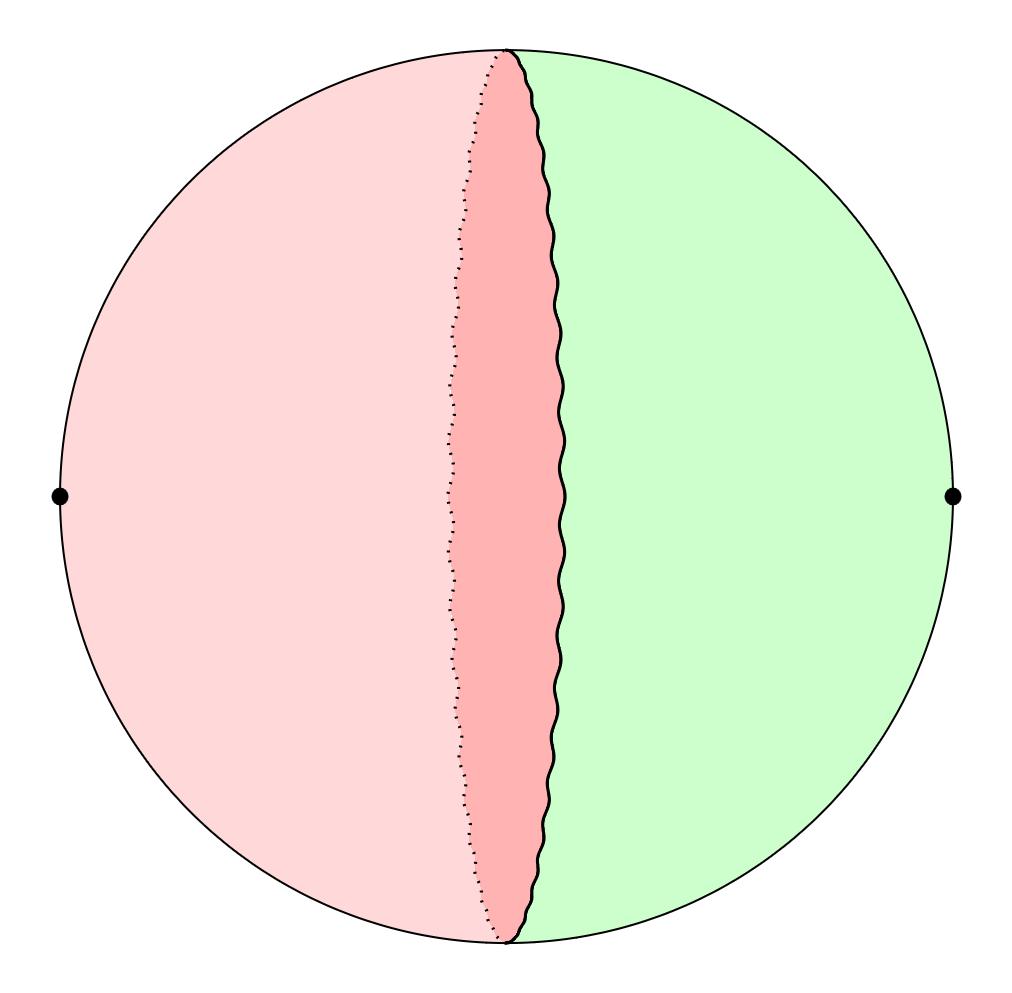}};
               \node (p) at (6.8,3.4) {\text{$p<1$
        }};\node (pt) at (3.3,3.4) {\text{$p >1$
        }};
        \node (left) at (3.5,7) {\text{$\sum | \chi_n\rangle \langle \tilde \chi_n |$
        }};\node (right) at (6.6,7) {\text{$\sum | \psi_n\rangle \langle \tilde \psi_n |$
        }};
            \end{tikzpicture}
            \caption[]%
            {{\small completeness of resonances}} 
        \end{subfigure}
        \hspace{0.8cm}
        \begin{subfigure}{0.32\textwidth}  
            \centering 
        \begin{tikzpicture}
              \node[inner sep = 0pt] at (5.,6.2) {\includegraphics[width=0.38\textwidth]{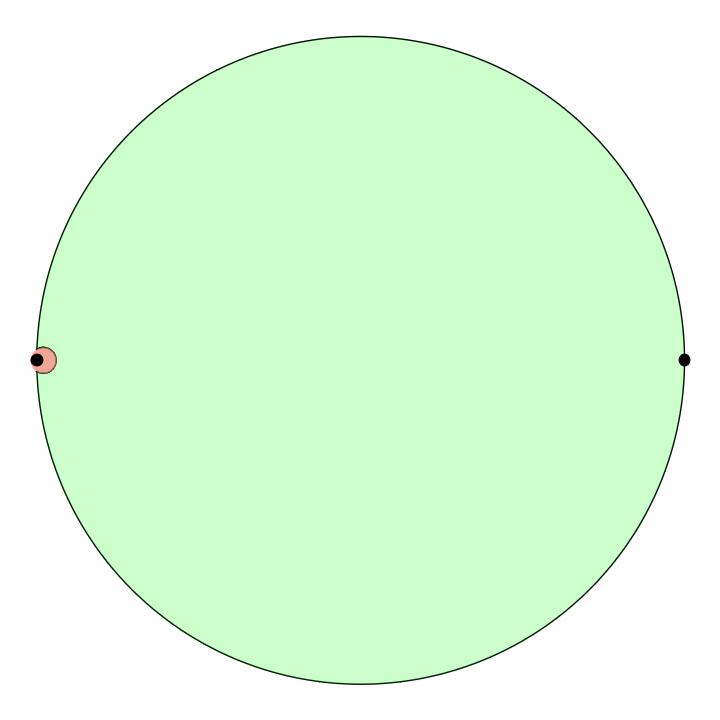}};
              \node[inner sep = 0pt] at (5.,4) {\includegraphics[width=0.38\textwidth]{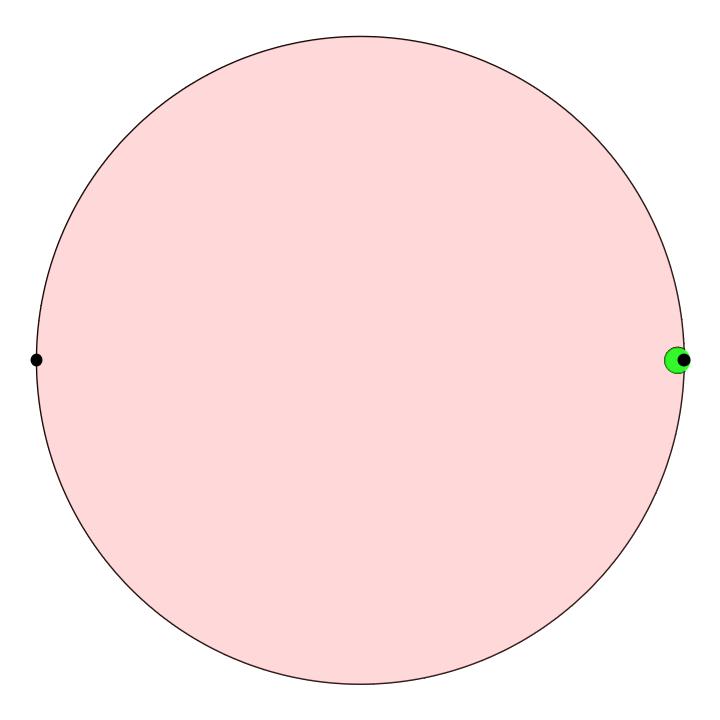}};
        \node (tinfty) at (2.8,6.2){\text{$\lim\limits_{t\to \infty} \rme^{-\rmi t H}$}};\node (tmin) at (2.8,4){\text{$\lim\limits_{t\to -\infty} \rme^{-\rmi t H}$}};
            \end{tikzpicture}
            \caption[]%
            {{\small  far future and past}}
        \end{subfigure}
    \caption{Illustration of \eqref{eq:dScomp} on the complex $p$-sphere. Southern excitations localized near $p=0$ are expanded in terms of the resonances $\psi_n$, and northern ones near $p=\infty$ in terms of the shadow resonances $\chi_n$. The radius of convergence of the expansions depends on time since $\rme^{-\rmi t H}\psi(p) = \rme^{-t\Delta}\,\psi(\rme^{-t} p)$.}
    \label{fig:completeness}
\end{figure}

\noindent where $\mathsf{p}$ is such that the Taylor series converges. For instance for the angular momentum eigenstates \eqref{eq:pbasis} the radius of convergence is $\mathsf{p}=1$. On the other hand, the time-evolved wave function $\rme^{-\rmi t H}\mathfrak{f}_n(p) = \rme^{-t\Delta}\,\mathfrak{f}_n(\rme^{-t} p)$ will have a rescaled radius of convergence $\mathsf{p}= \rme^{t}$, as is visualized in fig.\,\ref{fig:completeness}. For $t<0$ one should interchange the (anti-)resonances in \eqref{eq:dScomp}. 


\subsubsection{Flipping the Hamiltonian and evaluating the character}\label{sec:flipHam}

The north-south map \eqref{eq:mapNS} flips the sign of $H$. Indeed, boosts translate time in different directions on the northern and southern patches. On the other hand, the physics in either patch looks exactly the same. This symmetry implies that the character $  \chi(t) = \tr \rme^{-\rmi t H}$ must be even under time-reflection, and therefore real. We can see this explicitly in the angular momentum basis \eqref{eq:pbasis}. The north-south map $\varphi \to \varphi + \pi$ acts as $\rme^{-\rmi \pi J}\ket{n} = (-1)^n\ket{n}$ and $\rme^{-\rmi \pi J}H\, \rme^{\rmi \pi J} = -H$, so that:
\begin{equation}
    \chi(t) = \sum_n \bra{n}(-1)^n\rme^{-\rmi t H}(-1)^n \ket{n} = \sum_n \bra{n}\rme^{-\rmi \pi J}\rme^{-\rmi t H} \rme^{\rmi \pi J} \ket{n} = \chi(-t)\,.
\end{equation}
Here it was important that $J$ is a well-defined, self-adjoint operator, and that is where \eqref{eq:falloff} comes into play. The argument does not work when restricting to Gaussian wave packets on the real line, as we did for the upside-down oscillator. 

To actually compute the character, we can insert the decomposition \eqref{eq:dScomp} to arrive at 
\begin{equation}\label{eq:chards2}
    \chi(t) = \frac{\rme^{-t\Delta}}{|1-\rme^{-t}|} + \frac{\rme^{- t \bar{\Delta}}}{|1-\rme^{-t}|}\,,
\end{equation}
which is indeed the $\dS_2$ Harish-Chandra character \cite{HarishChandra1956TheCO}, featuring both QNM towers.

Alternatively, using position eigenstates $\ket{\varphi}$, one can directly calculate \cite{Grewal:2021bsu}
\begin{equation}\label{eq: char}
    \chi(t) = \int^{2\pi}_0 \rmd\varphi \bra{\varphi}\rme^{-\rmi t H} \ket{\varphi} = \int^{2\pi}_0 \rmd\varphi \;\rme^{\bar\Delta t}\;\frac{\braket{\varphi}{2 \arctan(\rme^t \tan\frac{\varphi}{2})}}{(\cos^2 \frac{\varphi}{2}+\rme^{2t}\sin^2\frac{\varphi}{2})^{\bar\Delta}}  \,,
\end{equation}
which localizes onto the fixed points of $H$,  $\varphi = 0$ and $\pi$, each of which yields a QNM tower.

\section{Quasinormal emergence}\label{sec:spin}
We can now apply what we learned in the previous section to the construction of toy versions of microscopic toy models, along the lines of sec.\,\ref{sec:intro}.  When exploring microscopic models, there are a few de Sitter avatars that one could look for :
\begin{enumerate}
    \item Excitations that locally behave like quasinormal modes, for a certain amount of time, which increases as the Hilbert space dimension grows large, until recurrences kick in.
    \item Logarithmic tails in $\rho(\omega)$, particular to the equidistant $\dS$ resonance spectrum.
    \item A symmetry $H\to - H$ which flips the sign of the Hamiltonian, interpreted as a north-south map, and which also gives rise to a second ``shadow'' tower of QNMs. 
\end{enumerate} 
Our aspiration then boils down to finding a finite quantum mechanical system within which $\dS$-like resonances emerge in the limit of large Hilbert space dimension $N$. As a diagnostic for this we will use the character $\chi(t)$ or equivalently the density of states $\rho(\omega)$ of the quantum system, and compare it with the exact $\dS$ results. 
Agreement of these quantities at the single-particle level then opens up possibilities towards the construction of multi-particle interacting microscopic models, as we will briefly discuss in sec.\,\ref{sec:disc}. 

\subsection{Spin resonances mimic de Sitter}\label{sec:spinmodel}
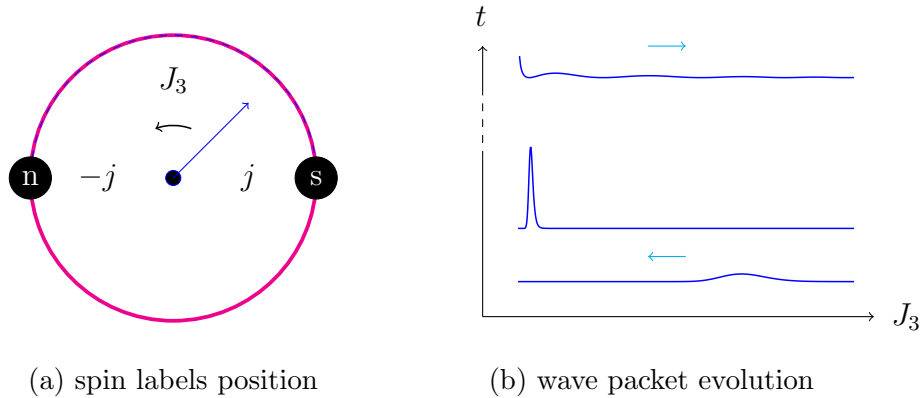
\begin{figure}
    \centering
  \begin{subfigure}{0.34\textwidth}
            \centering
            \begin{tikzpicture}[scale=2]
        \draw[magenta, line width = 0.50mm](0,2.2) circle (0.95);
        \draw[blue, fill= black] (0,2.2) circle (0.05);
        \draw[->, blue]  (0,2.2)   -- (0.5,2.7);
        \draw[blue, dashed] (0.938,2.3649) arc (10:175:0.95);
        \draw[black, fill= black] (-0.95,2.2) circle (0.14);
        \draw[black, fill= black] (0.95,2.2) circle (0.14);
        \node[white] at (-0.95,2.2) {\text{n}};
        \node[white] at (0.95,2.2) {\text{s}};
        \node[] at (0.5,2.2) {\text{$j$}};
        \node[] at (-0.5,2.2) {\text{$-j$}};
        \node[white] at (0,1.2) {\text{$.$}};
        \node[black] at (0,2.85) {\text{$J_3$}};
        \draw[->,semithick] (0.12,2.528) arc[radius=0.35, start angle=70, end angle=110];
    \end{tikzpicture}
            \caption[]%
            {{\small spin labels position}} 
        \end{subfigure}
        \hspace{0.8cm}
        \begin{subfigure}{0.3\textwidth}  
            \centering 
                        \begin{tikzpicture}
              \node[inner sep = 0pt] at (7,3) {\includegraphics[width=0.9\textwidth]{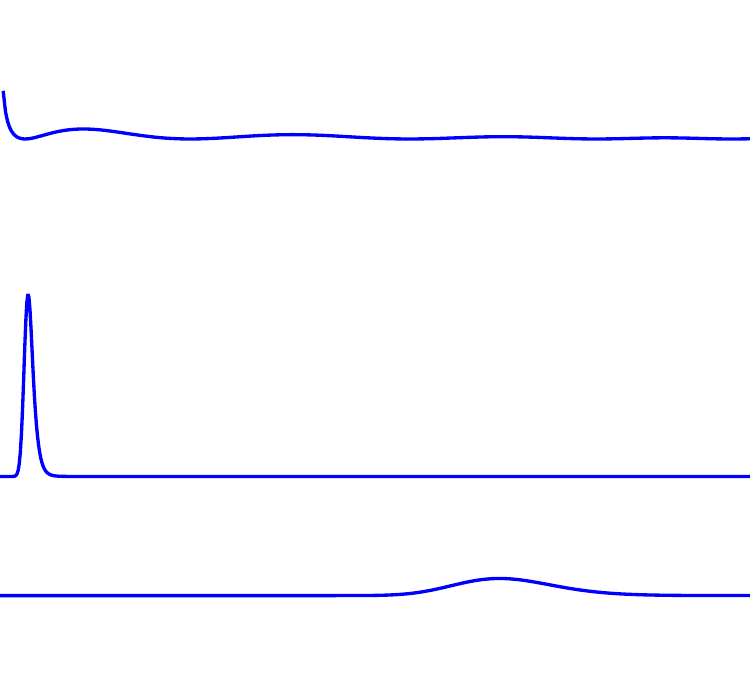}};
               \node (x2) at (4.35,5) {\text{$t$
        }};\node (jplus) at (10, 1) {\text{$J_3$
        }};\draw[->] (4.3,1) -- (9.5,1);
        \draw[->] (4.3,4) -- (4.3,4.6);
        \draw (4.3,1) -- (4.3,3.1);
        \draw[dashed](4.3,3.1) -- (4.3,4);
        \draw[->, cyan] (6.5,4.6) -- (7,4.6);
        \draw[-> ,cyan] (7,1.8) -- (6.5,1.8);
            \end{tikzpicture}
            \caption[]%
            {{\small wave packet evolution}}
        \end{subfigure}
    \caption{In the spin model with Hamiltonian \eqref{eq:HJJ} we think of $J_3$ as labeling the position on the boundary circle. In this interpretation, wave packets keep moving towards the north pole until they notice that $\dim \mathscr H = 2j+1$, and bounce back after a time of order $\log j$.}
    \label{fig:spinboundary}
\end{figure}

A first ambition would be to describe massive matter in $\dS_2$. 
For our single-particle model, we will consider states in the spin-$j$ representation of $\SU(2)$, so that the dimension of the Hilbert space is finite: $\dim \mathscr H =  2j+1$. By $J_i$ we denote the $\su(2)$ generators in the spin-$j$ representation, satisfying
\begin{equation}
    [J_i, J_j] = \rmi \epsilon_{ijk}J_k\,.
\end{equation}
Defining then $J_{\pm} = J_1 \pm \rmi J_2$, the Hamiltonian of interest will be:
\begin{equation}\label{eq:HJJ}
    H_j = \rmi\, \frac{J^2_- - J^2_+}{4j}+\nu\,\frac{J_3}{j}\,.
\end{equation}
One can immediately notice two properties of $H_j$ crucial to mimic matter excitations in $\dS_2$. First, close to maximal spin, $\su(2)$ behaves like the Heisenberg algebra:
\begin{equation}\label{eq:approxheis}
   J_3 \approx \pm j\implies [J_1, J_2] \approx \pm \rmi\, j\,,
\end{equation}
so that the first term of \eqref{eq:HJJ}, being proportional to $J_1J_2+J_2J_1$, behaves as the $p\partial_p$ in \eqref{eq:ham}, while the second one, with prefactor $\nu$, serves as the mass term. We thus expect resonance peaks that are at least qualitatively similar to those in $\dS_2$; the flipping of a spin mimicking a particle rolling down an upside-down harmonic potential. This analogy between $J_3$ and the boundary position is illustrated in fig.\,\ref{fig:spinboundary}, and will be made precise in the next subsection. The other relevant property is that $H_j$ possesses a time-reversal symmetry: conjugation by $\rme^{\rmi \pi J_2}$ maps $H_j \to - H_j$ and serves as a discrete analog of the north-south map.

The Hamiltonian \eqref{eq:HJJ} has a simple structure. In the $J_3$-eigenbasis $(J_3)_{n,n} = j-n$, where $n$ runs from $0$ to $2j$, its matrix elements are:
\begin{equation}\label{eq:Hjmat}
    (H_j)_{n,n} = \nu\,(1-\frac{n}{j})\,,\qquad  (H_j)_{n+2,n} = - (H_j)_{n,n+2} = \rmi \,a(n+1)\,,
\end{equation}
where
\begin{equation}
    a(n) = \frac{1}{4j}\sqrt{n(n+1)(2j-n)(2j-n+1)}\,.
\end{equation}
We will numerically diagonalize $H_j$ in sec.\,\ref{sec:exactmatch}. The surprise is that the full model, not limited to near-maximal spin, reproduces the exact $\dS_2$ density of states in the large-$j$ limit. 

Finally, at the level of dynamics, one observes that Gaussian spin wave packets move towards lower and lower spin as time moves on, before eventually bouncing back, as shown in fig.\,\ref{fig:spinboundary}. This generically happens on a timescale which grows like $\log j$. At finite $\nu/j$, the wave turns back slightly before reaching other side. The precise turning point depends on the choice of initial wave packet and size of $\nu/j$, and the effect goes away in the large-$j$ limit. These properties of $H_j$ can be understood analytically using the methods in sec.\,\ref{sec:semicl}.

\subsection{From spin to boundary Hamiltonian }\label{sec:direct}
The analogy, illustrated in fig.\,\ref{fig:spinboundary}, between $J_3$ and position on the future conformal boundary of $\dS_2$, can be made more precise. To this end, let us define a map\footnote{This is a double-cover, but to figure out the $\dS_2$ spectrum, looking at the $\varphi\in [0,\pi]$ semicircle suffices. }
\begin{equation}\label{eq:coordmap}
   y =  \frac12(1-\cos\varphi) = \sin^2(\frac{\varphi}{2})\,.
\end{equation}
This conformal transformation acts on $\dS_2$ boundary states in the following way: 
\begin{equation}\label{eq:ketmap}
    \ket{\varphi} =\Big(\frac{\rmd y}{\rmd \varphi}\Big)^{\bar\Delta}\ket{y} = \Big(\frac{\sin\varphi}{2}\Big)^{\bar\Delta}\ket{y} = \big(y(1-y)\big)^\frac{\bar\Delta}{2}\ket{y}.
\end{equation}
From this, we find that the $\dS_2$ Hamiltonian \eqref{eq:conformal} in the $\ket{y}$-basis takes the form
\begin{equation}\label{eq:yHam}
    H \ket{y} = 2\rmi \Big(y(1-y)\partial_y + \bar\Delta(1 - 2y)\Big)\ket{y}.
\end{equation}
The above is now seen to be the continuum limit of 
\begin{equation}\label{eq:Hfine}
    H_{N} \ket{n} \equiv \frac{2\rmi}{N}\Big((n+1)(N-n)S_+ - n(N-n+1)S_-  - \rmi \nu (N- 2n)\Big)\ket{n}, 
\end{equation}
after identifying $y =n N^{-1}$ for $n = 0,\, 1,\, \dots,\, N$. The
$S_\pm$ are shift operators acting on the states $\ket{n}$. Like its continuum counterpart, $H_N$ is Hermitian. At this point the relation to the spin model \eqref{eq:HJJ} is clear: at large $N=2j$ the matrix elements \eqref{eq:Hjmat} of $H_j$ are essentially the same as those of $H_N$. The main difference is that $H_j$ only shifts $n$ by multiples of two, meaning that the even- and odd-spin subspaces are invariant. 

A caveat is that not every naive discretization of \eqref{eq:yHam} reproduces the $\dS_2$ character. The resonance peaks in particular are very sensitive to the behavior near $y=0$ and $y=1$. In a way, the discretized model should not want to push wave packets beyond those points.

\subsection{Numerical diagonalization and large-spin limit }\label{sec:exactmatch}
We will now numerically diagonalize the spin model Hamiltonian \eqref{eq:HJJ}. The comparison of its spectrum with that of the $\dS_2$ Hamiltonian will be done at the level of the character $\chi(t)$, as given in \eqref{eq:chards2}, and equivalently, at the level of the $\dS_2$ density of states \cite{Anninos:2020hfj}:
\begin{equation}\label{eq:exactrho}\begin{split}
    \rho_{\dS_2}(\omega) &= \int^{\infty}_{\Lambda^{-1}} \frac{\rmd t}{2\pi} \,\big(\rme^{\rmi \omega t}+\rme^{-\rmi \omega t}\big)\, \chi(t)  \\[1.5ex]
    &= \frac{2}{\pi}\log\big(\rme^{-\gamma}\Lambda\big) - \frac{1}{2\pi}\sum_{\pm,\pm}\psi\big(\tfrac12 \pm \rmi \nu \pm \rmi \omega\big)\,,
    \end{split}
\end{equation}
where $\psi(x) = \Gamma'(x)/\Gamma(x)$ is the digamma function and $\gamma$ the Euler-Mascheroni constant. The UV-regulator $\Lambda$ shifts $\rho$ without affecting its shape; it is a constant to be matched\footnote{In the spin model for instance it will scale like $\Lambda \,\propto\; j$.}. The poles of $\rho$ are the quasinormal frequencies. Finally, it will be useful to recall that
\begin{equation}\label{eq:psisum}
    \psi(z) = -\gamma + \sum^{\infty}_{n=0}\Big(\frac{1}{n+1}-\frac{1}{n+z}\Big)\, ,
\end{equation}
so that we can split each $\psi$ in \eqref{eq:exactrho} into contributions coming from even and odd resonances:
\begin{equation}\label{eq:even/odd res}
    \psi\big(\tfrac12 \pm \rmi \nu \pm \rmi \omega\big) = \frac12\,\Big(\,\psi\big(\tfrac14 \pm  \tfrac{\rmi\nu}{2} \pm  \tfrac{\rmi\omega}{2}\big) + \psi\big(\tfrac34 \pm  \tfrac{\rmi\nu}{2} \pm  \tfrac{\rmi\omega}{2}\big)\,\Big)\,.
\end{equation}

\subsubsection{Peak splitting and averaging}
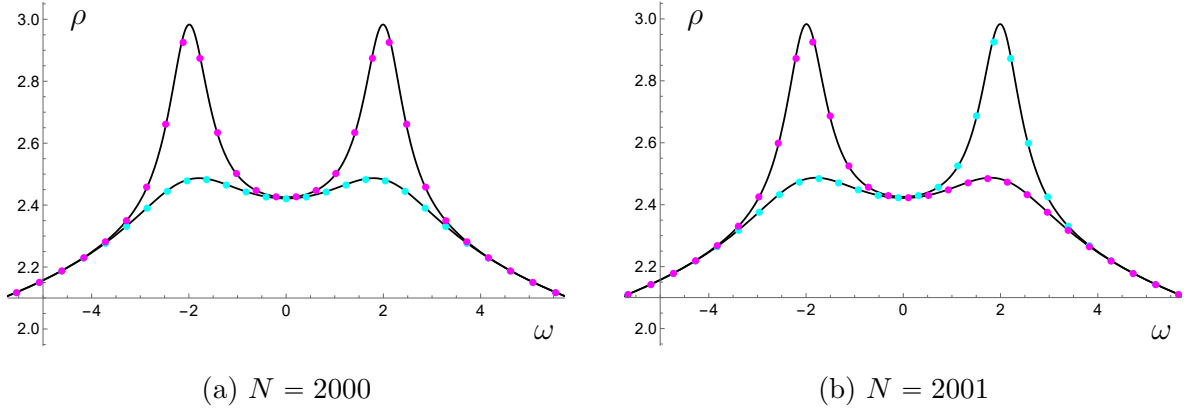
\begin{figure}
    \centering
  \begin{subfigure}{0.45\textwidth}
            \centering
            \begin{tikzpicture}
              \node[inner sep = 0pt] at (7,3) {\includegraphics[width=\textwidth]{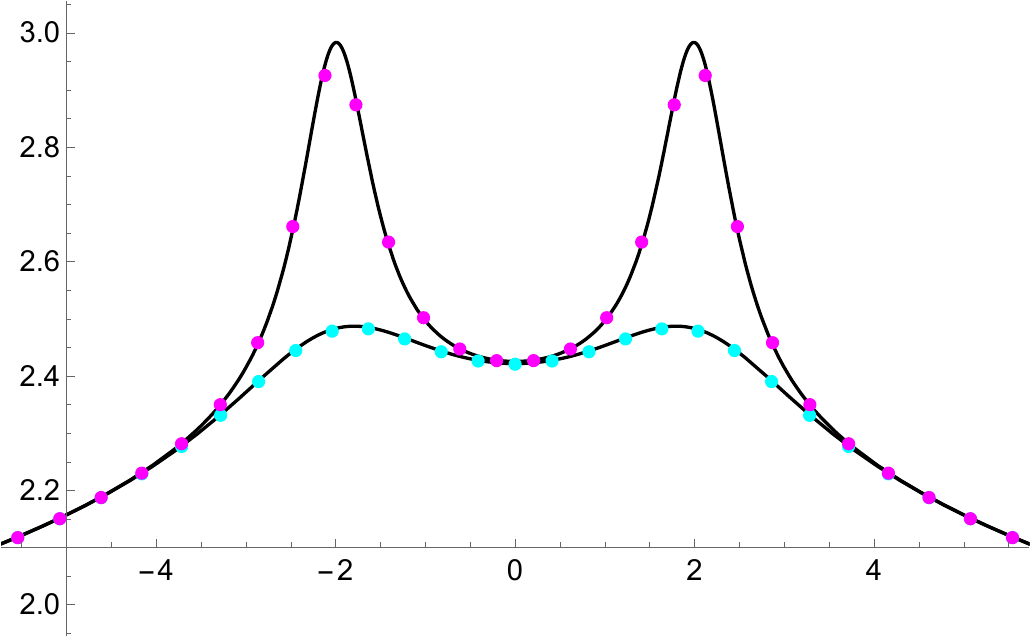}};
               \node (x2) at (4.3,5.05) {\text{$\rho$
        }};\node (jplus) at (10.5, 0.9) {\text{$\omega$
        }};
        \end{tikzpicture}
            \caption[]%
            {{\small  $N=2000$}} 
        \end{subfigure}
        \hspace{0.5cm}
        \begin{subfigure}{0.45\textwidth}  
            \centering 
            \begin{tikzpicture}
              \node[inner sep = 0pt] at (7,3) {\includegraphics[width=\textwidth]{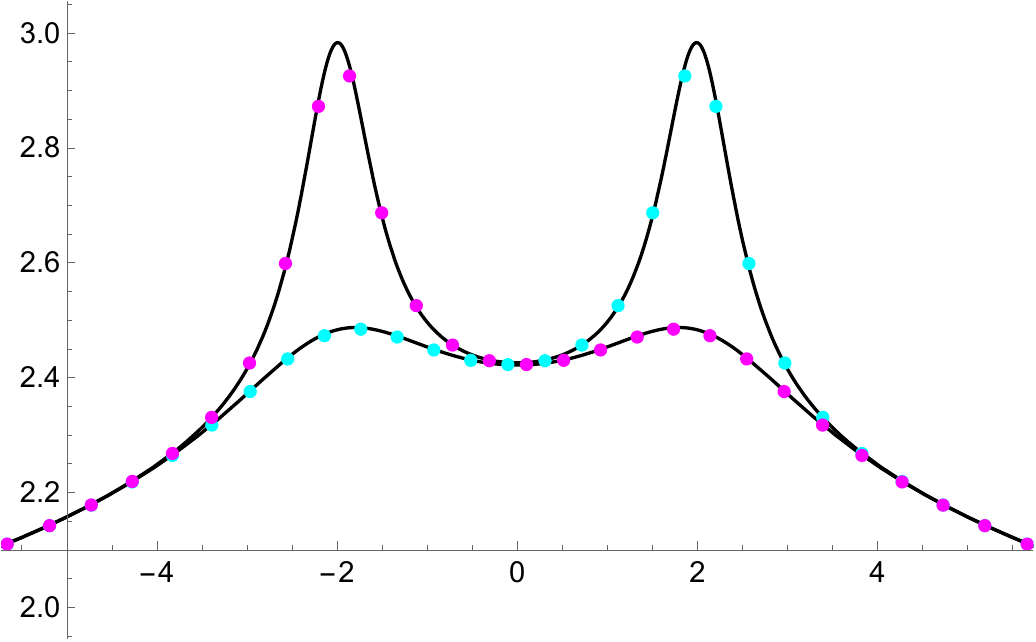}};
               \node (x2) at (4.3,5.05) {\text{$\rho$
        }};\node (jplus) at (10.5, 0.9) {\text{$\omega$
        }};
        \end{tikzpicture}
            \caption[]%
            {{\small  $N=2001$}}
        \end{subfigure}
    \caption{In black are the contributions to the exact density of states $\rho_{\dS_2}$ \eqref{eq:exactrho} coming from even and odd resonances, according to the split \eqref{eq:even/odd res}. The sharpest peaks are due to the even ones. In magenta is the numerical $\rho_{\text{even}}$ \eqref{eq:rho_odd/ev} coming from the even-spin eigenvalues of the spin model. In cyan we see $\rho_{\text{odd}}$. For larger $\omega$ these overlap almost perfectly, while for $\omega \lesssim \nu$ they appear nicely riffled. In the examples above we took $\nu=2$. The spin-model spectrum matches very well with that of the $\dS_2$ principal series.}
    \label{fig:evenodd}
\end{figure}

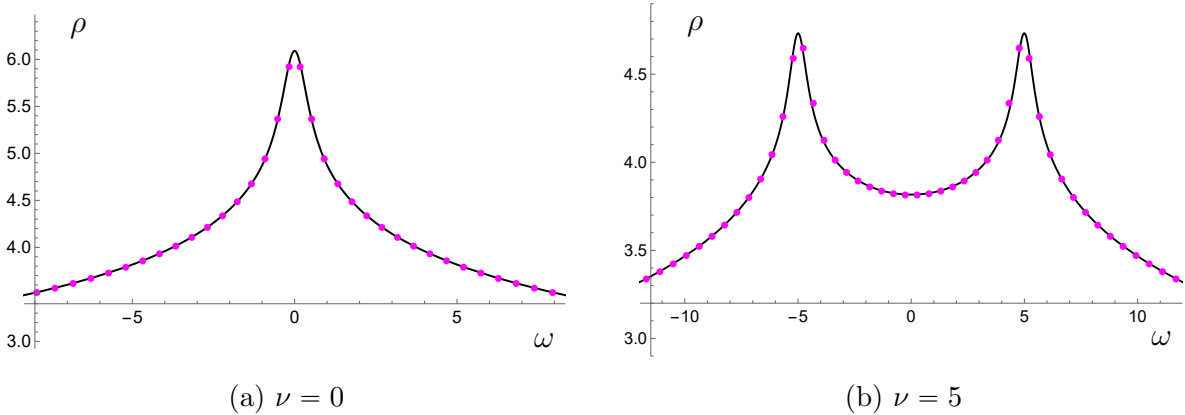
\begin{figure}
    \centering
  \begin{subfigure}{0.45\textwidth}
            \centering
            \begin{tikzpicture}
              \node[inner sep = 0pt] at (7,3) {\includegraphics[width=\textwidth]{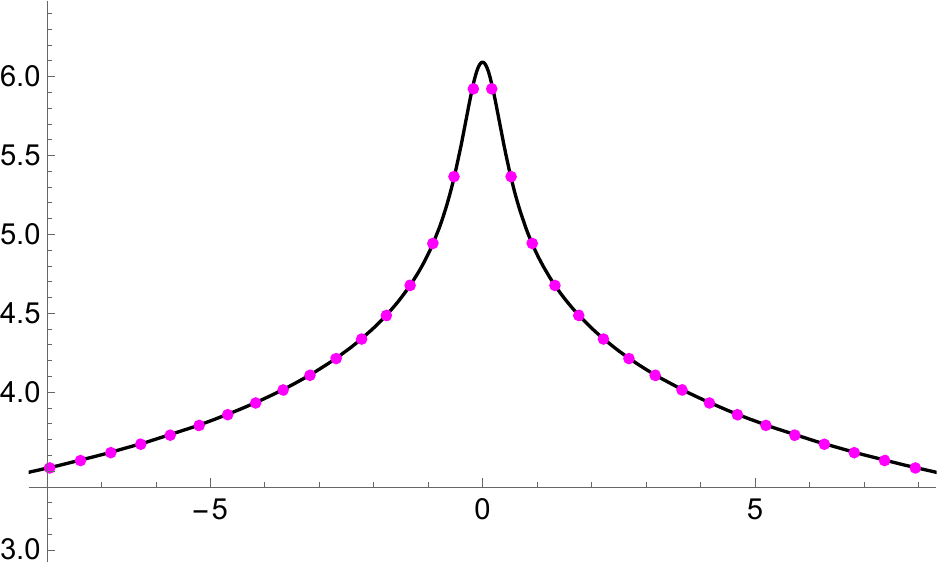}};
               \node (x2) at (4.3,5.05) {\text{$\rho$
        }};\node (jplus) at (10.5, 0.9) {\text{$\omega$
        }};
        \end{tikzpicture}
            \caption[]%
            {{\small  $\nu=0$}} 
        \end{subfigure}
        \hspace{0.5cm}
        \begin{subfigure}{0.45\textwidth}  
            \centering 
            \begin{tikzpicture}
              \node[inner sep = 0pt] at (7,3) {\includegraphics[width=\textwidth]{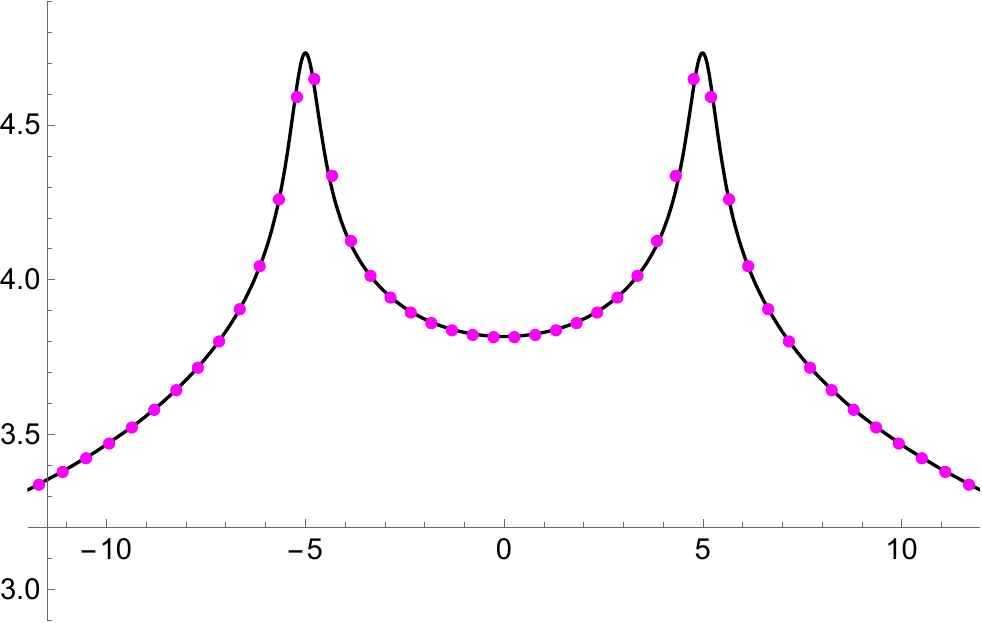}};
               \node (x2) at (4.3,5.05) {\text{$\rho$
        }};\node (jplus) at (10.5, 0.9) {\text{$\omega$
        }};
        \end{tikzpicture}
            \caption[]%
            {{\small  $\nu=5$}}
        \end{subfigure}
    \caption{The total density of states \eqref{eq:avdos} found by numerically diagonalizing the spin model at $N=1001$ is in excellent agreement with the analytic result \eqref{eq:exactrho} in black.}\label{fig:avmatch}
\end{figure}

 Grouping the ordered eigenvalues of $H_j$ into $\omega_{e,i}$ and $\omega_{o,i}$, according to the invariant subspaces of even and odd spin, we define the discretized densities to be the inverse level spacing:
\begin{equation}\label{eq:rho_odd/ev}
\rho_\text{even}\big(\,\tfrac12(\omega_{e,i+1}+\omega_{e,i})\,\big) \equiv (\omega_{e,i+1}-\omega_{e,i})^{-1}\,, \quad \rho_\text{odd}\big(\,\tfrac12(\omega_{o,i+1}+\omega_{o,i})\,\big) \equiv (\omega_{o,i+1}-\omega_{o,i})^{-1}\,.
\end{equation}
Fig.\,\ref{fig:evenodd} shows that, in the large-$j$ limit, these match rather seamlessly onto the contributions to $\rho_{\dS_2}$ coming from even and odd resonances. Given the identity \eqref{eq:even/odd res}, the full $\dS_2$ result \eqref{eq:exactrho} will then be obtained by considering the sum
\begin{equation}\label{eq:avdos}
\rho\big(\,\tfrac14(\omega_{e,i+1}+\omega_{e,i}+\omega_{o,i+1}+\omega_{o,i})\,\big) \equiv (\omega_{e,i+1}-\omega_{e,i})^{-1} + (\omega_{o,i+1}-\omega_{o,i})^{-1}\,.
\end{equation}
A glance at fig.\,\ref{fig:avmatch} reveals that this prescription works very well indeed. 

\subsubsection{Capturing overtones}
The previous figures focused on the behavior close to the resonance peaks. To get a sense for how well the overtones are captured, we should look at the logarithmic tails of $\rho_{\dS_2}$, since these arise due to the equidistant $\dS_2$ resonance spectrum. Increasing $j$ in the spin model increases the value of $|\omega|$ at which deviations from this tail become visible. 

\begin{figure}
    \centering
  \begin{subfigure}{0.45\textwidth}
            \centering
            \begin{tikzpicture}
              \node[inner sep = 0pt] at (7,3) {\includegraphics[width=\textwidth]{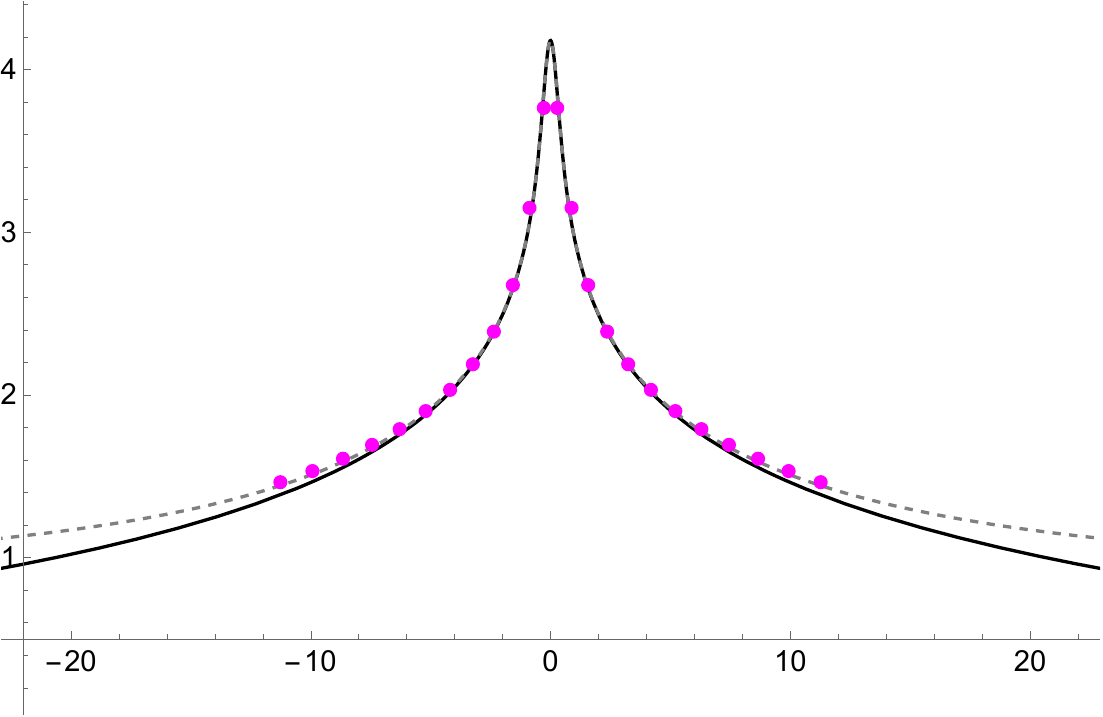}};
               \node (x2) at (4.1,5.05) {\text{$\rho$
        }};\node (jplus) at (10.7, 0.7) {\text{$\omega$
        }};
        \end{tikzpicture}
            \caption[]%
            {{\small  $N=50$}} 
        \end{subfigure}
        \hspace{0.5cm}
        \begin{subfigure}{0.45\textwidth}  
            \centering 
            \begin{tikzpicture}
              \node[inner sep = 0pt] at (7,3) {\includegraphics[width=\textwidth]{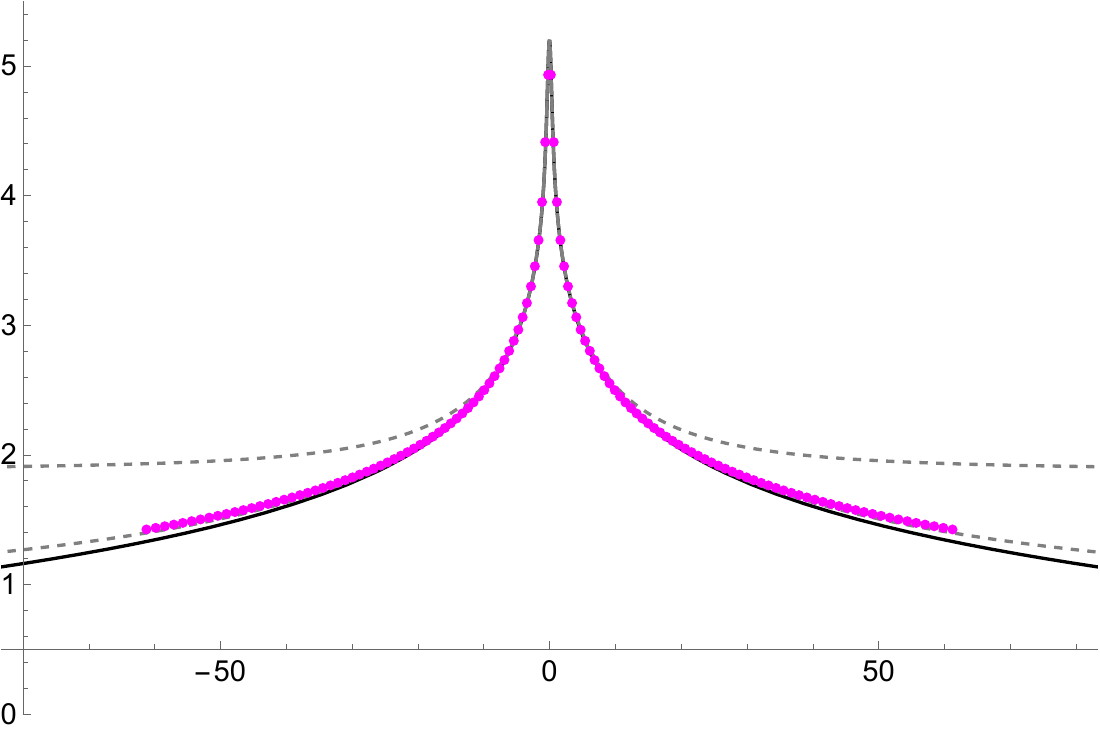}};
               \node (x2) at (4.1,5.05) {\text{$\rho$
        }};\node (jplus) at (10.7, 0.7) {\text{$\omega$
        }};
        \end{tikzpicture}
            \caption[]%
            {{\small  $N=250$}}
        \end{subfigure}
    \caption{In black is the exact density of states $\rho_{\dS_2}$ \eqref{eq:exactrho} for $\nu=0$, while the dashed lines include only those contributions coming from the first 25 and 125 resonances respectively. The dots in magenta instead represent the averaged density of states \eqref{eq:avdos} obtained by numerically diagonalizing the spin model Hamiltonian $H_j$ at $N=2j = 50$ and $250$ respectively.  The graphs thus indicate that $H_j$ captures roughly the first $j$ overtones.}\label{fig:tail}
\end{figure}

In fact, we can truncate the sum \eqref{eq:psisum} in $\rho_{\dS_2}$ to include only those contributions from the lowest $j$ overtones. In fig.\,\ref{fig:tail} we can see that it is this quantity which the spin model appears to reproduce. This makes it tempting to conclude that in the large-$j$ limit, the spin model will eventually resolve all resonances, and we will demonstrate this in sec.\,\ref{sec:semicl}.

    \begin{figure}[ht]
    \centering
  \begin{subfigure}{0.44\textwidth}
            \centering
        \begin{tikzpicture}
              \node[inner sep = 0pt] at (7,3) {\includegraphics[width=\textwidth]{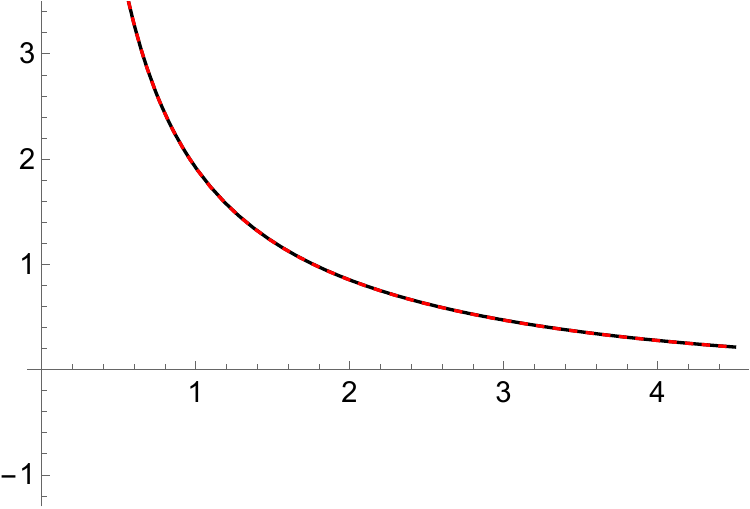}};
               \node (x2) at (4.2,5.2) {\text{$\chi$
        }};\node (jplus) at (10.5, 1.4) {\text{$t$
        }};
        \end{tikzpicture}
            \caption[]%
            {{\small  $\nu=0$}} 
        \end{subfigure}
        \hspace{0.8cm}
        \begin{subfigure}{0.44\textwidth}  
            \centering 
            \begin{tikzpicture}
              \node[inner sep = 0pt] at (7,3) {\includegraphics[width=\textwidth]{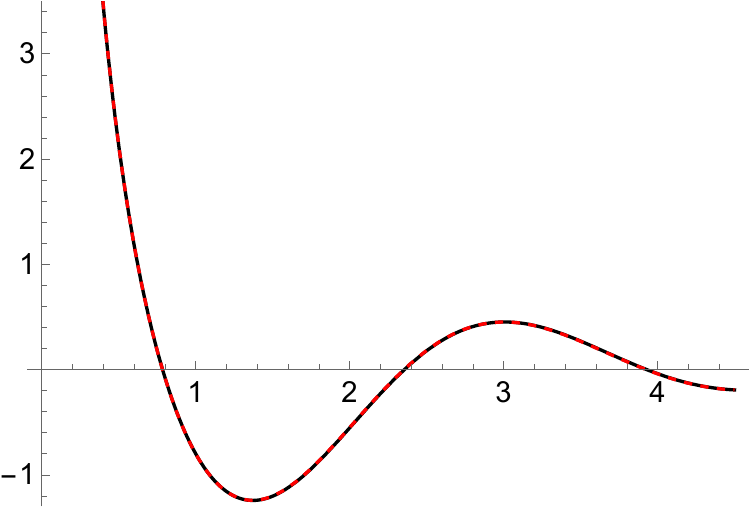}};
               \node (x2) at (4.2,5.2) {\text{$\chi$
        }};\node (jplus) at (10.5, 1.4) {\text{$t$
        }};
        \end{tikzpicture}
            \caption[]%
            {{\small  $\nu=2$}}
        \end{subfigure}
    \caption{The dashed red line is the spin model character $\eqref{eq:smearchar}$ for $j= 500$ with smearing $\epsilon=0.02$. Agreement with the $\dS_2$ Harish-Chandra character \eqref{eq:chards2} in black is excellent, as long as $t_{\text{\tiny{{UV}}}}<t<t_{\text{\tiny{IR}}}$. These cutoffs are discussed more in fig.\,\ref{fig:charcutoffs}. The $\epsilon^2$ and $1/j$ deviations are shown in fig.\,\ref{fig:numchar} and are obtained analytically in the main text.}
    \label{fig:leading}
\end{figure}
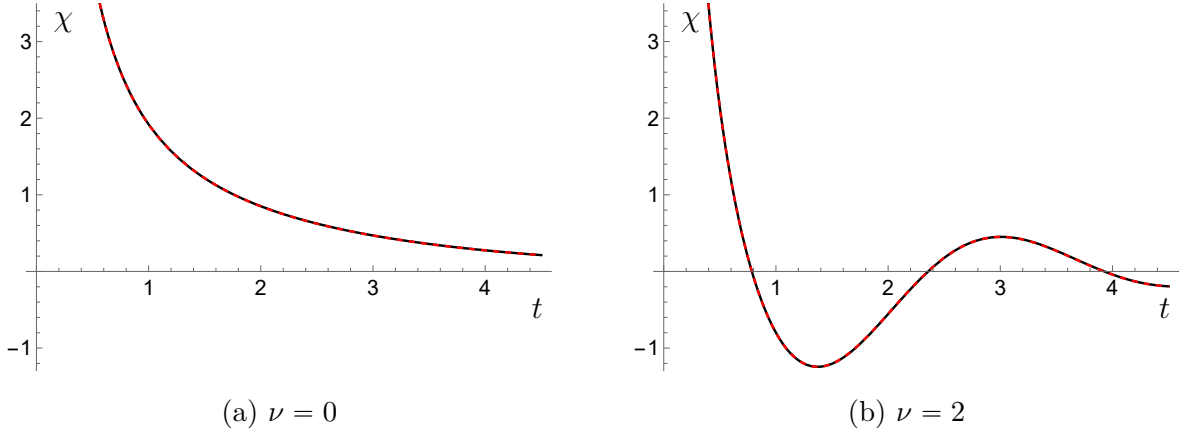

\subsubsection{Convergence of the coarse-grained character}
The comparison to the $\dS_2$ results can also be done at the level of the character. For the spin model it is simply a sum $\chi_j(t) = \sum \rme^{-\rmi \omega_i t}$ over all eigenvalues $\omega_i$ of $H_j$. As such, it avoids having to think about even- and odd-spin invariant subspace subtleties. At first sight $\chi_j(t)$ is rapidly oscillating and rather unwieldy, but we should keep in mind that it will be neither possible nor desirable to resolve arbitrarily small energy differences, or be sensitive to arbitrary UV-states. In fact this was already implicit when defining the discretized $\rho$ as the inverse level spacing, rather than a sum over $\delta$-spikes. 

In other words, test function overlaps $\int \rmd \omega \rho(\omega) f(\omega)$ should only agree when $f$ does not vary significantly over scales smaller than the eigenvalue spacings. For the character, this provides an IR-cutoff $t_{\text{\tiny{IR}}}$. Since the spectrum of $H_j$ is bounded, reasonable $f$ should also have compact support $|\omega| < \epsilon^{-1}$ for some UV-cutoff $\epsilon$. At the character level, this translates to a smearing in time of order $t_{\text{\tiny{UV}}} =\epsilon$. This leads us to define the coarse-grained character:
    \begin{equation}\label{eq:smearchar}
        \chi_{j,\epsilon}(t) \equiv \frac{1}{\sqrt{2\pi}\epsilon} \int^\infty_{-\infty}\rmd t' \,\rme^{-(t-t')^2/2\epsilon^2} \chi_j(t') = \sum_{i}\, \rme^{-\rmi \omega_i t-\epsilon^2 \omega^2_i/2}\,.
    \end{equation}
It is this quantity which at large $j$, and $t_{\text{\tiny{UV}}}< t < t_{\text{\tiny{IR}}}$, converges to the $\dS_2$ character \eqref{eq:chards2}. Numerical evidence of this provided in fig.\,\ref{fig:leading}. An analytic demonstration, and more precise understanding of the timescales $t_{\text{\tiny{UV}}}$ and $t_{\text{\tiny{IR}}}$, will be the topic of the next section.

\section{Quasinormal semiclassics }\label{sec:semicl}
In the previous section we found strong numerical hints that the spectrum of the spin model converges in the large-$j$ limit to that of a massive particle in  $\dS_2$. To get an analytic handle on this problem,  the coherent spin state formalism is ideally suited. It is reviewed in app.\,\ref{app:cohspin}. In app.\,\ref{app:symbols} we recall how the large-$j$ limit can then be understood as a semiclassical one. Starting from the associated phase space path integral, we will prove the claims made in sec.\,\ref{sec:exactmatch}, and find the $1/j$ corrections. 

\subsection{Spin-model spectrum and Heun polynomials}\label{sec:heun}
The $2j+1$ eigenvalues of the spin Hamiltonian $H_j$ in \eqref{eq:HJJ} are in principle found as roots of its characteristic polynomial\footnote{For matrices with constant off-diagonal bands, these polynomials sometimes take on a simple form. This happens for the mass matrices of $A_n$ Toda field theories, where Chebyshev polynomials appear \cite{Koca:1990cf}.}. Since in our case it does not take on any particularly pleasant form, we will proceed differently. In holomorphic polarization, see app.\,\ref{app:cohspin}, $H_j$ acts as  
\begin{equation}\label{eq:holoH}
    H_j\, \psi(z) = \Big(\, \tfrac{\rmi}{4j }\big(\,(1-z^4)\partial_z^2 +(4j-2 )z^3\partial_z + (2j-4j^2)z^2 \,\big) +\tfrac{\nu}{j}( z\partial_z-j)\, \Big) \psi(z)\,,
\end{equation}
 Solutions of the finite-$j$ eigenvalue equation are then polynomials $p(z)$ satisfying
\begin{equation}\label{eq:eigeqn}
    H_j\, p(z) = \lambda\, p(z)\,,\qquad \text{deg}(p)\leq 2j\,.
\end{equation}
In fact, deg$(p)$ must be either $2j$ or $2j-1$; otherwise one finds from \eqref{eq:eigeqn} that $p(z)$ vanishes in its entirity. Not restricting to polynomials, one finds eigenfunctions for every $\lambda$:
\begin{equation}\begin{split}
    f_1(z) &= \mathsf{H}\ell(-1,\;\rmi j (\nu+\lambda),\; \frac{1}{2}-j,\; -j,\;\frac12,\;\frac{1}{2}-j+\rmi \nu,\;z^2)\,,\\
    f_2(z) &= z\;\mathsf{H}\ell(-1,\rmi j (\nu+\lambda)-\rmi\nu,\; \frac{1}{2}-j,\;1- j,\; \frac{3}{2},\; \frac{1}{2}-j+\rmi \nu,\; z^2)\,.
    \end{split}
\end{equation}
In the above, $\mathsf{H}\ell$ is the Heun function. The finite-$j$ eigenvalues are then those $\lambda$ for which one of these functions reduces to a polynomial of degree $2j$ or $2j-1$. Moreover, these two options correspond to the even- and odd-spin subspaces discussed in sec.\,\ref{sec:spin}. Although this will be of limited practical use, it is good to have at least a characterization of the exact spin-model spectrum. It is also noteworthy that in the large-$j$ limit, it relates the $\dS_2$ density of states to a distribution of Heun polynomials.

 
\subsection{Large-spin emergent phase space and density of states}\label{sec:spinsemicl}
The real value of coherent spin states lies in them making manifest the emergence of a classical $S^2$ phase space in the large-$j$ limit \cite{berezin}. This is reviewed in app.\,\ref{app:symbols}. 

This emergent classical dynamics on the spin sphere, with symplectic form \eqref{eq:symp}, is governed by the coherent state expectation value, see \eqref{eq:symbol}, of the Hamiltonian $H_j$:
\begin{equation}\label{eq:hamsymbol}
    H_j(z,\bar{z}) = \rmi\, (j - \tfrac12) \,\frac{(\bar{z}^2-z^2)}{\,(1+z\bar{z})^2} + \nu\, \frac{z\bar{z}-1}{1+z\bar{z}}\,. 
\end{equation}

\begin{figure}
    \centering
  \begin{subfigure}{0.22\textwidth}
            \centering
            \includegraphics[width=\textwidth]{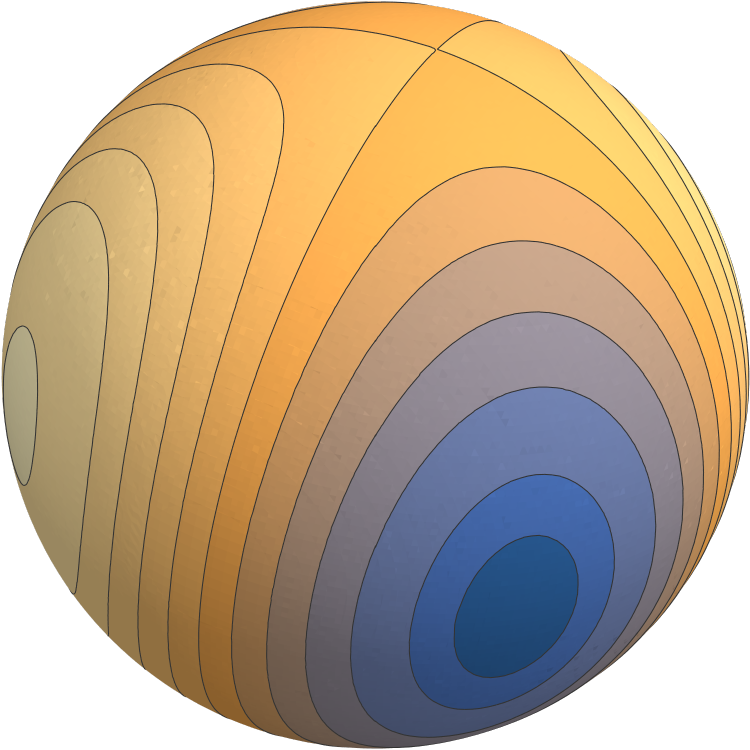}
            \caption[]%
            {{\small  $\frac{\nu}{j}=0$}} \label{fig:orbit1}
        \end{subfigure}
        \hfill
        \begin{subfigure}{0.22\textwidth}  
            \centering 
            \includegraphics[width=\textwidth]{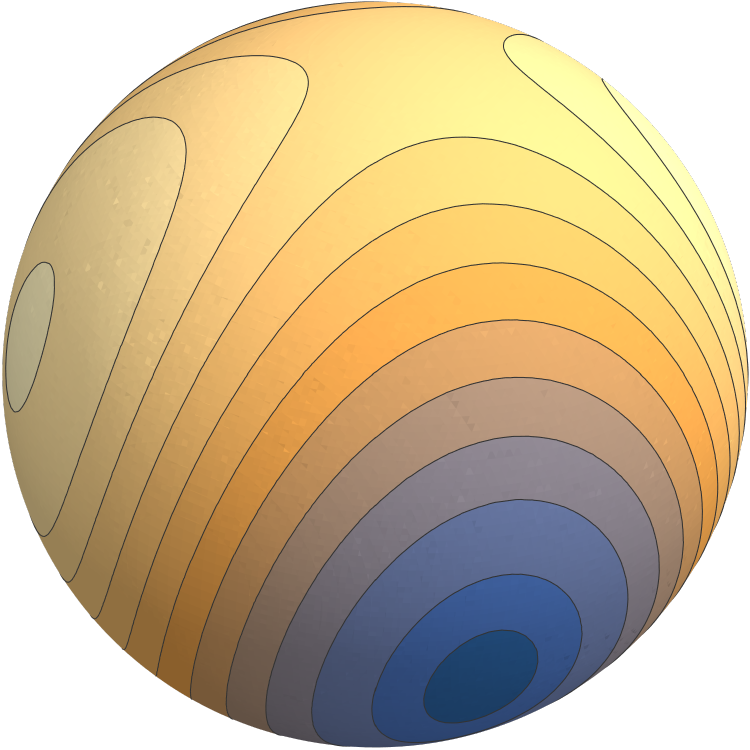}
            \caption[]%
            {{\small  $\frac{\nu}{j}=\frac14$}}
        \end{subfigure}
        \hfill
        \begin{subfigure}{0.22\textwidth}  
            \centering 
            \includegraphics[width=\textwidth]{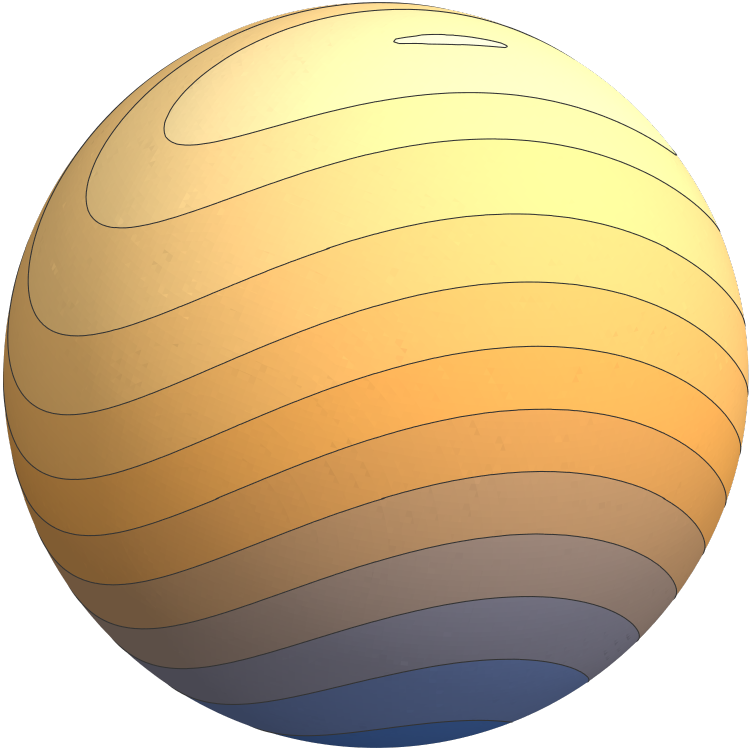}
            \caption[]%
            {{\small  $\frac{\nu}{j}=1$}}
        \end{subfigure}
        \hfill
        \begin{subfigure}{0.22\textwidth}  
            \centering 
            \includegraphics[width=\textwidth]{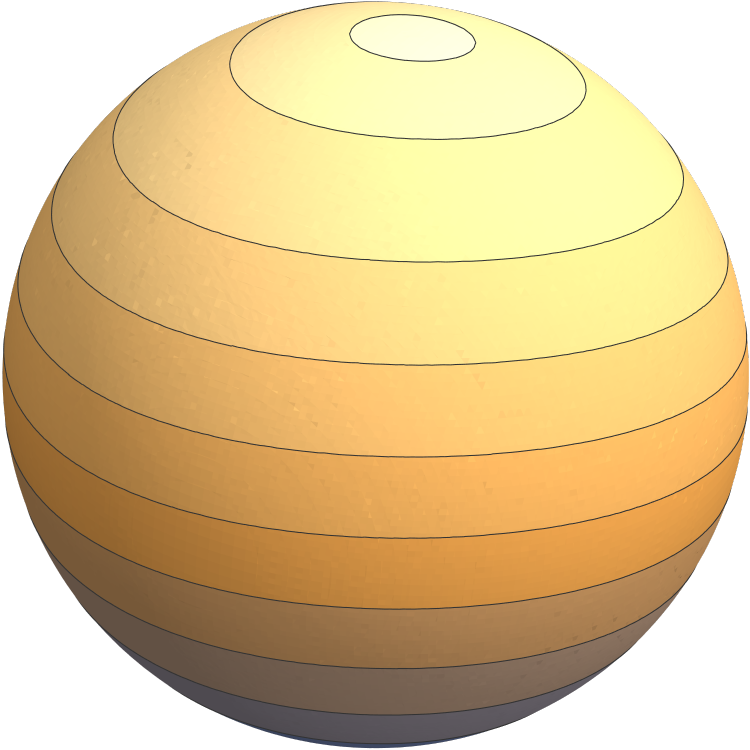}
            \caption[]%
            {{\small  $\frac{\nu}{j}=10$}}
        \end{subfigure}
    \caption{The Hamiltonian \eqref{eq:cartH} generates orbits on the compact phase space $S^2$, which we can think of as the spin sphere. We see that (a) at $\frac{\nu}{j}=0$ phase space is split into 4 regions, (b) when $\nu\neq 0$ these regions merge in pairs as orbits start crossing over, (c) eventually the hyperbolic fixed points disappear and (d) when $\nu\gg j$ the orbits become circles at fixed $J_3$.}\label{fig:orbits}
\end{figure}

In the current section we will only consider the leading large-$j$ behavior, so for now we can drop the $-\tfrac12$ above. The classical Hamiltonian then takes the elegant form 
\begin{equation}\label{eq:cartH}
    H_j = j XY + \nu Z\,, \qquad X^2+ Y^2 + Z^2 = 1\,,
\end{equation}
in Cartesian coordinates. Phase space orbits for different values of $\frac{\nu}{j}$ are shown in fig.\,\ref{fig:orbits}. For small $\frac{\nu}{j}$, there are 2 hyperbolic and 4 elliptic fixed points. The energy of the elliptic ones grows with $j$. The  hyperbolic ones, at the north and south poles, instead have fixed energy $\pm \nu$. Close to the poles, the orbits look just like those of the upside-down harmonic oscillator in fig.\,\ref{fig:hyporb}. The hyperbolic fixed points are therefore responsible for the emergent de-Sitter-like dynamics in the spin model at large $j$. In the figure one also sees that for larger $\frac{\nu}{j}$ the orbits start crossing over and eventually we are left with 2 elliptic fixed points at the poles. That the number of elliptic minus hyperbolic fixed points always equals 2, the Euler characteristic of the sphere, is a cute instance of Morse theory.

Now we can try to find the logarithmic tails of \eqref{eq:exactrho} in semiclassical quantization\footnote{As one can check explicitly for the upside-down harmonic oscillator,  with this method we should indeed only expect to be able reproduce the behavior far away from the resonance peaks.}, where the density of states at energy $\omega$ is given in terms of the periods as:
\begin{equation}\label{eq:semirho}
    \rho(\omega) = 2\cdot\frac{T(\omega)}{2\pi \hbar} \, .
\end{equation}
A factor $2$ was included since there will be two distinct orbits contributing the same $T(\omega)$. To calculate the periods, we can use angular coordinates $\phi \in [0,2\pi),\,\varphi\in[0,\pi]$ on the sphere. The Hamiltonian and symplectic form \eqref{eq:symp} then become\footnote{Already here one could try and draw the conclusion that the small-$\phi$ limit $\sin 2\phi \approx 2\phi$ yields the de Sitter $H$ in \eqref{eq:conformal}. We should be careful with that argument though, since $\phi$ is not conserved by the dynamics.}: 
\begin{equation}\label{eq:classSpin}
    H_j = \tfrac{j}{2} \sin^2\varphi \sin 2\phi + \nu \cos \varphi\,, \qquad \Omega = j\hbar \;\rmd \phi \wedge \rmd\cos \varphi \,.
\end{equation}
In what follows we will put $\hbar=1$ again. The Hamilton equations are
\begin{equation}\label{eq:classeom}
   \dot{\varphi} =  \sin \varphi \cos 2\phi, \quad \dot{\phi} = -\cos \varphi \sin2\phi + \tfrac{\nu}{j}\, .
\end{equation}
Since the energy $H_j = \omega$ is conserved, we can further elimate $\varphi$:
\begin{equation}\label{eq:phidotsq}
    \dot{\phi}^2 = \sin^2 2\phi - \tfrac{2\omega}{j}\sin2\phi+\tfrac{\nu^2}{j^2}\,.
\end{equation}
If we then define $a,b$ through the relations
\begin{equation}\label{eq:sinasinb}
    \sin a \,\sin b \equiv \tfrac{\nu^2}{j^2}\,, \quad \sin a + \sin b \equiv \tfrac{2\omega}{j}\,, \qquad a< b\leq \tfrac{\pi}{2}\,,
\end{equation}
we find that the period is given by
\begin{equation}\label{eq:period}
  T(\omega) = 2\int^{\frac{\pi-b}{2}}_{\frac{b}{2}} \frac{\rmd \phi}{\sqrt{(\sin 2\phi - \sin a)( \sin 2\phi-\sin b)}}=\frac{4\,\text{K}\Big(\frac{(1+\sin a)(1-\sin b)}{(1-\sin a)(1+\sin b)}\Big)}{\sqrt{(1-\sin a)(1+\sin b)}}\,,
\end{equation}
in terms of the elliptic $\text{K}$-function\footnote{Using either Mathematica or identities like Abramowitz \& Stegun 17.4.8.}. We are interested in the high-energy tails of the spectrum at large $j$, meaning $j\gg \omega \gg \nu$. Then it follows from \eqref{eq:sinasinb} that $\sin a \to \frac{\nu^2}{2j \omega}$ and $\sin b \to \frac{2\omega}{j}$. Combining this with \eqref{eq:semirho} and \eqref{eq:period} yields the difference in density of states
\begin{equation}
  \rho(\omega)-\rho(\text{\textomega})\, \to \,\frac{4\,\text{K}\big(1-\frac{4\omega}{j}\big)}{\pi\,\sqrt{1+\frac{2\omega}{j}}} - \frac{4\,\text{K}\big(1-\frac{4\text{\textomega}}{j}\big)}{\pi\,\sqrt{1+\frac{2\text{\textomega}}{j}}} \,\to \,\frac{2}{\pi}\,\log\frac{\text{\textomega}}{\omega} \;,
\end{equation}
compared to a reference scale \textomega. This matches the large-$\omega$ behavior of the exact \eqref{eq:exactrho}.

The goal now is to also get analytic control near the resonance peaks. As explained in sec.\,\ref{sec:poschl}, for potentials which approximate an upside-down harmonic oscillator,  changes in the resonance spectrum can be calculated perturbatively. This will prove to be powerful in combination with the coherent spin path integral, as we shall discuss next.

\subsection{Coherent spin path integral and saddle points}
We will now compute the large-$j$ limit of the coarse-grained character \eqref{eq:smearchar}.
Inserting the completeness relation \eqref{eq:cohspin2} for coherent spin states, it takes the form
\begin{equation}\label{eq:spinchar}
  \chi_{j,\epsilon}(t) = \frac{1}{\sqrt{2\pi}\epsilon}\,\int^\infty_{-\infty} \rmd t' \,\rme^{-(t-t')^2/2\epsilon^2}\int \rmd^2 z \,\mu(z, \bar{z}) \,\bra{z}\rme^{-\rmi t' H_j}\ket{\bar{z}}\,.
\end{equation}
To see the semiclassical limit, we can use the path integral representation \cite{VIEIRA1995331, Anninos:2015eji, stone_00}
\begin{equation}\label{eq:spinpath}
   (z_f|\rme^{-\rmi T H}|\bar{z}_i)  = \int [\CD z\CD\bar{z} \,\mu(z,\bar{z})]\, \rme^{\rmi S(z,\bar{z})}\,,
\end{equation}
with the measure $\mu(z,\bar z)$ defined in \eqref{eq:cohspin2}, and the action:
\begin{equation}
    S =  \rmi \,j \int^T_0 \rmd t \,\frac{z \dot{\bar{z}}- \dot{z}\bar{z}}{1+z\bar{z}} - \int^T_0 \rmd t\, H_j(z,\bar{z}) - S_{\text{bdy}}\,.
\end{equation}
The symbol $H_j(z,\bar z)$ was given in \eqref{eq:hamsymbol} and the boundary term takes the form
\begin{equation}
    S_{\text{bdy}} = \rmi j\, \big(\log (1+ z(0)\bar{z}_i) +  \log (1+\bar{z}(T)z_f)\big)
\end{equation}
The boundary conditions are $\bar{z}(0)= \bar{z}_i$ and $z(T)= z_f$. It is instructive to compare the above expressions to \eqref{eq:cohpath} for the standard coherent states, especially in the limit \eqref{eq:spintoHOlimit}. 

When $j\gg 1$, a saddle-point approximation becomes appropriate. In this limit, the dominant contributions to the character come from classical solutions\footnote{As in sec.\,\ref{sec:cohstate}, the dominant saddle for the propagator (rather than the character) generically gets complexified. For the character this matters when taking into account fluctuations, as we will do below. } which after a time $t$ end up back where they started  \cite{Combescure, HELLER1975321, MWilkinson_1987, LITTLEJOHN1986193}. See also the discussion at the end of sec.\,\ref{sec:cohstate}. Fixed points of the Hamiltonian flow satisfy this requirement for any $t$. Apart from them, there is also a pair of non-trivial orbits with period $T(\omega)=t$, as in fig.\,\ref{fig:orbits}.  The leading contribution from each of these saddles is discussed below. The first perturbative correction to the dominant saddle is calculated in sec.\,\ref{sec:spincorr}.

\subsubsection{Elliptic fixed points}
When $j\gg\nu$, there are 4 elliptic fixed points. To leading order they are located at $z = \rmi^{n+\frac12}$, as can be seen in fig.\,\ref{fig:orbit1}. Their on-shell action $S_{\text{cl}}= \pm\frac{\rmi \,j}{2} t$ grows with $j$. The coarse-graining in \eqref{eq:spinchar} then renders their rapidly oscillating contributions exponentially small:
    \begin{equation}
        \chi_{j,\epsilon, \text{elliptic}} \sim \int^{\infty}_{-\infty}\rmd t'\, \rme^{-(t-t')^2/2\epsilon^2}\,\rme^{\pm \rmi j t/2} \sim \rme^{-(\epsilon j)^2/8\,\pm \,\rmi j t/2}\,.
    \end{equation}
A suitable coarse-graining at large $j$ will then consist in having both $\epsilon \ll 1$ and $\epsilon j \gg 1$. 
    
\subsubsection{Periodic orbits} 
The on-shell action of the non-trivial orbits with period $T(\omega)=t$ also scales linearly with $j$.\newline Fortunately, we do not require its explicit form to see when these orbits start contributing to \eqref{eq:spinchar}. To get a rough estimate, note that coarse-graining has removed contributions coming from $\omega \gtrsim \epsilon^{-1}$. At the same time, we know from \eqref{eq:period} that the period scales like $T(\omega) \sim \log j - \log \omega$. This means that the remaining orbits\footnote{Though beyond the scope of the current work, it should in principle be possible to apply the Gutzwiller trace formula \cite{Gutzwiller:1971fy, Combescure} to determine the periodic orbit contributions at fixed $j$. However, the form of the on-shell action \eqref{eq:Sonshell} (in angular coordinates) does not look very encouraging. } with $\omega \lesssim \epsilon^{-1}$ can only start contributing when $t \gtrsim \log \epsilon j$.  For a numerical confirmation of this scaling, see fig.\,\ref{fig:charcutoffs}. 

\subsubsection{Hyperbolic fixed points} 
Finally, we have the fixed points at $z=0$ and $z=\infty$, which exist for any value of $j$ and $\nu$,\newline see fig.\,\ref{fig:orbits}. In the case of interest, $j\gg\nu$, they are hyperbolic. Moreover, since $H_j(z,\bar z)=\mp \nu$ there, the on-shell action $S_{\text{cl}} = \pm  \nu t$ is independent of $j$, in contrast to the other saddles.

To evaluate how fluctuations around the $z=0$ saddle contribute to the character, we go to Darboux coordinates $u$ \eqref{eq:darboux}, which trivialize the measure. Zooming in near $z=0$ is done by keeping $u$ fixed and sending $j\to \infty$. As shown in \cite{Arecchi:1972td} and reviewed in app.\,\ref{app:cohspin}, this also turns coherent spin states into harmonic oscillator coherent states. The leading limit of the symbol $H_j(z,\bar z)\to \tfrac{\rmi}{2}(\bar{u}^2-u^2) - \nu$, is then recognized as the symbol of the upside-down harmonic oscillator, which we already solved in sec.\,\ref{sec:cohstate}. Alternatively, we could see this by scaling $z\to \frac{u}{\sqrt{2j}}$ in \eqref{eq:holoH}, leading at large $j$ to the ``free" Hamiltonian     
    \begin{equation}\label{eq:H0}
    H_0 =  \frac{\rmi}{2}(\partial_u^2 - u^2) - \nu\,.
    \end{equation} 
For the $z=\infty$ saddle $\tilde H _0 = -H_0$: the now familiar sign flip in the north-south map \eqref{eq:mapNS}. 

The leading contribution then follows from the upside-down oscillator result \eqref{eq:cohstatechi}:
    \begin{equation}\label{eq:saddlechi}
    \lim_{j\to\infty} \chi_{j,\text{hyp}}(t) =  \int \frac{\rmd^2 u}{\pi} \, \rme^{-u\bar{u}} \, (u|\rme^{-\rmi t H_0}|\bar{u}) + \text{c.c.} = \frac{\rme^{-t \bar\Delta}+\rme^{-t\Delta}}{|1- \rme^{-t}|} = \chi_{\dS_2}(t) \,.
    \end{equation}
Thus, the two hyperbolic saddles in the large-$j$ limit of the spin model yield the $\dS_2$ character. 

\begin{figure}
    \centering
  \begin{subfigure}{0.424\textwidth}
            \centering
            \begin{tikzpicture}
              \node[inner sep = 0pt] at (7,3) {\includegraphics[width=\textwidth]{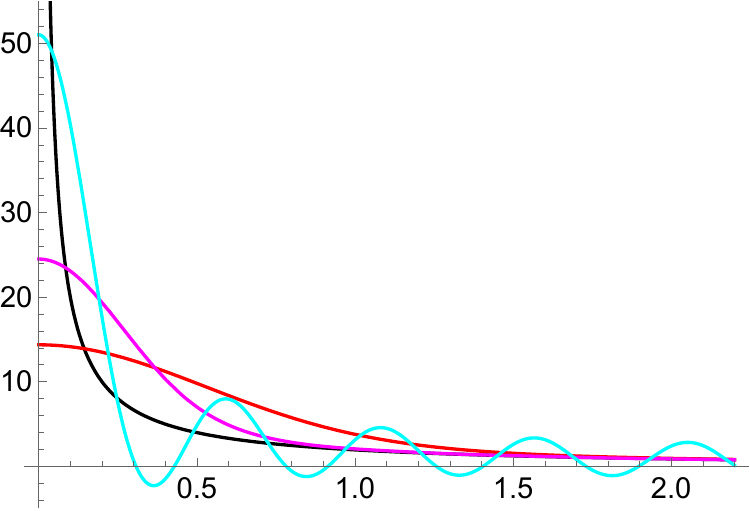}};
               \node (x2) at (4.25,5.25) {\text{$\chi$
        }};\node (jplus) at (10.55, 1.45) {\text{$t$
        }};
        \end{tikzpicture}
            \caption[]%
            {{\small  UV cutoff at small t}} 
        \end{subfigure}
        \hspace{0.5cm}
        \begin{subfigure}{0.424\textwidth}  
            \centering 
            \begin{tikzpicture}
              \node[inner sep = 0pt] at (7,3) {\includegraphics[width=\textwidth]{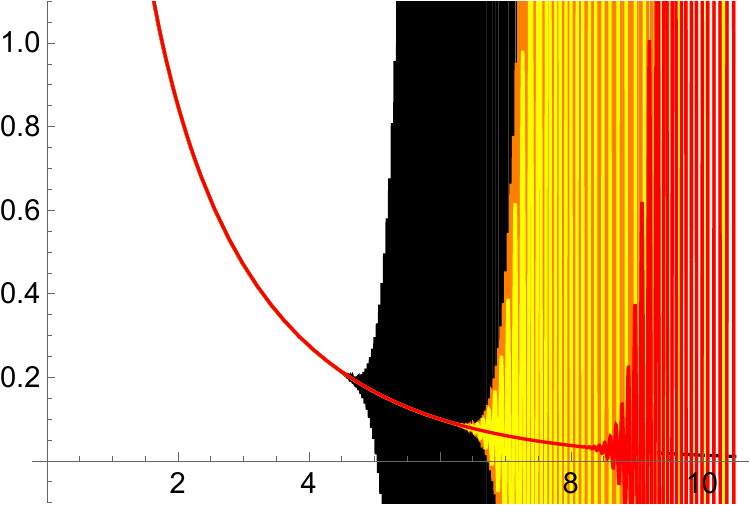}};
               \node (x2) at (4.25,5.25) {\text{$\chi$
        }};\node (jplus) at (10.85, 1.45) {\text{$t$
        }};
        \end{tikzpicture}
            \caption[]%
            {{\small  IR cutoff at large t}}
        \end{subfigure}
    \caption{In (a) we take $j=25$ and  $\nu=0$, plotting the character \eqref{eq:smearchar} with coarse-graining $\epsilon=0,\,0.25,\, 0.5$ in cyan, magenta and red. For $\epsilon=0$ the state count at $t=0$ is correct, but oscillations around the $\dS_2$ result in black are not suppressed. Non-zero $\epsilon$ tames these oscillations, at the expense of modifying the behavior for $t < t_{\text{\tiny UV}}$, which scales linearly with $\epsilon$. In (b) we see how for $t>t_{\text{\tiny IR}}$ periodic orbit contributions kick in, leading to rapid oscillations.  We show $j=500$ with $\epsilon = 0.02,\,0.02 \rme$ in black and yellow, and $j=\lfloor500\rme \rfloor$ with the same $\epsilon$ in orange and red. The results are consistent with $t_{\text{\tiny IR}}\sim \log \epsilon j$.}
    \label{fig:charcutoffs}
    \end{figure}
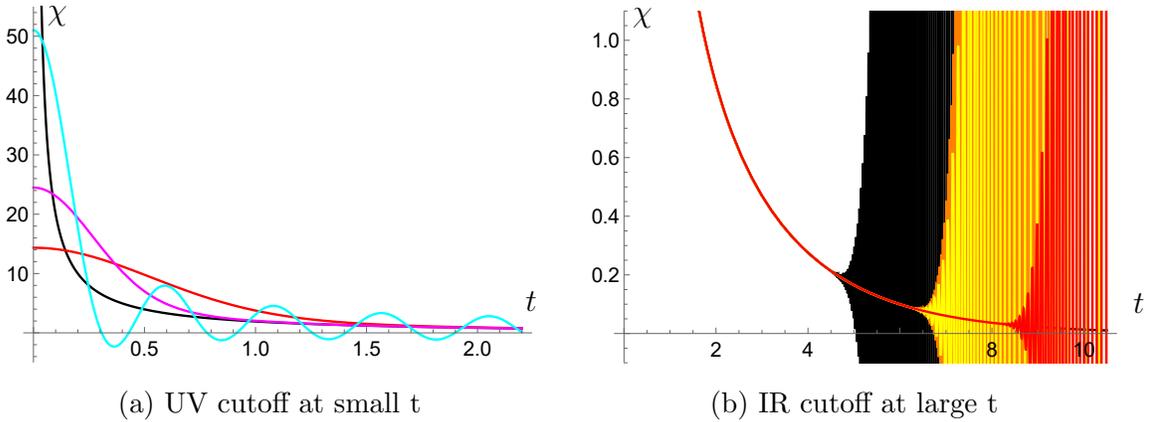

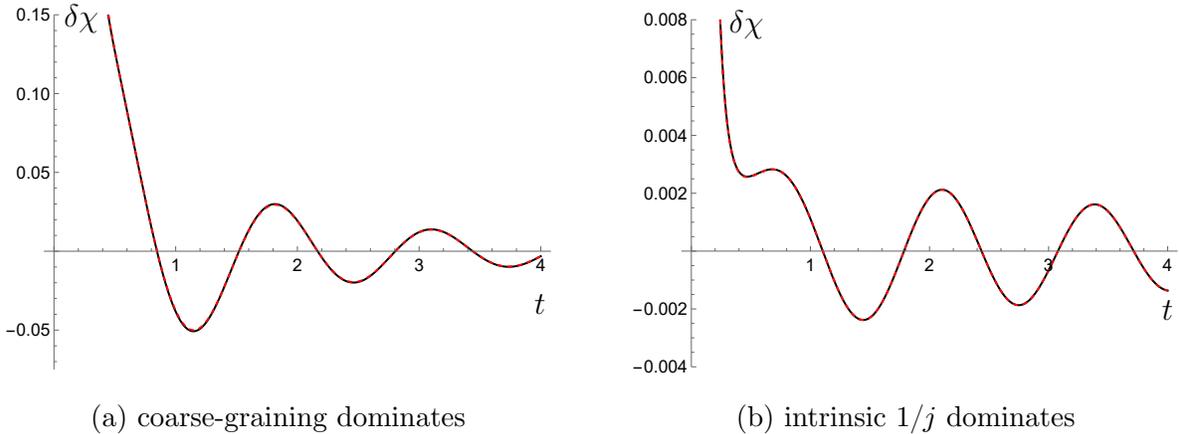
\begin{figure}
    \centering
  \begin{subfigure}{0.44\textwidth}
            \centering
            \begin{tikzpicture}
              \node[inner sep = 0pt] at (7,3) {\includegraphics[width=\textwidth]{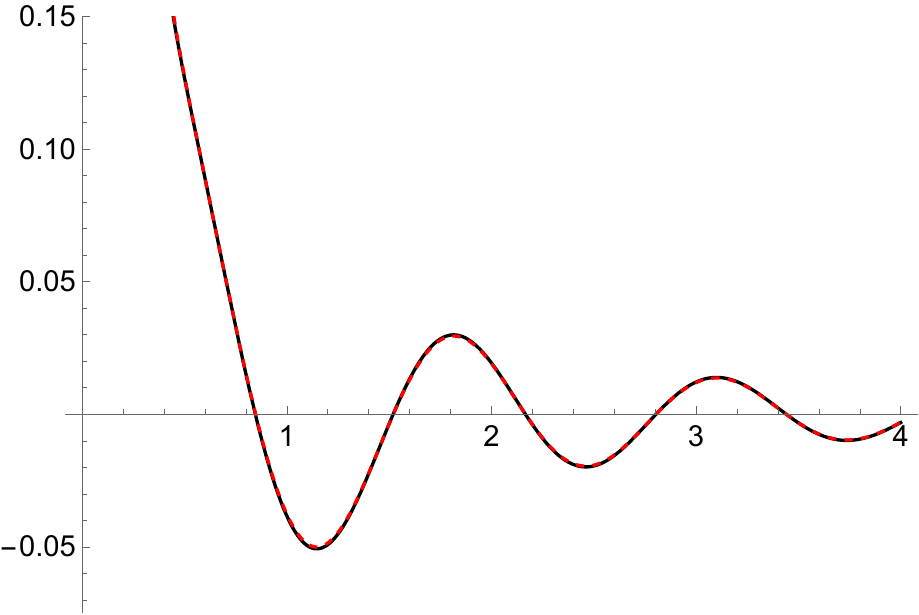}};
               \node (x2) at (4.45,5.25) {\text{$\delta\chi$
        }};\node (jplus) at (10.55, 1.45) {\text{$t$
        }};
        \node[white!10]  at (10.55, 0.63) {\text{$.$
        }};
        \end{tikzpicture}
            \caption[]%
            {{\small  coarse-graining dominates}}\label{fig:smearing} 
        \end{subfigure}
        \hspace{0.8cm}
        \begin{subfigure}{0.44\textwidth}  
            \centering 
            \begin{tikzpicture}
              \node[inner sep = 0pt] at (7,3) {\includegraphics[width=\textwidth]{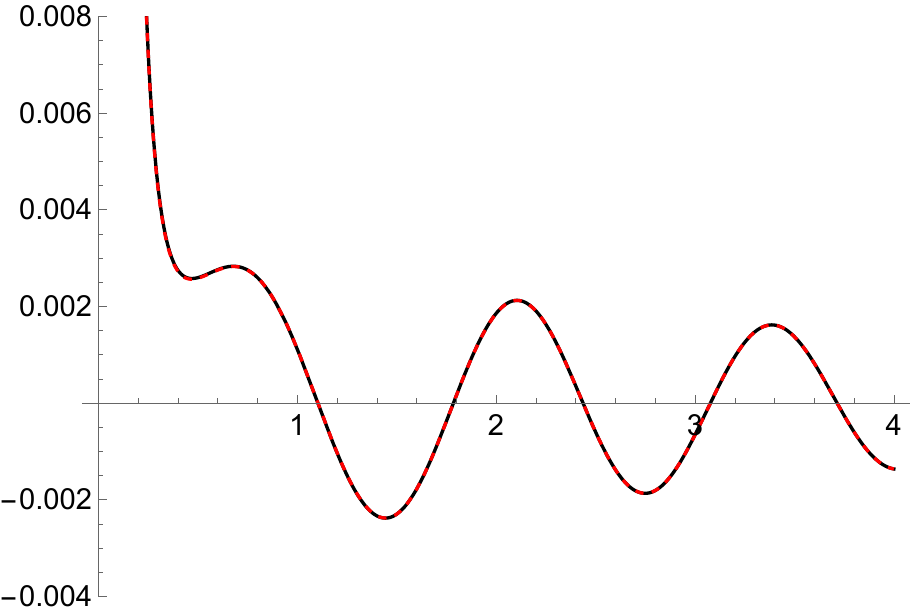}};
               \node (x2) at (4.95,5.25) {\text{$\delta\chi$
        }};\node (jplus) at (10.55, 1.45) {\text{$t$
        }};
        \end{tikzpicture}
            \caption[]%
            {{\small  intrinsic $1/j$ dominates}}
        \end{subfigure}
    \caption{The dashed red graph shows the difference $\chi_{j,\epsilon} - \chi_{\dS_2}$ between the numerical evaluation of the coarse-grained character \eqref{eq:smearchar} at $j=1000$ and $\nu=5$ and the exact $\dS_2$ character \eqref{eq:chards2}. The black lines are the large-$j$ analytic prediction $\eqref{eq:1overj} + \eqref{eq:smearingerr}$ for this quantity, which matches the numerical result very well. In (a) we took a rather large $\epsilon=0.05$, for which the coarse-graining error \eqref{eq:smearingerr} dominates. In (b)  $\epsilon=0.008$ so that the intrinsic correction \eqref{eq:1overj} is the larger one. The errors vanish when both $\epsilon\to 0$ and $\epsilon j \to \infty$.  }
    \label{fig:numchar}
\end{figure}

It remains to discuss the coarse-graining in \eqref{eq:spinchar}. Besides exponentially suppressing the other saddles, it introduces a correction to the above hyperbolic contribution, since for $\epsilon\ll 1$:
\begin{equation}\label{eq:smearingerr}
    \delta\chi_\epsilon(t) = \chi_\epsilon(t)-\chi(t) \approx \frac{\epsilon^2}{2} \chi''(t)\,.
\end{equation}
For \eqref{eq:saddlechi}, the relative size of this correction becomes of order one on scales $t\lesssim t_{\text{\tiny UV}}= \epsilon$. Further, at a much smaller timescale $t\lesssim j^{-1}$ the saddle-point approximation breaks down. In particular, at $t=0$ the entire phase contributes: the fine-grained character $\chi_{j}(0)= 2j+1$ then simply counts the total number of states. Finally, recall that periodic orbits enter the picture when $t\gtrsim t_{\text{\tiny IR}} = \log \epsilon j$. These different effects can be seen in figs.\,\ref{fig:charcutoffs} and \ref{fig:smearing}.

The take-away message is then that for $t_{\text{\tiny UV}}\lesssim t\lesssim t_{\text{\tiny IR}}$, the large-$j$ limit of the coarse-grained spin model character \eqref{eq:spinchar} will be dominated by the contribution coming from the hyperbolic fixed points, which, as we just showed,  converges to the $\dS_2$  result \eqref{eq:chards2}. Moreover, taking $\epsilon = j^{-\alpha}$ with $\alpha \in (0,1)$, we get an increasingly large window of convergence.

Finally, we can ask which spin states are the emergent QNMs? Manipulating (C.8) in \cite{Arecchi:1972td} for the associated Legendre polynomials $\mathsf{P}^m_l$, we find that, again with $z=\frac{u}{\sqrt{2j}}$:  
\begin{equation}
  (4\pi j)^{\frac14} \Big(\frac{(4j-n)!}{n!}\Big)^{\frac12}\, \mathsf{P}^{n-2j}_{2j}(z) \xrightarrow{\makebox[0.8cm]{$\scriptscriptstyle{j\to \infty}$}} \psi_n(u) \, ,
\end{equation}
where $\psi_n(u)$ is the QNM defined in \eqref{eq:cohresbas}. This follows from the Rodrigues formulas
\begin{equation}
  \mathsf{P}^m_l = \frac{\;(-1)^{l+m}}{2^l \,l!} \,(1-u^2)^{\frac{m}{2}}\, \partial^{l+m}_u \,(1-u^2)^l \, ,\qquad \mathsf{H}_n(u) = (-1)^n\,\rme^{u^2} \,\partial^n_u\,\rme^{-u^2}\,.
\end{equation}

\newpage
\subsection{Finding the \texorpdfstring{$1/j$}{} corrections }\label{sec:spincorr}
To find the first subleading correction to the hyperbolic saddles, we again expand \eqref{eq:holoH} in Darboux coordinates, now keeping the $1/j$ terms: $z\to \tfrac{u}{\sqrt{2j}}(1+\tfrac{u\partial_u}{4j})$ and $\partial_z \to (1-\tfrac{u\partial_u}{4j}){\scriptstyle\sqrt{2j}}\partial_u$. This has the following effect on the spin operators \eqref{eq:holo}: 
\begin{equation}
  J_- \to\, \big(1-\frac{u\partial_u}{4j}\big){\sqrt{2j}}\,\partial_u, \qquad   J_+ \to \sqrt{2j}\,u\,\big(1-\frac{u\partial_u}{4j}\big)\, , \qquad J_3 \to  u\partial_u - j \,,
\end{equation}
which is the same as found by expanding the Holstein-Primakoff operators \cite{Holstein:1940zp} to this order. Plugging this into $\eqref{eq:holoH}$ then gives
  \begin{equation}\label{eq:Hdecomp}
    H_j= H_0  + {H_\text{int}}\,,
    \end{equation}
where $H_0$ was found in \eqref{eq:H0} and the first subleading correction takes the form 
\begin{equation}\label{eq:H1int}
     H^{(1)}_{\text{int}} =\frac{\rmi}{8j} \Big(u^2-\partial_u^2 \Big)+  \frac{\rmi}{4j} \Big(  u^3\partial_u-u\partial^3_u \Big) + \nu \frac{u\partial_u}{j}\,.
\end{equation}
This $H^{(1)}_{\text{int}}$ introduces a correction to the spectrum, which we want to evaluate perturbatively. As in sec.\,\ref{sec:poschl}, we will do so by making use of the QNM basis. This is further simplified by noting that, with $c_\pm$ as defined in \eqref{eq:raiselower},
\begin{equation}\label{eq:H1corr}
  H^{(1)}_{\text{int}} =  \frac{1}{2j}H_0 + \frac{1}{4j}(c_+ H_0\, c_+ - c_-H_0 \,c_-) + \frac{\nu}{4j}(c^2_- - c^2_+)\,.
\end{equation}
Since $c_\pm$ are the QNM raising and lowering operators, only the first term contributes to the $1/j$ correction $\langle{\tilde\psi_n}|H^{(1)}_{\text{int}}\ket{\psi_n}$, which is then simply proportional to the original  spectrum:
\begin{equation}
    \delta \omega_n =\frac{\omega_n}{2j}= -\frac{\rmi}{2j}(n+\frac12) - \frac{\nu}{2j}\,,\qquad  n\in \mathbb{N}\, .
\end{equation}
To this order then, adding also the $\nu \to -\nu$ contribution of the $z=\infty$ saddle, the overall effect on the character is a shift in its time-dependence:
\begin{equation}\label{eq:1overj}
    \delta \chi_j(t) = \tr\rme^{-\rmi t (H_0+ H^{(1)}_{\text{int}})}-  \tr\rme^{-\rmi t H_0} = \chi(t(1+\tfrac{1}{2j}))- \chi(t) \approx \frac{t}{2j}\,\chi'(t)\, .
\end{equation}
This intrinsic $1/j$ correction is (sub)dominant compared to the coarse-graining one \eqref{eq:smearingerr}, depending on whether $(\epsilon > j^{-1/2})$ or $\epsilon < j^{-1/2}$. By varying this relative size, both effects can be distinguished in the numerical results, as  shown in fig.\,\ref{fig:numchar}. 
Since the final result has the simple effect of rescaling time by $1+\tfrac{1}{2j}$, one may wonder if there exists a simpler way of arriving at this conclusion. At least at the level of \eqref{eq:H1int} this does not appear trivial\footnote{The same shift occurs for $\tr \rme^{\rmi t J_3}$. Seeing it from the path integral requires adding a so-called Solari-Kochetov phase \cite{stone_00}. It was clarified in \cite{pletyukhov04} that this is due to the difference between Weyl and principal symbols, and that the Holstein-Primakoff approach reproduces this phase.  Despite these complications, a path-integral approach may still prove useful to evaluate higher-order corrections directly from \eqref{eq:spinpath}.
}.  

\newpage
\section{Quasinormal discussion }\label{sec:disc}

We have encountered a rather simple and finite quantum mechanical model with Hermitian Hamiltonian \eqref{eq:HJJ} acting on the spin-$j$ representation of $\SU(2)$, and showed that its coarse-grained character converges at large $j$ to that of a massive particle in $\dS_2$, indicating the emergence of de Sitter quasinormal modes.

In fact, it turns out\footnote{I am thankful to Ana Asenjo-Garcia and Stuart Masson for bringing this to my attention.} that in the quantum optics literature, the Hamiltonian \eqref{eq:HJJ}, at least when $\nu = 0$, is known as the two-axis twisting Hamiltonian. It can be used to generate squeezed states in the sense of \cite{kitagawa}. Thinking of experimental realizations, it is interesting to note a relatively recent proposal for engineering two-axis twisting in cavity QED \cite{Borregaard_2017}. 

As the introduction suggested, we are ultimately in pursuit of de Sitter quantum gravity. It will then not have escaped the reader's notice that we have yet to talk about black holes or indeed gravity. The idea is that perhaps our toy model could arise as a matter subsector in a more mature microscopic description. As a first step in that direction, we could begin by associating a pair of fermionic creation and annihilation operators $c_\omega, c^\dagger_\omega$ to any of the $2j+1$ spin-model frequencies. This leads to finite-dimensional multi-particle Hilbert space. Along the lines of \cite{Balasubramanian:2002zh}, we can then split these oscillators into southern ones with $\omega >0$ and northern ones with $\omega <0$, and further impose the constraint $H=0$ on the physical states. Similar steps were recently taken for the double-scaled SYK model \cite{Narovlansky:2023lfz}, reproducing QNMs for the complementary rather than principal series. It will be interesting to see whether, building further on our ideas, we end up having to go in a similar direction, or not.

Ideally, the picture outlined above would provide a unified description of particles and black holes in the spirit of \cite{Banks:2006rx}. Black holes in de Sitter are not stable, so we might expect also these to eventually show up as huge resonances in the microscopic model. 

Besides the generalization to other fields, the treatment of the higher-dimensional case also remains an open question. Moreover, since the analysis in sec.\,\ref{sec:semicl} suggests that the crux lies in the emergence of two hyperbolic fixed points, the result does not appear terribly sensitive to other details, and one may wonder what happens when considering ensembles of Hamiltonians like \eqref{eq:HJJ}. All of these are ongoing efforts on which we hope to report before disappearing in the horizon \cite{denef2}. 

\newpage
\subsection*{Acknowledgements}
I wish to thank Frederik Denef for initial collaboration, and many helpful discussions. I am also grateful to Dio Anninos for insightful suggestions, to Ruben Monten, Zimo Sun and Herman Verlinde for useful comments on the draft, to the other participants of the DAMTP workshop  ``Quantum de Sitter Universe" and PCTS workshop ``Towards the Beginning of Time'' for stimulating interactions, and finally, to Ana Asenjo-Garcia and Stuart Masson for bringing to my attention the quantum optics literature on the two-axis twisting Hamiltonian. This work was supported in part by the US DOE (DE-SC011941). 

\appendix

\section{Appendix}
\subsection{More general decompositions }\label{app:gendec}
To improve our understanding of the resonance decompositions in sec.\,\ref{sec:qm}, let us consider more abstractly a family of decompositions, containing the upside-down and ordinary harmonic oscillator as subcases, starting from the following data:
\begin{align} \label{data}
 [c_-,\,c_+] = 1 \, , \qquad 
 c_- |\psi_0\rangle = 0 \, , \qquad \langle \tilde\psi_0| c_+ = 0 \, , \qquad \langle \tilde\psi_0|\psi_0\rangle = 1 \, .
\end{align} 
The commutation relation is obtained for any $\mathrm c_- = \mathrm a \, \hat p + \mathrm b \, \hat x$, $c_+ = \mathrm c  \, \hat p + \mathrm d \, \hat x$ with $\mathrm a,\mathrm b,\mathrm c,\mathrm d$ complex constants satisfying $\mathrm a \mathrm d-\mathrm b \mathrm c = \rmi$. 
Defining then \label{descendantdef}
\begin{align} \label{phichidef}
 |\psi_n \rangle = \frac{c_+^n}{\sqrt{n!}} \, |\psi_0 \rangle \, , \qquad 
 \langle \tilde\psi_n| = \langle \tilde\psi_0| \, \frac{c_-^n}{\sqrt{n!}} \,,
\end{align}
we get the relations
\begin{align} \label{decompgen}
  \langle \tilde\psi_n|\psi_m \rangle = \delta_{nm} \, , \qquad \one = \sum_n |\psi_n \rangle \langle \tilde\psi_n| \, .
\end{align}
The first equation follows directly from \eqref{data}.
The second one is to be understood as saying $\langle \alpha|\beta \rangle = \sum \langle \alpha|\psi_n\rangle \langle \tilde\psi_n|\beta\rangle$ for suitable $|\alpha \rangle$ and $|\beta\rangle$, meaning that their wave functions have sufficiently fast fall-offs. This follows from the observation that the operator 
\begin{align}
  \CI = \sum_n |\psi_n\rangle \langle \tilde\psi_n| = \sum_n \frac{1}{n!} \, c_+^n |\psi_0 \rangle \langle \tilde\psi_0 | c_-^n 
\end{align}
commutes with both $c_\pm$. Given that the Hilbert space under consideration is an irreducible representation of the Heisenberg algebra $[c_-,\,c_+]=1$, Schur's lemma implies that $\CI \,\propto \, \one$. The matrix element $\langle \tilde\psi_0|\CI|\psi_0\rangle =  1$ then fixes the proportionality constant to be $1$.

\subsection{P\"oschl-Teller exact solution}\label{app:ptexact}
It is well-known that the stationary Schr\"odinger equation \eqref{eq:ptschrod} with energy $E$ is solved by
\begin{equation}
    \psi_\pm(t,\, x) =  \rme^{-\rmi E t \pm \rmi \sqrt{2E}x} \,_2\text{F}_1\Big(\Delta,\, \bar{\Delta},\, 1\mp \frac{\rmi }{\lambda} \sqrt{2E} \,;\, \frac12(1-\tanh \lambda x)\Big)\, , 
\end{equation}
where we defined
\begin{equation}
    \Delta = \frac12 + \rmi \,\sqrt{\lambda^{-2} - \frac14}\,, \qquad \bar{\Delta} = 1- \Delta\,.
\end{equation}
Essentially the same solution arises for the Klein-Gordon equation of a scalar in $\dS_2$; only there the time-dependence will be $\rme^{\rmi \sqrt{E} t}$, due to the second-order time-derivative. The resonance or quasinormal mode frequencies are found by looking at the asymptotic expansion\footnote{ Using a similar approach one can also see that for $\lambda^{-2} =  l(l+1)$ with $l\in \mathbb{N}$, the usual attractive P\"oschl-Teller potential is reflectionless, which is due to it being the superpartner of the flat potential \cite{Cooper:1994eh}.} $\psi(t,\, x\to\pm \infty)$ and imposing either purely outgoing boundary conditions, for the resonances, or ingoing ones, for the anti-resonances.
The frequency corresponding to the $n$-th resonance is then found to be
\begin{equation}\label{eq:PTspectrum}
    E_n = -\frac12\lambda^2(\Delta + n)^2\,.
\end{equation}
As long as $\lambda n \ll 1$, the $E_n$ are approximately evenly spaced, as is the case for the upside-down oscillator, since \eqref{eq:PTspectrum} can then be expanded as
\begin{equation}\label{eq:PTspectrum2}\begin{split}
    E_n &= \frac12 -\frac12\lambda^2(n^2+n+\frac12) - \rmi \lambda(n+\frac12)\,\sqrt{1-\frac{\lambda^2}4}.\\
    &=\frac12 - \rmi \lambda (n+\frac12) - \frac12 \lambda^2(n^2+n+\frac12)+\frac{\rmi}{8}\lambda^3(n+\frac12)  + \dots\, .
    \end{split}
\end{equation}
In the same limit, the resonance wavefront at large $|x|$ looks like 
\begin{equation}
    \psi(t,\,x\to \pm \infty) \; \propto \;\, \rme^{-\frac{\rmi}2( t \mp 2x)- (t\mp x)\lambda(n+\frac12)}\,.
\end{equation}

\subsection{Single-particle phase space in \texorpdfstring{$\dS$}{}}\label{app:phase}
We want to describe a relativistic particle in embedding space $\mathbb{R}^{1,D}$, constrained to the $\dS_D$ hyperboloid. In this appendix, we will do so using first-class constraints. Besides making quantization straightforward, this also relates the $\dS$ propagator to the embedding space one by group averaging, as will be shown elsewhere \cite{bandaru}. For the purposes of this paper, we can stick to $D=2$, although the generalization to arbitrary $D$ will be clear.

\subsubsection{Constraint action and phase space}

Let $(X,P)\in \IR^{1,2}\times \IR^{1,2}$ be the standard phase space coordinates in embedding space. After defining new coordinates $(X,\CP)$ through the canonical transformation 
\begin{equation}\label{eq:canon}
    (X, \CP) = (X,\, P + mX)\,,\qquad m\in \IR \,,
\end{equation}
we can consider the constraint action
\begin{equation}\label{eq:action}
    \CS = \int \CP \cdot \dot{X}- H\,, \qquad H = L^0 \CG_0 + L^1 \CG_1\,,
\end{equation}
with Lagrange multipliers $L$ and first-class constraints $\CG$:
\begin{equation}\label{eq:constraints}
    \CG_0 = \CP\cdot \CP\,,\qquad \CG_1 = X\cdot \CP -m\,.
\end{equation}
These constraints generate the gauge symmetry
\begin{equation}\label{eq:gauge}
    (X, \,\CP) \sim (\,\rho\,(X+\alpha\, \CP),\, \rho^{-1}\CP\,)\,,\qquad \alpha,\, \rho \in \IR\,.
\end{equation}
At this point we can fix the gauge $X\cdot X=1$. Then \eqref{eq:canon} and \eqref{eq:constraints} imply $X\cdot P=0$, as well as $P\cdot P+m^2=0$, so that we are indeed describing a relativistic particle on $\dS_2$.

We have the freedom to fix another gauge though; one more adapted to a southern static patch observer. This can be done by defining light-cone coordinates
\begin{equation}
    X^\pm = X^2 \pm X^0\,, \qquad \CP^\pm = \CP^2 \pm \CP^0\,,
\end{equation}
and imposing the following gauge-fixing:
\begin{equation}\label{eq:Sgauge}
  \CP^+ = 1   \,, \qquad X^+  = 0\,.
\end{equation}
Solutions to the above can be parametrized by $(x, p)\in \mathbb{R}^{2}$ as follows:
\begin{equation}\label{eq:Spoint}
   \CP = \big(\,\frac{1+p^2}{2},\; p\, ,\; \frac{1-p^2}{2}\,\big)\,, \qquad X = (x p  - m,\; x ,\; m - x p)\,,
\end{equation}
and the symplectic quotient parametrized by $(x, p)$ comes with the standard symplectic form $\Omega = \rmd x \wedge \rmd p$. Further, the $\so(1,2)$ generators $J^{IL} = X^I \CP^L - X^L \CP^I$ are given by
\begin{equation}\label{eq:HQK}
    \CH \equiv J^{20} = m - xp\, ,\qquad
    \CQ \equiv J^{1+} = x\,,\qquad
    \CK \equiv J^{1-} = p^2 x - 2m p \,.
\end{equation}
These dilate, translate and special conformally transform the momentum $p$, which is crucial for the discussion of conformal boundary states in sec.\,\ref{sec:boundaryQM}. 

It is important to realise that the chart \eqref{eq:Spoint} is missing a point at infinity. To resolve this, we need to consider also the northern analog of \eqref{eq:Sgauge}:
\begin{equation}\label{eq:Ngauge}
  \CP^- = -1   , \quad X^-  = 0\,.
\end{equation}
Repeating the analysis then leads to solutions parametrized by $(\tilde{x}, \tilde{p})\in \mathbb{R}^{2}$ as follows:
\begin{equation}\label{eq:Npoint}
   \CP = (\frac{1+\tilde{p}^2}{2},\; -\tilde{p} ,\; \frac{\tilde{p}^2-1}{2}) , \quad X = (\tilde{x}\tilde{p} - m,\; -\tilde{x} ,\; \tilde{x}\tilde{p} -m)\,.
\end{equation}
Now we need to figure out how the northern and southern patches are glued together, as this will determine the global structure of phase space.  The most straightforward way is to find the gauge transformation \eqref{eq:gauge} which maps \eqref{eq:Sgauge} to \eqref{eq:Ngauge}. We can also take a shortcut by defining the northern counterparts of \eqref{eq:HQK}:
\begin{equation}
    \tilde \CH \equiv m -\tilde x \tilde p = - J^{20}\,, \qquad \tilde \CQ \equiv \tilde x = J^{1-}\,, \qquad \tilde \CK \equiv \tilde{p}^2 \tilde x - 2m \tilde p = J^{1+}\,.
\end{equation}
Since the $J^{IL}$ must agree between the two charts, we find 
\begin{equation}\label{eq:mapNS}
    \tilde{\CH} = - \CH\,,\qquad
    \tilde \CQ =   \CK\,, \qquad
    \tilde \CK = \CQ\,,
\end{equation}
which in turn determines
\begin{equation}\label{eq:sympcoord}
   (\tilde{x}, \tilde p) = \big(\,p^2 x-2mp ,\, -\frac{1}{p}\,\big)\,.
\end{equation}
This is the desired symplectomorphism, since $\Omega = \rmd x \wedge \rmd p = \rmd \tilde x \wedge \rmd \tilde p$. Moreover, it takes the form of an inversion, as may have been expected. 

\subsubsection{Particle trajectories and phase space coordinates}\label{sec:trajectories}
 To arrive at a more physical understanding of the coordinates \eqref{eq:Spoint} and \eqref{eq:Npoint}, we can take a closer look at the particle trajectories, since these are in one-to-one correspondence with points in phase space. We will begin by adopting a global point of view. 

 Let $\Omega_{i,f}$ be any pair of points on the past and future conformal boundary circles of $\dS_2$. Generically, there is a unique geodesic connecting them. The exception occurs when $\Omega_{i,f}$ are antipodal, in which case they are connected by an entire light-cone. In $\dS_2$ these are simply the left- and right-moving rays wrapping around to the other side of the hyperboloid.

The geodesic is the intersection in $\IR^{1,2}$ of the plane spanned by the light rays 
\begin{equation}
L_i = \alpha(-1,\; \Omega_i)\,, \qquad L_f = \beta(1,\;\Omega_f)\,, \qquad \alpha, \beta \in \IR\,,
\end{equation}
with the $\dS_2$ hyperboloid. This leads to the requirement
\begin{equation}
    \alpha \beta = \frac{1}{2+2\,\Omega_i\cdot \Omega_f} = \frac{1}{4 \cos^2\theta}\,,\qquad \Omega_i\cdot \Omega_f\equiv \cos 2\theta\,,
\end{equation}
which we can solve by parametrizing
\begin{equation}
  \alpha = \frac12\, \rme^{-\tau} \sec\theta  \,, \qquad \beta = \frac12\, \rme^{\tau} \sec\theta\,.
\end{equation}
The geodesic is then given explicitly by
\begin{equation}\label{eq:globaltraj}
    X = \sec\theta \,\big(\,\sinh\tau\,,\; \frac12(\rme^\tau \Omega_f + \rme^{-\tau} \Omega_i)\,\big)\,.
\end{equation}
Since $(\partial_\tau X)^2=-1$, $\tau$ is the proper time on the geodesic. It is also clear now that the particle indeed comes in from the point $\Omega_i$ on the past conformal boundary and moves towards $\Omega_f$ on the future boundary.  

We now want to relate these geodesics to the phase space coordinates $(x,p)$ of the previous section. The result is summarized in fig.\,\ref{fig:phasespace}. We get there by considering the point $(X,\CP)$ determined by $(x,p)$ as in \eqref{eq:Spoint}, and intersecting its gauge orbit \eqref{eq:gauge} with $\dS_2$: 
\begin{equation}
    \alpha X+ \beta \,\CP \in \dS_2 \implies \alpha^2 x^2 + 2\alpha\beta m = 1,
\end{equation}
which we can solve by parametrizing
\begin{equation}
    \alpha = \rme^{-\tau}\,, \qquad \beta = \frac{\rme^\tau - x^2 \rme^{-\tau}}{2m}\,.
\end{equation}
This determines for us the particle trajectory
\begin{equation}\begin{split}\label{eq:trajectory}
    X = \Big(\,&\frac{1+p^2}{4m}\rme^\tau+ \big(\,(xp-m) -\frac{1+p^2}{4m}x^2\,\big)\,\rme^{-\tau}\,,\;\frac{p}{2m}\rme^{\tau} + (x -\frac{p x^2}{2m} )\, \rme^{-\tau}\, ,\\
    &\frac{1-p^2}{4m}\rme^\tau + \big(\,(m-xp) -\frac{1-p ^2}{4m}x^2\,\big)\,\rme^{-\tau}\,\Big)\,.
    \end{split}
\end{equation}
One checks that once again, $\tau$ is the proper time. Now we can make a few observations:
\begin{enumerate}
    \item Comparing \eqref{eq:trajectory} to \eqref{eq:globaltraj}, we find that
    \begin{equation}\label{eq:omegatop}
        \Omega_f = (\frac{2p}{1+p^2},\, \frac{1-p^2}{1+p^2})\,.
    \end{equation}
    Evidently, $p$ are coordinates on the boundary circle obtained via stereographic projection through the north pole, and label the asymptotic direction of particle trajectories. Similarly so, up to a minus sign, for $\tilde{p}$ and projection through the south pole. This minus sign ensures that points labeled by the same $p$ and $\tilde p$ are antipodal. The above also shows that $\CP$ lies on the light ray $L_f$.
    \item The geodesic \eqref{eq:trajectory} reaches the past horizon $X^+= 0$ when $\rme^{\tau}=|x|$, corresponding to the embedding space time $X^0 = (xp - m)/|x|$. In general dimensions it is reached from the direction $x^i/|x|$. In $\dS_2$ this is simply a sign; the only options are `from the left' or `from the right'.
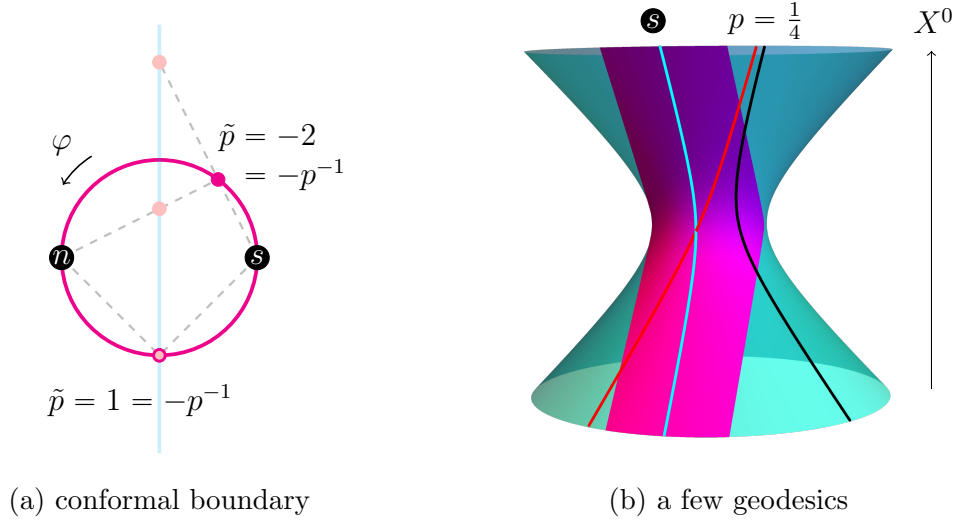
\begin{figure}
    \centering
  \begin{subfigure}{0.45\textwidth}
            \centering
                        \begin{tikzpicture}[remember picture, overlay,scale=1.,transform canvas={shift={(0cm,-0.7cm)}}]
        \draw[magenta, line width = 0.50mm](0,3.4) circle (1.3);
        \draw[cyan, line width = 0.50mm, opacity=0.2](0,0.8) -- (0,6.5);
        \draw[gray, dashed, line width = 0.30mm, opacity=0.5](-1.3,3.4) -- (0,2.1);
        \draw[gray, dashed, line width = 0.30mm, opacity=0.5](1.3,3.4) -- (0,6);
        \draw[gray, dashed, line width = 0.30mm, opacity=0.5](-1.3,3.4) -- (0.78,4.44);
        \draw[gray, dashed, line width = 0.30mm, opacity=0.5](1.3,3.4) -- (0,2.1);
        \draw[black, fill= black] (-1.3,3.4) circle (0.16);
        \draw[black, fill= black] (1.3,3.4) circle (0.16);
        \draw[magenta, fill= magenta] (0,2.1) circle (0.09);
        \draw[magenta, fill= magenta] (0.78,4.44) circle (0.09);
        \draw[pink, fill= pink] (0,6) circle (0.09);
        \draw[pink, fill= pink] (0,4.05) circle (0.09);
        \draw[pink, fill= pink] (0,2.1) circle (0.05);
        \node[white] at (-1.3,3.4) {\text{$n$}};
        \node[white] at (1.3,3.4) {\text{$s$}};
        \node[black] at (-0.25,1.5) {\text{$\tilde p = 1 = -p^{-1}$}};
        \node[black] at (1.44,5.) {\text{$\tilde p = -2 $}};
        \node[black] at (1.8,4.5) {\text{$=- p^{-1}$}};
        \node[black] at (-1.3,4.9) {\text{$\varphi$}};
        \draw[->,semithick] (-0.9,4.75) arc[radius=1.2, start angle=122, end angle=148];
        \end{tikzpicture}
            \caption[]%
            {{\small  conformal boundary}} 
        \end{subfigure}
        \begin{subfigure}{0.45\textwidth}  
            \centering 
            \begin{tikzpicture}              \node[inner sep = 0pt] at (6,3.5) {\includegraphics[trim= 1.2cm 0.5cm 0 0,width=0.8\textwidth]{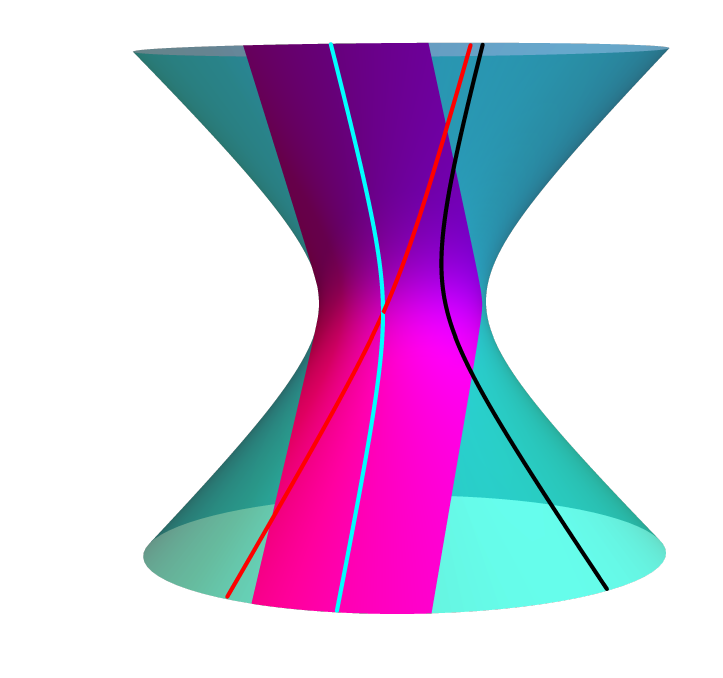}};
               \node (x2) at (9.1,6.4) {\text{$X^0$
        }};
        \draw[->] (9.,1.5) -- (9,6);
        \draw[black, fill= black] (5.3,6.4) circle (0.16);
        \node[black] at (6.8,6.4) {\text{$p=\frac14$}};
        \node[white] at (5.3,6.4) {\text{$s$}};
            \end{tikzpicture}
            \caption[]%
            {{\small a few geodesics}}
        \end{subfigure}
    \caption{Points on the future conformal boundary (a) are be labeled by $p,\, \tilde p$ or $\varphi$, which are related by inversion \eqref{eq:sympcoord} and stereographic projection \eqref{eq:phiandJ}. These points correspond to the asymptotic direction of particle trajectories,
    as seen in (b), where we plot geodesics \eqref{eq:trajectory} for $(x,p)=(0,0)$, $(-\frac12,\frac14)$ and $(1,\frac14)$ in cyan, red and black respectively. The first one is the worldline of the southern observer, the static patch of which is highlighted in magenta. The coordinate $x$ determines when and from which direction the past horizon is entered.}
    \label{fig:phasespace}
\end{figure}
    \item Similarly, the future horizon $X^-= 0$ is reached when $\rme^{\tau}= |x - \frac{2m}{p}|$, corresponding to $X^0 = (m- \tilde{x}\tilde{p})/|\tilde x|$. In general dimensions, it is reached in the direction determined by $\tilde{x}^i \equiv 2 p^i x\cdot p- p^2x^i - 2m p^i$. For $\dS_2$ this reduces to $\tilde{x}$ as previously defined in \eqref{eq:sympcoord} and the direction corresponds to the sign of $\tilde{x}$.
\end{enumerate}

 To conclude this section, note that if we write $\Omega_f = (\sin \varphi,\,\cos \varphi)$, and let $\mathcal{J}$ be conjugate to $\varphi$, then \eqref{eq:omegatop} gives us the relation
\begin{equation}\label{eq:phiandJ}
    \varphi = \arcsin(\frac{2p}{1+p^2})\,, \qquad \mathcal{J}= \frac12(1+p^2)\,x-m p\, ,
\end{equation} 
between {\it planar} coordinates $p$ and {\it global} coordinates $\varphi$ on the conformal boundary. Under the north-south map \eqref{eq:sympcoord} we get $\varphi \to \varphi + \pi$, while $\mathcal{J}$ is invariant. Both coordinate systems play a useful role. The planar one makes the relation to the upside-down oscillator manifest. On the other hand, from the global perspective, it becomes natural to understand phase space as the cotangent bundle $T^*(S^{1})$. Upon quantization this results in the Hilbert space $L^2(S^1)$ of square-normalizable wave functions on the future conformal boundary. Both of these points are discussed more in sec.\,\ref{sec:desitter}.

\newpage
\subsection{Coherent states and holomorphic wave functions}
In this appendix we collect several definitions and useful formulas related to coherent states, both for the harmonic oscillator, and for coherent spin states.
\subsubsection{Harmonic oscillator states}\label{app:cohho}
Let $a_\pm= \frac{\hat p \pm \rmi\hat x}{\sqrt{2}}$ be the raising and lowering operators for the standard harmonic oscillator, and denote the ground state by $\ket{0}$. Coherent states are then defined for any $u\in \IC$ as
\begin{equation}\label{eq:harmcoh}
    |\bar{u}) = \rme^{\bar{u} a_+}\ket{0}\,, \qquad (u| = \bra{0}\rme^{u a_-}\,.
\end{equation}
They satisfy the following overlaps and completeness relation:
\begin{equation}\label{eq:coherentcomp}
    (v|\bar{u}) = \rme^{v \bar{u}}\,, \qquad \int \frac{\rmd^2 u}{\pi} \, \rme^{-u \bar{u}}\,|\bar{u})(u| = \one\,.
\end{equation}
For any other state $\ket{\psi}$, we can now define its holomorphic wave function as
\begin{equation}\label{eq:holodef}
\psi(u) = (u\ket{\psi}\,.
\end{equation}
Given \eqref{eq:coherentcomp}, the inner product can then be written as
\begin{equation}\label{eq:cohinner}
  \braket{\xi}{\psi} = \int\frac{\rmd^2 u}{\pi}\,\rme^{-u\bar u}\, \bar\xi(\bar u)\, \psi(u) \,.
\end{equation}
The $a_\pm$ have elegant holomorphic representations:
\begin{equation}
     (u| a_+\ket{\psi}=u\, \psi(u)\, ,\qquad  (u| a_-\ket{\psi}=\partial_u \psi(u)\,,
\end{equation}
which also imply that position and momentum operators are represented by
\begin{equation}
   (u|\hat{x}\ket{\psi}=\tfrac{\rmi}{\sqrt{2}}(\partial_u-u)  \psi(u)\,,\qquad  (u|\,\hat{p}\ket{\psi}=\tfrac{1}{\sqrt{2}}(\partial_u + u) \psi(u) \,.
\end{equation}
The time evolution of coherent states can be written as a path integral
\begin{equation}\label{eq:cohpath}
 (u_f|\rme^{-\rmi T H} |\bar{u}_i)   = \int \CD \bar u\, \CD u\,  \rme^{\rmi S}, \quad S = \frac{\rmi}{2} \int^T_0  \rmd t\,\big(\dot{\bar{u}} u- \bar u \dot{u} \big) - \int^T_0  \rmd t\, H(u,\bar{u})  - S_{\text{bdy}} \, .
\end{equation}
One usually calls $H(u,\bar{u})$ the symbol of $H$. It is simply the normalized expectation value 
\begin{equation}
    H(u,\bar u ) = \frac{(u| H |\bar u )}{(u|\bar u)}\,.
\end{equation}
The boundary term $S_{\text{bdy}}$ is given by
 \begin{equation}
   S_{\text{bdy}} =    \frac{\rmi }{2}\big (u(0)\bar{u}_i + \bar{u}(T)u_f\big ) \,.
 \end{equation}
 The path integral \eqref{eq:cohpath} has to be evaluated with boundary conditions $u(T) = u_f$ and $\bar{u}(0) = \bar{u}_i$. This generically complexifies the paths, meaning that $u$ and $\bar{u}$ are no longer each others complex conjugate. An example is given in sec.\,\ref{sec:cohstate}.

\subsubsection{Coherent spin states}\label{app:cohspin}
Replacing the $a_\pm$ with spin-raising and lowering operators $J_\pm$, we can give a completely parallel discussion for coherent spin states. More details can be found in \cite{radcliffe_1971, Arecchi:1972td, VIEIRA1995331, stone_00}. 

Any given $z\in \IC$ defines an unnormalized coherent spin state as follows:
\begin{equation}\label{eq:cohspin}
    |\bar{z}) = \rme^{\bar{z}J_+}\ket{0}\,, \qquad (w|\bar{z}) = (1+ w\bar{z})^{2j} \,,
\end{equation}
where now $\ket{0}$ is the state annihilated by $J_-$ in the spin-$j$ representation of $\SU(2)$. It will also be convenient to introduce notation for the normalized counterpart of $|\bar z)$:
\begin{equation}
    \ket{\bar{z}} = \frac{|\bar{z})}{(1 + z\bar{z})^j}\, .
\end{equation}
Coherent spin states are again an overcomplete set of minimal uncertainty states, satisfying the following completeness relation:
\begin{equation}\label{eq:cohspin2}
     \int  \rmd^2z \,\mu(z,\bar{z}) \ket{\bar{z}}\bra{z} = \one\,, \qquad \mu(z,\bar{z}) = \frac{2j+1}{\pi (1+z\bar{z})^2}\,.
\end{equation}
As before, holomorphic wave functions are defined by the overlap
\begin{equation}
    \psi(z) = (z\ket{\psi}\,,
\end{equation}
such that $J_\pm$ and $J_3$ are represented by holomorphic differential operators:
\begin{equation}\label{eq:holo}
     (z| J_- \ket{\psi}=\partial_z \psi(z)\,, \quad  (z| J_+\ket{\psi}=(2zj - z^2 \partial_z)\psi(z)\,, \quad  (z| J_3\ket{\psi}=(z\partial_z - j )\psi(z)\,.
\end{equation}
From the inner product \eqref{eq:cohspin} it also follows that the harmonic oscillator coherent states $|u)_\text{\tiny{HO}}$ defined in \eqref{eq:harmcoh} are recovered in the large-$j$ limit by zooming in near the origin:
\begin{equation}\label{eq:spintoHOlimit}
    (z,w)= \tfrac{1}{\sqrt{2j}}(u,v) \quad \implies \quad (w|\bar{z}) = \big(1 + \tfrac{v\bar u}{2j}\big)^{2j} \;\;\xrightarrow{\makebox[0.8cm]{$\scriptscriptstyle{j\to \infty}$}} \;\;\rme^{v \bar{u}} = (v|\bar u)_\text{\tiny{HO}}\,. 
\end{equation}

\subsection{Large-spin emergent phase space}\label{app:symbols}
Following \cite{berezin}, let us begin by defining the symbol of an operator $A$:
\begin{equation}\label{eq:symbol}
    A(z,\bar{z}) = \frac{(z|A|\bar z)}{(z|\bar z)} =  \bra{z}A\ket{\bar{z}}\,.
\end{equation} 
It is the middle form which we will use to analytically continue to arguments where $z$ is not the complex conjugate of $\bar z$. As an example, the symbols of the spin matrices are given by
\begin{equation}
    \big(\,J_+ ,\, J_- ,\, J_3\,\big)(z,\bar{z}) = j\, \Big(\,\frac{2z}{1+z\bar{z}},\, \frac{2\bar{z}}{1+z\bar{z}},\, \frac{z\bar{z}-1}{1+z\bar{z}} \,\Big)\,.
\end{equation}
In the semiclassical limit, these coherent state expectation values become well-defined functions on phase space. The same thing happens in the large-$j$ limit. To see this, we need to look at
operator products. At the level of the symbols, these define a star product:
\begin{equation}
    (A\star B)(z, \bar{z}) = \bra{z}AB\ket{\bar{z}}\,. 
\end{equation}
Inserting the completeness relation \eqref{eq:cohspin2},  we then have
\begin{equation}
  (A\star B)(z, \bar{z})=  \int \rmd^2 w\, \mu(w,\bar{w}) |\braket{w}{\bar{z}}|^2 A(z,\bar{w})B(w,\bar{z})\,.
\end{equation}
At large $j$ the overlap $|\braket{w}{\bar{z}}|^2$ becomes exponentially peaked around $w\approx z$. This means that the operators will commute, up to $1/j$ corrections. In particular one finds:
\begin{equation}
    (A\star B)(z,\bar{z}) = A(z,\bar{z})B(z,\bar{z}) + \tfrac{1}{2j}(1+z\bar{z})^2\partial_{\bar{z}}A(z,\bar{z})\partial_z B(z,\bar{z})+ \dots.
\end{equation}
In the large-$j$ limit, commutators will therefore behave as Poisson brackets:
\begin{equation}\label{eq:poisson}
  [A,B]_\star = A\star B - B\star A  = \tfrac{1}{2j}(1+z\bar{z})^2(\partial_{\bar{z}}A\partial_z B -\partial_{z}A\partial_{\bar{z}} B)\, .
\end{equation}

In this limit we can then replace perturbative operators\footnote{These are defined as those operators which are sums and products of the spin operators $J_i$, but do not have more than $\sqrt{j}$ such factors, since otherwise commutators would grow large again. For instance, exponentials of spin operators are excluded as they involve arbitrarily many $J_i$ factors.} by their symbols, up to $1/j$ corrections. Given the $\star$-commutators \eqref{eq:poisson}, their time evolution becomes:
\begin{equation}\label{eq:symp}
    \dv{A}{t}  = \Omega^{\alpha \beta} \,\partial_\alpha A\, \partial_\beta H\,, \qquad \Omega = \frac{2\rmi j \hbar}{(1+z\bar{z})^2} \,\rmd z \wedge \rmd \bar{z}\,.
\end{equation}
This is the Hamiltonian flow for a function $A$ on the phase space with symplectic form $\Omega$. Looking at \eqref{eq:symp} we see that the phase space is compact and spherical -- we can think of it as the spin sphere. Moreover, taking $j \to \infty$ and $\hbar \to 0$ with $j\hbar$ fixed corresponds to the semiclassical limit at fixed phase space volume. The symplectic volume is $4\pi j \hbar$, correctly implying the existence of approximately $2j$ states at large $j$. In conclusion, we arrived at an emergent classical mechanics on the spin sphere.

Finally, it will be useful to introduce Darboux coordinates $u, \bar u$, via the relation
\begin{equation}\label{eq:darboux}
    z = \frac{u}{\sqrt{2j-u\bar{u}}}\,, \qquad \bar{z} = \frac{\bar{u}}{\sqrt{2j-u\bar{u}}}\,.
\end{equation}
In Darboux coordinates the symplectic form \eqref{eq:symp} is flat:
\begin{equation}
   \Omega = \rmi \hbar\, \rmd u \wedge \rmd \bar{u}\,.
\end{equation}
The symplectic volume is unaltered, since these coordinates are restricted to the disk $u\bar{u} < 2j$. Darboux coordinates also make the limit \eqref{eq:spintoHOlimit} more explicit. The relation between regular and spin coherent states can be understood more precisely as a group contraction \cite{Arecchi:1972td}.

\subsection{Periodic orbit on-shell action}\label{app:onshell}
 The on-shell action as function of the period $T$, for orbits of \eqref{eq:classSpin}, is given by the integral 
 \begin{equation}
     S_j(T) = \int^T_0 \rmd t\, (j \cos\varphi \,\dot\phi- H_j) = \int^T_0 \rmd t\, \Big(\, \frac{\nu - j\dot{\phi}}{\sin2\phi}\,\dot\phi - \omega\,\Big) 
 \end{equation}
In the regime of interest $j \gg \omega \gg \nu$, we can put $\nu = 0$. Then using \eqref{eq:phidotsq}, we end up with:
\begin{equation}\label{eq:Sonshell}
    S_j(T) = - 2 j \int^j_{2\omega} \rmd x \Big(\frac{x-2\omega}{x(j^2-x^2)}\Big)^\frac12- \omega\, T = j\,\Big(\tfrac{\omega}{j}\, T - 2(\tfrac{j}{\omega})^\frac12\,\text{Im}\big(\Pi(1+\tfrac{j}{2\omega},\tfrac{1}{2}+\tfrac{j}{4\omega})\big)\Big)\, ,
\end{equation}
in terms of the elliptic integral of the third kind $\Pi$. Since the period is a function only of $\frac{\omega}{j}$, $T(\omega) = 4(1+\tfrac{2\omega}{j})^{-\frac{1}{2}}\,\text{K}\big(\frac{j-2\omega}{j+2\omega}\big)$ \eqref{eq:period}, the on-shell action is of the form $S_j(T)= jS(T)$.

{\fontsize{12}{15}\selectfont
\bibliographystyle{jhep}
\bibliography{spin.bib}}

\end{document}